\documentclass{jfm}

\usepackage{amsmath}
\usepackage{graphicx}
\usepackage{natbib}
\usepackage{psfrag}
\usepackage{fixmath}

\allowdisplaybreaks

\ifCUPmtlplainloaded \else
  \checkfont{eurm10}
  \iffontfound
    \IfFileExists{upmath.sty}
      {\typeout{^^JFound AMS Euler Roman fonts on the system,
                   using the 'upmath' package.^^J}%
       \usepackage{upmath}}
      {\typeout{^^JFound AMS Euler Roman fonts on the system, but you
                   dont seem to have the}%
       \typeout{'upmath' package installed. JFM.cls can take advantage
                 of these fonts,^^Jif you use 'upmath' package.^^J}%
       \providecommand\upi{\pi}%
      }
  \else
    \providecommand\upi{\pi}%
  \fi
\fi

\ifCUPmtlplainloaded \else
  \checkfont{msam10}
  \iffontfound
    \IfFileExists{amssymb.sty}
      {\typeout{^^JFound AMS Symbol fonts on the system, using the
                'amssymb' package.^^J}%
       \usepackage{amssymb}%
       \let\le=\leqslant  \let\leq=\leqslant
         \let\geq=\geqslant
      }{}
  \fi
\fi

\ifCUPmtlplainloaded \else
  \IfFileExists{amsbsy.sty}
    {\typeout{^^JFound the 'amsbsy' package on the system, using it.^^J}%
     \usepackage{amsbsy}}
    {\providecommand\boldsymbol[1]{\mbox{\boldmath $##1$}}}
\fi

\newcommand{\figref}[1]{figure~\ref{#1}}
\newcommand{\figrefa}[1]{\figref{#1}\,({\it a})}
\newcommand{\figrefb}[1]{\figref{#1}\,({\it b})}

\newcommand{\taga}{\tag{\theequation\,{\it a}}}
\newcommand{\tagb}{\tag{\theequation\,{\it b}}}
\newcommand{\tagc}{\tag{\theequation\,{\it c}}}
\newcommand{\tagd}{\tag{\theequation\,{\it d}}}
\newcommand{\tage}{\tag{\theequation\,{\it e}}}
\newcommand{\tagf}{\tag{\theequation\,{\it f}}}
\newcommand{\tagg}{\tag{\theequation\,{\it g}}}
\newcommand{\tagh}{\tag{\theequation\,{\it h}}}
\newcommand{\tagi}{\tag{\theequation\,{\it i}}}
\newcommand{\tagab}{\tag{\theequation\,{\it a,\,b}}}
\newcommand{\tagcd}{\tag{\theequation\,{\it c,\,d}}}
\newcommand{\tagac}{\tag{\theequation\,{\it a--c}}}
\newcommand{\tagbc}{\tag{\theequation\,{\it b,\,c}}}
\newcommand{\tagde}{\tag{\theequation\,{\it d,\,e}}}
\newcommand{\tagef}{\tag{\theequation\,{\it e,\,f}}}
\newcommand{\eqrefa}[1]{(\ref{#1}\,{\it a})}
\newcommand{\eqrefb}[1]{(\ref{#1}\,{\it b})}
\newcommand{\eqrefc}[1]{(\ref{#1}\,{\it c})}
\newcommand{\eqrefd}[1]{(\ref{#1}\,{\it d})}
\newcommand{\eqrefe}[1]{(\ref{#1}\,{\it e})}
\newcommand{\eqrefi}[1]{(\ref{#1}\,{\it i})}
\newcommand{\eqrefac}[1]{(\ref{#1}\,{\it a--c})}
\newcommand{\eqrefab}[1]{(\ref{#1}\,{\it a,\,b})}

\newcommand{\eqrefef}[1]{(\ref{#1}\,{\it e,\,f})}

\newcommand{\homogu}{u^{(h)}}
\newcommand{\homogb}{b^{(h)}}
\newcommand{\partintu}{u^{(p)}}
\newcommand{\partintb}{b^{(p)}}
\newcommand{\colvec}[1]{\boldsymbol{#1}}
\newcommand{\mat}[1]{\mathsfbi{#1}}

\newcommand{\ran}{\operatorname{ran}} 
\newcommand{\sech}{\operatorname{sech}} 
\newcommand{\csch}{\operatorname{csch}} 

\newcommand\metre{\nobreak\mbox{$\;$m}}

\newcommand\second{\nobreak\mbox{$\;$s}}
\newcommand\newton{\nobreak\mbox{$\;$N}}

\providecommand\bnabla{\boldsymbol{\nabla}}

\newcommand\Real{\mbox{Re}} 
\newcommand\Imag{\mbox{Im}} 
\newcommand\Rey{\mbox{\textit{Re}}}  
\newcommand\Prm{\mbox{\textit{Pm}}} 
\newcommand\Fro{\mbox{\textit{Fr}}} 
\newcommand\Har{Ha} 
\newcommand\Harz{H_z}
\newcommand\Harx{H_x}

\newcommand\Reyc{\mbox{\textit{Re}$_c$}}
\newcommand\Reyb{\mbox{\textit{Re}$_b$}}
\newcommand\Reym{\mbox{\textit{Rm}}}

\newcommand\Reyinv{\mbox{\Rey$^{-1}$}}
\newcommand\Alf{\mbox{\textit{Al}}}

\newcommand\Gal{\mbox{\textit{Ga}}}
\newcommand\Capil{\mbox{\textit{Ca}}}

\newcommand\Web{\mbox{\textit{We}}}
\newcommand\alphac{\mbox{$\alpha_c$}}

\newsavebox{\astrutbox}
\sbox{\astrutbox}{\rule[-5pt]{0pt}{20pt}}

\newcommand\tti{\ensuremath{\rightarrow\infty}}

\newcommand\ttz{\ensuremath{\rightarrow 0}}

\newcommand\ord{\ensuremath{O}}
\newcommand\ii{\mathrm{i}}
\newcommand\ee{\mathrm{e}}
\newcommand\DD{\operatorname{D}}
\newcommand\dd{\mbox{d}}

\newcommand\eg{e.g.\ }
\newcommand\ie{i.e.\ }
\newcommand\cf{cf.\ }

\newcommand\dt{\ensuremath{\partial_t}}
\newcommand\dx{\ensuremath{\partial_x}}
\newcommand\dy{\ensuremath{\partial_y}}

\newcommand{\dotp}[2]{#1\boldsymbol{\cdot} #2}
\newcommand{\crossp}[2]{#1\boldsymbol{\times} #2}

\newcommand{\lapl}{\upDelta}
\newcommand{\curl}[1]{\crossp{\bnabla}{#1}}
\newcommand{\divr}[1]{\dotp{\bnabla}{#1}}

\newcommand\vx{\ensuremath{\boldsymbol{x}}}
\newcommand\vy{\ensuremath{\boldsymbol{y}}}
\newcommand\vz{\ensuremath{\boldsymbol{z}}}
\newcommand\vpos{\ensuremath{\boldsymbol{r}}} 

\newcommand{\vnorm}{\ensuremath{\boldsymbol{n}}}
\newcommand{\vtanx}{\ensuremath{\boldsymbol{t}^{(x)}}}
\newcommand{\vtany}{\ensuremath{\boldsymbol{t}^{(y)}}}

\newcommand{\altvb}{\skew3\bar{\boldsymbol{\mathcal{B}}}}
\newcommand{\vu}{\boldsymbol{\mathcal{U}}}
\newcommand{\vb}{\boldsymbol{\mathcal{B}}}
\newcommand{\vj}{\boldsymbol{\mathcal{J}}}
\newcommand{\ve}{\boldsymbol{\mathcal{E}}}
\newcommand{\vfl}{\boldsymbol{\mathcal{F}}}
\newcommand{\vbext}{\boldsymbol{\mathcal{B}}'}

\newcommand{\vg}{\boldsymbol{g}}
\newcommand{\phyd}{\mathcal{P}'}
\newcommand{\ptot}{\mathcal{P}} 
\newcommand{\stresst}{\mathcal{T}} 
\newcommand{\stresstext}{\mathcal{T}'} 
\newcommand{\straint}{\mathcal{S}} 

\newcommand{\basee}{\boldsymbol{E}}

\newcommand{\basej}{\boldsymbol{J}}
\newcommand{\baseu}{\boldsymbol{U}}
\newcommand{\baseb}{\boldsymbol{B}}
\newcommand{\basebext}{\boldsymbol{B}'}
\newcommand{\basebind}{\boldsymbol{I}}
\newcommand{\basebextunit}{\skew3\hat{\boldsymbol{B}}'}
\newcommand{\basep}{P}
\newcommand{\basebx}{B_x}
\newcommand{\baseby}{B_y}
\newcommand{\basebz}{B_z}
\newcommand{\basebextx}{B'_{x}}
\newcommand{\basebexty}{B'_{y}}
\newcommand{\basebextz}{B'_{z}}
\newcommand{\basebindx}{I_x}
\newcommand{\basebindy}{I_y}

\newcommand{\pgrad}{\varPi}

\newcommand{\pertu}{\boldsymbol{u}}
\newcommand{\pertb}{\boldsymbol{b}}

\newcommand{\pertp}{p}
\newcommand{\pertbext}{\boldsymbol{b}'}
\newcommand{\pertj}{\boldsymbol{j}}
\newcommand{\perte}{\boldsymbol{e}}
\newcommand{\pertfl}{\boldsymbol{f}}
\newcommand{\pertfj}{\boldsymbol{f}_J}

\newcommand{\pertux}{u_x}

\newcommand{\pertuz}{u_z}
\newcommand{\pertbx}{b_x}
\newcommand{\pertby}{b_y}
\newcommand{\pertbz}{b_z}

\newcommand{\bbind}{B}
\newcommand{\bux}{U}

\newcommand\modeux{\skew3\hat{u}_x }
\newcommand\modeuy{\skew3\hat{u}_y }
\newcommand\modeuz{\skew3\hat{u}_z }
\newcommand\modebx{\skew3\hat{b}_x }
\newcommand\modeby{\skew3\hat{b}_y }
\newcommand\modebz{\skew3\hat{b}_z }
\newcommand\modeui{\skew3\hat{u}_i}
\newcommand\modebi{\skew3\hat{b}_i}
\newcommand\modepsi{\skew3\hat{\psi}}
\newcommand\modep{\skew3\hat{p}}
\newcommand\modea{\skew3\hat{a}}
\newcommand\modeu{\skew3\hat{u}}
\newcommand\modeb{\skew3\hat{b}}
\newcommand\modebext{\skew3\hat{b}'}

\newcommand\squirealpha{\skew3\tilde{\alpha}}
\newcommand\squiregamma{\skew3\tilde{\gamma}}
\newcommand\squireU{\skew3\tilde{U}}
\newcommand\squireBx{\skew3\tilde{B}_x}
\newcommand\squireBz{\skew3\tilde{B}_z}
\newcommand\squireux{\skew3\tilde{u}_x}

\newcommand\squireuz{\skew3\tilde{u}_z}
\newcommand\squirebx{\skew3\tilde{b}_x}

\newcommand\squirebz{\skew3\tilde{b}_z}
\newcommand\squireui{\skew3\tilde{u}_i}
\newcommand\squirebi{\skew3\tilde{b}_i}
\newcommand\squireBi{\skew3\tilde{B}_i}
\newcommand\squirep{\skew3\tilde{p}}
\newcommand\squirea{\skew3\tilde{a}}
\newcommand\squirepsi{\skew3\tilde{\psi}}
\newcommand\squirerey{\skew3\widetilde{\Rey}}

\newcommand\squirecapil{\skew3\widetilde{\Capil}}
\newcommand\squireprm{\skew3\widetilde{\Prm}}
\newcommand\squiretheta{\skew3\tilde{\theta}}

\newcommand\squirereym{\skew3\widetilde{\Reym}}
\newcommand\squiregal{\skew3\widetilde{\Gal}}

\newcommand\squirefro{\skew3\widetilde{\Fro}}
\newcommand\squireweb{\skew3\widetilde{\Web}}
\newcommand\squirehar{\skew3\widetilde{\Har}}

\newcommand\modeeu{\skew3\hat{E}_u}
\newcommand\modeeb{\skew3\hat{E}_b}

\newcommand\modegr{\skew3\hat{g}_R}
\newcommand\modegm{\skew3\hat{g}_M}
\newcommand\modegj{\skew3\hat{g}_J}
\newcommand\modegnu{\skew3\hat{g}_\nu}
\newcommand\modegeta{\skew3\hat{g}_\eta}

\newcommand{\dmn}{\Omega}
\newcommand{\dmnplus}{{\Omega_+}}
\newcommand{\dmnminus}{{\Omega_-}}
\newcommand{\dmns}{{\partial\Omega_s}}
\newcommand{\zw}{z_w} 

\title[Large-wavelength instabilities in free-surface Hartmann flow]{Large-wavelength instabilities in free-surface Hartmann flow at low magnetic Prandtl numbers}
\author[D. Giannakis, R. Rosner and P. F. Fischer]%
{D.\ns G\ls I\ls A\ls N\ls N\ls A\ls K\ls I\ls S$^1$,\ns R.\ns  R\ls O\ls S\ls N\ls E\ls R$^{1,2,3}$ \and P.\ns F.\ns F\ls I\ls S\ls C\ls H\ls E\ls R$^3$}

\affiliation{$^1$Department of Physics, University of Chicago, Chicago, IL 60637, USA\\[\affilskip]
$^2$Department of Astronomy and Astrophysics, University of Chicago, Chicago, IL 60637, USA\\[\affilskip]
$^3$Argonne National Laboratory, Argonne, IL 60439, USA}

\pubyear{1996}
\volume{538}
\pagerange{119--126}
\date{?? and in revised form ??}

\begin{document}

\maketitle

\begin{abstract}
We study the linear stability of the flow of a viscous electrically conducting capillary fluid on a planar fixed plate in the presence of gravity and a uniform magnetic field, assuming that the plate is either a perfect electrical insulator or a perfect conductor. We first confirm that the Squire transformation for magnetohydrodynamics is compatible with the stress and insulating boundary conditions at the free surface, but argue that unless the flow is driven at fixed Galilei and capillary numbers, respectively parameterising gravity and surface tension, the critical mode is not necessarily two-dimensional. We then investigate numerically how a flow-normal magnetic field, and the associated Hartmann steady state, affect the soft and hard instability modes of free-surface flow, working in the low-magnetic-Prandtl-number regime of laboratory fluids ($ \Prm \leq 10^{-4} $). Because it is a critical-layer instability (moderately modified by the presence of the free surface), the hard mode is found to exhibit similar behaviour to the even unstable mode in channel Hartmann flow, in terms of both the weak influence of $ \Prm $ on its neutral-stability curve, and the dependence of its critical Reynolds number $ \Reyc $ on the Hartmann number $ \Har $. In contrast, the structure of the soft mode's growth-rate contours in the $(\Rey, \alpha)$ plane, where $ \alpha $ is the wavenumber, differs markedly between problems with small, but nonzero, $ \Prm $, and their counterparts in the inductionless limit. As derived from large-wavelength approximations, and confirmed numerically, the soft mode's critical Reynolds number grows exponentially with $ \Har $ in inductionless problems. However, when $ \Prm $ is nonzero the Lorentz force originating from the steady-state current leads to a modification of $ \Reyc( \Har ) $ to either a sublinearly increasing, or decreasing function of $ \Har $, respectively for problems with insulating and conducting walls. In the former, we also observe pairs of counter-propagating Alfv\'en waves, the upstream-propagating wave undergoing an instability driven by energy transfered from the steady-state shear to both of the velocity and magnetic degrees of freedom. Movies are available with the online version of the paper.    
\end{abstract}

\section{\label{sec:introduction}Introduction}

Free-surface shear magnetohydrodynamic (MHD) flows arise in a variety of industrial and astrophysical contexts, including liquid-metal blankets in fusion devices \cite[][]{AbdouEtAl01,Buhler07}, liquid-metal forced-flow targets \cite[][]{ShannonEtAl98}, plasma oceans on white dwarfs and neutron stars \cite[][]{BildstenCutler95,AlexakisEtAl02}, and accretion discs around stellar remnants \cite[][]{RudigerEtAl99,BalbusHenri08}. Flows of this type typically take place at high Reynolds number $\Rey \gtrsim 10^4$ and within strong background magnetic fields ($\Har \gtrsim 10^2$, where $ \Har $ is the Hartmann number). Broadly speaking, one is interested in characterising their stability properties either because of engineering requirements (\eg in a fusion-device blanket), or a free-surface instability may be involved in the observed phenomena (such as classical novae and neutron star X-ray bursts). 

When the magnetic Prandtl number $ \Prm $ of the working fluid is small, the effect of an external magnetic field is generally known to be stabilising and weakly dependent on $ \Prm $ \cite[][and references therein]{MullerBuhler01}. In particular, \cite{Takashima96} has numerically studied the stability of plane Poiseuille flow modified by a flow-normal magnetic field, hereafter called \emph{channel Hartmann flow}, under insulating boundary conditions, and has determined that the critical Reynolds number $ \Reyc $ for instability increases monotonically with the Hartmann number $ \Har $. Moreover, for $ \Prm \leq 10^{-4} $ (an interval encompassing all known laboratory fluids)  $ \Reyc $ was found to experience a mild, $ \ord( 10^{-3} ) $, decrease compared to its value in the inductionless limit $ \Prm \searrow 0 $.

In the present work, we pursue a similar stability analysis for \emph{free-surface Hartmann flow}, \ie the parallel steady flow of a viscous electrically conducting capillary fluid on an inclined planar surface (assumed to be either a perfect electrical conductor, or a perfect insulator), subject to gravity and a flow-normal external magnetic field. Our main result is that for large wavelengths ($ \alpha \lesssim 1 $, where $ \alpha $ is the modal wavenumber), the spectrum of free-surface Hartmann flow contains two types of normal modes, neither of which is present in channel problems, and whose stability depends strongly on $ \Prm $, even in the $ \Prm = \ord( 10^{-5} ) $ regime of liquid metals. 

The first of these modes is related to the unstable gravity wave, oftentimes referred to as the \emph{soft instability mode}, encountered in free-surface Poiseuille flow \cite[][]{Yih63,Yih69,LamBayazitoglu86,FloryanDavisKelly87}. In inductionless problems, the gravity wave becomes stabilised when $ \Har $ is increased, but in the strong-field limit its growth rate $ \Gamma $ does not obey the characteristic $ | \Gamma | \propto \Har^2 $ Lorentz damping of the shear modes in the spectrum. Instead, it transitions to an asymptotically neutral phase, where the Lorentz force nearly balances gravity, and the decay rate $ | \Gamma | \propto \Har^{-2} $ decreases towards zero. In problems with nonzero, yet small, $ \Prm $, Lorentz forces originating from the Hartmann current profile are found to be sufficient to alter the near-equilibrium attained in the inductionless limit, leading to high sensitivity of the mode's stability contours in the $ (\Rey, \alpha )$ plane to the magnetic Prandtl number. 

The second type of modes in question is travelling Alfv\'en waves, which we have only encountered in flows with an insulating lower wall. When the background fluid is at rest, these modes appear at sufficiently large Hartmann numbers as a pair of counter-propagating waves, whose phase speed and decay rate increase linearly with $ \Har $. Their kinetic and magnetic energies are nearly equal, but in the large-wavelength cases studied here the majority of the magnetic energy is carried by the magnetic field penetrating into the region exterior to the fluid. When a steady-state flow is established, the upstream propagating Alfv\'en mode becomes unstable due to positive Reynolds and Maxwell stresses as the Alfv\'en number $ \Alf $ is increased. 
 
Aside from instabilities associated with gravity and Alfv\'en waves, free-surface flow also exhibits an instability of the critical-layer type (modified by the presence of the free surface), called the \emph{hard instability} \cite[][]{Lin67,DeBruin74,FloryanDavisKelly87}. Sharing a common origin with the even unstable mode in channel Hartmann flow \cite[][]{Lock55,PotterKutchey73,Takashima96}, the critical parameters of the hard mode at small $ \Prm $ are close to the corresponding ones in the inductionless limit. However, in light of the presence of gravity and Alfv\'en waves, our analysis suggests that the inductionless approximation must be used with caution when dealing with free-surface MHD. 

The plan of this paper is as follows. In \S\ref{sec:problemFormulation} we formulate the governing equations and boundary conditions of our stability problems, and discuss the validity of the Squire transformation. In \S\ref{sec:energyBalance} we derive an energy-conservation law for temporal normal modes in free-surface MHD. We present our results in \S\ref{sec:discussion}, and conclude in \S\ref{sec:conclusions}. Appendix~\ref{app:pertTheory} contains a discussion of large-wavelength ($\alpha \searrow 0$) perturbation theory. Movies illustrating the behaviour of the modal eigenvalues on the complex plane as $ \Har $ or $ \Prm $ are varied are available with the online version of the paper. 
 
\section{\label{sec:problemFormulation}Problem formulation}

\subsection{\label{sec:geometry}Geometrical configuration}

\begin{figure}
\psfrag{z}{\small{$z$}}
\psfrag{0}{\small{$0$}}
\psfrag{x}{\small{$x$}}
\psfrag{U(z)}{\small{$U(z)$}}
\psfrag{B(z)}{\small{$B(z)$}}
\psfrag{z=l}{\small{$z=l$}}
\psfrag{z=-l}{\small{$z=-l$}}
\psfrag{z=0}{\small{$z=0$}}
\psfrag{a}{\small{$a(x,y,t)$}}
\psfrag{theta}{\small{$\theta$}}
\begin{center}
\includegraphics{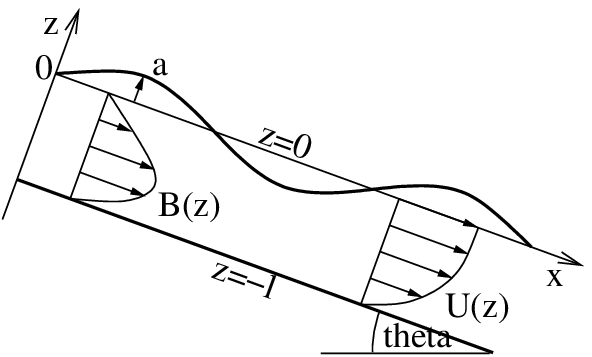}
\hfill
\includegraphics{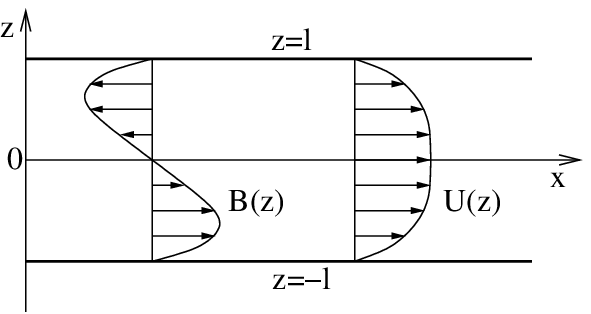} 
\caption{\label{fig:modelGeometry}Geometry of free-surface (left) and channel (right) Hartmann flow. The steady-state velocity and induced magnetic field profiles are $ \bux( z ) $ and $ \bbind( z ) $, respectively (see \S\ref{sec:steadyState}). The $ y $ axis is directed into the plane of the paper.}
\end{center}
\end{figure}

Using $ x $, $ y $, and $ z $ to respectively denote the streamwise, spanwise, and flow-normal coordinates, oftentimes collected in the position vector $ \vpos := ( x, y, z ) $, and $ t $ to denote time, we consider the flow geometries shown in figure~\ref{fig:modelGeometry}. In free-surface problems the lower planar surface $ z= - l $ is inclined at an angle $ \theta $ with respect to the horizontal, and the upper fluid boundary has oscillation amplitude $ z = a( x, y, t ) $, with $ a = 0 $ in the steady-state, while in channel problems the flow takes place between two fixed parallel plates located at $ z = \pm l $. In both cases, the working fluid is incompressible, and has density $ \rho $, dynamic viscosity $ \mu $, and electrical conductivity $ \lambda $. Additionally, the free surface has surface tension $ \sigma $, and is also acted upon by a gravitational field $ \vg := g ( \sin(\theta) \vx - \cos( \theta ) \vz ) $. The fixed plates are treated either as perfect electrical insulators or perfect electrical conductors.
   
For future convenience we introduce the function $ A( \vpos, t ) := z - a( x, y, t ) $, which vanishes on the free surface, and leads, through its gradient, to the expression    
\begin{equation}
\label{eq:freeSurfaceNormal}
\vnorm := \bnabla A / || \bnabla A || = ( -\dx a, -\dy a, 1 ) + \ord( a^2 )
\end{equation}
for the free-surface outward unit normal (for our purposes it suffices to work at linear order in $ a $). Moreover, we choose $\vtanx := ( 1, 0, \dx a ) $ and  $ \vtany := ( 0, 1, \dx a ) $ as mutually orthogonal unit vectors tangent to the free surface ($ \dotp{ \vtanx }{ \vnorm } = \dotp{ \vtany }{ \vnorm } = \dotp{ \vtanx }{ \vtany } = \ord( a^2 ) $). The divergence of $ \vnorm $ is equal to twice the mean surface curvature $ \kappa $, for which we compute 
\begin{equation}
\label{eq:freeSurfaceCurvature}
2 \kappa = \divr{ \vnorm } = - ( \dx^2 a + \dy^2 a ) + \ord( a^2 ).
\end{equation} 
   
\subsection{\label{sec:governingEquations}Governing equations}

Our starting point is the equations for incompressible resistive MHD \cite[e.g.][]{MullerBuhler01},
\begin{equation}
\label{eq:goveqs}
\begin{gathered}
\dt \vu + \dotp{ \vu }{ \bnabla } \vu 
= -\bnabla \phyd + \vfl + \Rey^{-1} \lapl \vu, \\
\vfl := \Reym \crossp{ \vj }{ \vb }, \quad \vj := \Reym^{-1} \curl{ \vb } = \ve + \crossp{ \vu }{ \vb }, \\
\dt \vb + \dotp{ \vu }{ \bnabla } \vb = \dotp{ \vb }{ \bnabla } \vu + \Reym^{-1} \lapl \vb, \\
\quad \divr{ \vu } = 0, \quad \divr{ \vb } = 0,
\end{gathered}
\end{equation}
obeyed by the velocity field $ \vu( \vpos, t ) $, the hydrodynamic pressure $ \phyd( \vpos, t ) $, the Lorentz force $ \vfl( \vpos, t ) $, the current $ \vj( \vpos, t ) $, and the magnetic and electric fields in the interior of the fluid, respectively $ \vb( \vpos, t ) $ and $ \ve( \vpos, t ) $. Here velocity has been non-dimensionalised by its steady-state value at $ z = 0 $, $ U_* $, and the characteristic values for the remaining dynamical variables are $ P_* := \rho U_*^{2} $, $ B_* := (\mu_0 \rho)^{1/2} U_* $, $ E_* := U_* B_* $, and $ J_* := \lambda E_* $, where each *-subscripted symbol denotes the characteristic value for the corresponding variable in script type. Choosing $ l $ as the characteristic length (for both free-surface and channel problems), the resulting hydrodynamic and magnetic Reynolds numbers are $ \Rey := U_* l / \nu $ and $ \Reym := U_* l / \eta $, where $ \nu := \mu / \rho $ and $ \eta := 1/(\mu_0 \lambda) $ are the viscous and magnetic diffusivities. In the following, we frequently substitute for $ \Reym $ using the magnetic Prandtl number $ \Prm := \nu / \eta = \Reym / \Rey $.

We consider solutions of the form 
\begin{equation}
\label{eq:decomp}
\begin{gathered}
\vu( \vpos, t ) = \baseu( z ) + \pertu( \vpos, t ), \quad 
\phyd( \vpos, t ) = P'( x, z ) + p'( \vpos, t ), \\
\vb( \vpos, t ) = \baseb( z ) + \pertb( \vpos, t ), \quad \vj( \vpos, t ) = \basej( z ) + \pertj( \vpos, t ), \quad \ve( \vpos, t ) = \basee( z ) + \perte( \vpos, t ),
\end{gathered}
\end{equation}
consisting of steady-state components and linear perturbations, respectively denoted by uppercase and lowercase symbols. The steady-state flow $ \baseu( z ) := ( U( z ), 0, 0 ) $ is assumed to be streamwise-invariant and unidirectional, and to take place within a uniform, externally applied magnetic field $ \basebext := ( \basebextx, \basebexty, \basebextz ) $, which, for the time being, is allowed to be of arbitrary direction. The applied field permeates the fluid, and assuming its components perpendicular to $ \baseu $ are non-zero, the associated induced current generates a secondary internal magnetic field $ \basebind := ( \basebindx( z ), \basebindy( z ), 0 ) $. Thus, in the interior of the fluid the steady-state magnetic field is $ \baseb := \basebext + \basebind $. We remark that since pressure affects the dynamics only through its gradient, we have allowed $  P' $ to depend on the streamwise coordinate $ x $, but, in light of the streamwise-invariance of the steady state, that dependence can be at most linear. Moreover, the flow-normal component of the induced magnetic field has been set to zero in order to meet the divergence-free condition $ \divr{ \basebind } = 0 $ (a constant nonzero $ I_z $ can be absorbed in $ \basebextz $). We also note that with our choice of characteristic magnetic field,  $ \vu $ and $ \vb $ are naturally additive. In particular, using $ \basebextunit $ and $ B'_* $ to respectively denote a unit vector in the direction of $ \basebext $ and the magnitude of the dimensionful external magnetic field, $ \basebext $ can be expressed as  $ \basebext = \basebextunit / \Alf $, where $ \Alf := U_* ( \mu_0 \rho )^{1/2} / B'_* $ is the Alfv\'en number of the flow (see \eg \cite{Shercliff65} for an overview of dimensionless groups in MHD).  An alternative option for non-dimensionalisation, frequently encountered in the literature for Hartmann flow \cite[e.g.][]{Takashima96,Takashima98,MullerBuhler01,Buhler07}, is to set $ B_* = B'_* $, in which case the resulting dimensionless magnetic field $ \altvb $ is related to ours via $ \altvb = \Alf \, \vb $. In the ensuing analysis, we mainly parameterise the background magnetic field strength by means of the \emph{Hartmann number} $ \Har := B'_* l ( \lambda / \mu )^{1/2} = ( \Rey \Reym ) ^{1/2} / \Alf $, where $ \Har ^ 2 $ measures the ratio of Lorentz to viscous stresses, rather than $ \Alf $.

In problems with insulating boundaries, a further dynamical variable is the magnetic field $ \vbext( \vpos, t ) := \basebext + \pertbext( \vpos, t ) $ in the region exterior to the fluid. As follows from Amp\`ere's law,  $ \pertbext $ is expressible as the gradient of the magnetic potential $ \psi( \vpos, t ) $, which, in light of the solenoidal condition $ \divr{ \pertbext } = 0 $, obeys Laplace's equation; \ie
\begin{subequations}
\label{eq:goveqsExt}
\begin{equation}
\tagab
\pertbext = -\bnabla \psi, \quad \text{with} \quad \lapl \psi = 0.
\end{equation} 
\end{subequations}

The equations governing the steady state and the perturbations follow by substituting~\eqref{eq:decomp} into~\eqref{eq:goveqs}, and neglecting quadratic terms in the perturbed fields. Using $ \DD $ to denote differentiation with respect to $ z $, the nonzero components of the time-independent equations read 
\begin{subequations}
\label{eq:baseGovEqs}
\begin{gather}
\tagac
\Rey^{-1} \DD^2 U + \basebextz \, \DD \basebindx - \dx P = 0, \quad \Reym^{-1} \DD^2 \basebindx + \basebextz \DD U = 0, \quad \DD P = 0, \\
\tagde
\basebextz \DD \basebindy = 0, \quad \DD^ 2 \basebindy = 0,
\end{gather}
\end{subequations}
where $ P(x,z) := P'(x,z) + \dotp{ \baseb( z ) }{\baseb( z ) } / 2 $ is the total steady-state pressure, consisting of hydrodynamic and magnetic contributions. As for the perturbations, we obtain
\begin{subequations}
\label{eq:linearizedGovEqs}
\begin{gather}
\taga
\dt \pertu + U \dx \pertu + \pertuz \DD U \vx = - \bnabla \pertp' + \pertfl + \Rey^{-1} \lapl \pertu, \\
\tagbc
\pertfl := \Reym ( \crossp{ \pertj }{ \baseb } + \crossp{ \basej }{ \pertb } ), \quad \pertj =  \Reym^{-1} \curl{ \pertb } = \perte - U \pertbz \vy + \crossp{ \pertu }{ \baseb }, \\
\tagd
\dt \pertb + U \dx \pertb + \pertuz \DD \baseb  = \dotp{ \baseb }{ \bnabla \pertu } + \pertbz \DD U \vx + \Reym^{-1} \lapl \pertb, \\
\tagef
\divr{ \pertu } = 0, \quad \divr{ \pertb } = 0
\end{gather}
\end{subequations}
where $ \pertfl $ is the Lorentz force acting on the perturbed velocity field. 

From the linearised Amp\`ere and Ohm laws~\eqrefc{eq:linearizedGovEqs}, one can make the heuristic estimate $ || \pertb || / || \pertu || = \ord( ( \Prm \Har )^{1/2} ) $ (see \cite{Hunt66} for similar scaling arguments), where $ || \cdot || $ denotes suitable norms for the velocity and magnetic field perturbations. This suggests that as $ \Prm \searrow 0 $ with $ \Rey $ and $ \Har $ fixed, \ie in the so-called \emph{inductionless limit} \cite[][]{MullerBuhler01}, $ \pertb $ is negligible and, in consequence, electromagnetic forces only arise from currents generated by the perturbed electric field $ \perte $, and from currents induced by the perturbed fluid motions within the steady-state magnetic field. Moreover, as follows from Faraday's law, $ \curl{ \perte } = -\dt \pertb \approx \boldsymbol{0} $, the electric field can be determined from the gradient of a potential $ \phi( \vpos, t ) $. These observations suggest that for sufficiently small $ \Prm $, (\ref{eq:linearizedGovEqs}\,{\it b--d,\,f}) can be replaced by the approximate relations
\begin{subequations}
\label{eq:zeroPm}
\begin{equation}
\tagac
\pertfl = \Har^2 \Rey^{-1} \crossp{ ( \perte + \crossp{ \pertu }{ \basebextunit } ) }{ \basebextunit }, \quad \perte = - \bnabla \phi, \quad \lapl \phi = 0.
\end{equation}
\end{subequations}
If, in addition, the flow is two-dimensional, $ \phi $ must be set to zero (as $ \phi \neq 0 $ would give rise to spanwise forces), resulting in a significant reduction of the analytical and computational complexity of the stability problem \cite[][]{Stuart54}. The fact that all known conducting laboratory fluids have small magnetic Prandtl numbers ($ \Prm \lesssim 10^{-5} $) has led to a widespread adoption of the inductionless approximation. However, the small-$ || \pertb || $ assumption is not guaranteed to hold a priori, and the full problem must be solved to confirm that the scheme is valid in the parameter regime of interest \cite[][]{Takashima96}.

\subsection{\label{sec:boundaryConditions}Boundary conditions}

The governing equations presented in the preceding section must be solved subject to appropriate initial and boundary conditions. In the temporal stability analysis that follows initial conditions, as well as periodic boundary conditions on the streamwise and spanwise domain boundaries, are implicitly assumed. However, care must be taken in the choice of boundary conditions on the non-periodic boundaries, as this has led to errors in the past (\cite{Lin67} and \cite{PotterKutchey73}, as indicated by \cite{DeBruin74} and \cite{Takashima96}, respectively).

Let $ \zw $ collectively denote the flow-normal wall coordinates (in the dimensionless representation, $ \zw := \{-1\} $ for free-surface problems and $ \zw := \{\pm 1\} $ for channel problems). In insulating-wall problems we assume that no surface charges and surface currents are present at the fluid--wall interface. Then, in accordance with Maxwell's equations and charge--current conservation \cite[e.g.][]{Shercliff65}, we set  
\begin{equation}
\label{eq:insulatingWall}
\vb |_{ z = \zw} = \vbext|_{z=\zw }, \quad \dotp{ \vz }{ \vj }|_{z=\zw} = 0,
\end{equation}
which leads to the boundary conditions
\begin{subequations}
\label{eq:insulatingWallBase}
\begin{equation}
\tagab
\basebindx( \zw ) = 0, \quad \basebindy( \zw ) = 0
\end{equation}
\end{subequations}
and
\begin{subequations}
\label{eq:insulatingWallPert}
\begin{equation}
\tagab
( \pertb + \bnabla \psi )|_{ z = \zw  } = \boldsymbol{ 0 }, \quad \dotp{ \vz }{ \curl{ \pertb } }|_{ z = \zw } = ( \dx \pertby - \dy \pertbx )|_{ z = \zw }= 0,
\end{equation}
\end{subequations}
respectively for the steady-state fields and the perturbations. If, on the other hand, the wall is conducting, the tangential electric-field components are required to vanish at the boundary, and the wall-normal magnetic field $ \dotp{ \vz }{ \vb } $ is set to the externally imposed value $ B'_z $, giving
\begin{subequations}
\label{eq:conductingWallBase}
\begin{equation}
\tagab
\DD \basebindx( \zw ) = 0, \quad \DD \basebindy( \zw ) = 0,
\end{equation}
\end{subequations}
and
\begin{equation}
\label{eq:conductingWallPert}
\begin{gathered}
\pertbz |_{ z = \zw } = 0, \quad
\dotp{ \vx }{ \curl{ \pertb } }|_{ z = \zw } = ( \dy \pertbz - \DD \pertby )|_{z=\zw } = 0, \\ 
\dotp{ \vy }{ \curl{ \pertb } }|_{z = \zw } = ( \DD \pertbx - \dx \pertbz )|_{z=\zw } = 0,
\end{gathered}     
\end{equation}
where we have used~\eqrefc{eq:linearizedGovEqs} to substitute for the electric field $ \perte $ in terms of $ \pertb $. Turning now to the free surface, we assume throughout that the exterior region $ z > a $ is electrically insulating. Then, on the basis of similar electrodynamic arguments as those used to write down~\eqref{eq:insulatingWall}, we demand
\begin{equation}
\label{eq:insulatingSurf}
\dotp{ \vtanx } ( \vb - \vbext )|_{ z = a } = \dotp{ \vtany } ( \vb - \vbext )|_{ z = a } = \dotp{ \vnorm } ( \vb - \vbext )|_{ z = a } = 0, \quad
\dotp{ \vnorm }{ \curl{ \vj } }|_{ z = a } = 0. 
\end{equation}
In order to evaluate the expressions above for the perturbed quantities, we analytically continue the induced magnetic field $ \basebind $ in the region $ 0 < z \le a $. A Taylor expansion to linear order in $ a $ then yields the boundary conditions
\begin{subequations}
\label{eq:insulatingSurfBase}
\begin{equation}
\tagab
\basebindx( 0 ) = 0, \quad \basebindy( 0 ) = 0,
\end{equation}
\end{subequations}
and
\begin{equation}
\label{eq:insulatingSurfPert}
\begin{gathered}
( \dx \psi + \pertbx )|_{ z = 0 } + \DD \basebx( 0 ) \dx a = 0, \quad
( \dy \psi + \pertby )|_{ z = 0 } + \DD \baseby( 0 ) \dy a = 0, \\
( \DD \psi + \pertbz ) |_{ z= 0 } = 0, \quad
( \dx \pertby - \dy \pertbx )|_{ z = 0 } + \DD \baseby( 0 ) \dx a - \DD \basebx( 0 ) \dy a = 0, 
\end{gathered}
\end{equation}
which now involve the free-surface amplitude (\cf \eqref{eq:insulatingWallPert}). The requirement that, whenever present, the external magnetic field perturbations vanish at infinity completes the specification of boundary conditions for the magnetic field.

Regarding the velocity field, we impose, as usual, no-slip conditions  
\begin{subequations}
\label{eq:noSlip}
\begin{equation}
\tagab
U( \zw ) = 0, \quad \pertu|_{ z=\zw } = \boldsymbol{ 0 },
\end{equation}
\end{subequations}
at the solid walls, and consider the kinematics and stress balance to establish boundary conditions at the free surface \cite[see e.g.][for a discussion on interfacial dynamics]{Batchelor67}. First, noting that the free surface is, by definition, a material surface, leads to the kinematic boundary condition,  $ \dd A / \dd t := ( \dt + \dotp{ \vu|_{z=a} }{ \bnabla } ) A = 0 $, which, upon linearisation, becomes
\begin{equation}
\label{eq:kinematic}
\dt a + U( 0 ) \dx a = \pertuz |_{z = 0 }.
\end{equation}
In order to formulate the stress conditions, we introduce the stress tensors in the fluid and exterior domains, whose components in the $ ( x, y, z ) $ coordinate system are given by
\begin{equation}
\label{eq:tau}
\stresst_{ i j } := - \ptot \delta_{ij} + \mathcal{ B }_i \mathcal{ B }_j + \Rey^{-1 } \straint_{ i j }, \quad
\stresstext_{ i j } := - ( \varPhi + \mathcal{ P }_{ B' } ) \delta_{ i j } + \mathcal{ B }'_i \mathcal{ B }'_j, 
\end{equation}
respectively. Here $ \ptot := \phyd + \dotp{ \vb }{ \vb } / 2 $ is the resultant of the hydrodynamic and magnetic pressures,  $ \straint_{ i j } := \partial_i \mathcal{ U }_j + \partial_j \mathcal{ U }_i $ are the components of the rate-of-strain tensor, $ \mathcal{P}_{ B' }( \vpos, t )  := \dotp{ \vbext }{ \vbext } / 2 $ is the external magnetic pressure, and  $ \varPhi( \vpos ) := ( -x \sin \theta + z \cos \theta )  / \Fro^2 $ is the gravitational potential, expressed in terms of the Froude number $ \Fro := U_* / ( g l )^{1/2} $. Using  $ \Web :=  \rho l U_*^2 / \sigma $ to denote the Weber number, the free-surface curvature $ \kappa $~\eqref{eq:freeSurfaceCurvature}, in conjunction with surface tension, introduces a discontinuity $ \Sigma := 2 \kappa / \Web $ in the normal stress, such that
\begin{equation}
\label{eq:stressJump}
n_j ( \stresstext_{ i j } - \stresst_{ i j } )|_{ z = a } = \Sigma n_i,
\end{equation}
where $ n_i $ are the components of the normal vector $ \vnorm $~\eqref{eq:freeSurfaceNormal}, and summation is assumed over repeated indices. Forming the contraction of~\eqref{eq:stressJump} with the orthonormal vectors $ \vnorm $, $ \vtanx $, and $ \vtany $ then leads to three stress conditions that we enforce at the free surface, namely the normal-stress boundary condition
\begin{equation}
\label{eq:normalStressBC}
n_i n_j ( \stresstext_{ i j } - \stresst_{ i j } ) |_{ z = a } = \Sigma, 
\end{equation}
and the shear-stress conditions
\begin{equation}
\label{eq:shearStressBC}
t^{(x)}_i n_j ( \stresstext_{ i j } - \stresst_{ i j } ) |_{ z = a } = t^{ ( y ) }_i n_j ( \stresstext_{ i j } - \stresst_{ i j } ) |_{ z = a } = 0.
\end{equation} 
Evaluating \eqref{eq:normalStressBC} and \eqref{eq:shearStressBC} to linear order in the perturbed quantities, and eliminating $ b'_i |_{z=a} $ using~\eqref{eq:insulatingSurfPert} yields
\begin{subequations}
\label{eq:stressBCBase}
\begin{equation}
\tagab
\basep( x, 0 ) = - \sin(\theta) \Fro ^{-2} x + \dotp{ \baseb( 0 ) }{ \baseb( 0 ) } /2, \quad
\DD U( 0 ) = 0,
\end{equation}
\end{subequations}
and
\begin{equation}
\label{eq:stressBCPert}
\begin{gathered}
\begin{aligned}
( p - p_B - 2 \Rey^{-1} \DD u_z )|_{z=0} &= ( \cos(\theta) \Fro^{-2} + \dotp{ \baseb( 0 ) }{ \DD \baseb( 0 ) } ) a \\
& \quad - 2 \Rey^{-1} \DD U( 0 )  \dx a - \Web^{-1} ( \dx^2 a + \dy^2 a ), 
\end{aligned}\\
( \dx  u_z + \DD u_x )|_{ z = 0 } = -\DD^2 U( 0 )  a , \quad ( \dy u_z + \DD u_y )|_{ z= 0 } = 0,
\end{gathered}
\end{equation}
where $ p_B := \dotp{ \baseb }{ \pertb } $ is the internal magnetic-pressure perturbation. 

Besides $ \Web $ and $ \Fro $, alternative dimensionless groups for the capillary and normal gravitational forces are the \emph{capillary number} $ \Capil := \mu U_* / \sigma = \Web / \Rey $ and the \emph{Galilei number} $ \Gal := g l^3 \cos\theta / \nu^2 = \Rey^2 \cos( \theta )/ \Fro^2 $. As we will see in \S\ref{sec:3dNormalModes}, unlike $ \Web $, $ \Fro $ and $ \theta $, the parameters $ \Gal $ and $ \Capil $ are invariant under the Squire transformation from three-dimensional to two-dimensional normal modes, and for this reason we have opted to perform the calculations in \S\ref{sec:discussion} working in the latter representation.

\subsection{\label{sec:steadyState}Steady-state configuration}

In the present linear-stability analysis we employ the physically motivated Hartmann velocity and magnetic field profiles, which are solutions to~\eqrefac{eq:baseGovEqs} (in anticipation of the Squire transformation in \S\ref{sec:3dNormalModes}, we do not explicitly consider the spanwise induced magnetic field $ I_y $). For convenience we make the substitutions  $ \basebindx( z ) = B'_z \Reym B( z ) = \Prm^{1/2} \Harz B( z ) $ and, taking into account~\eqrefc{eq:baseGovEqs}, $ \basep = - \pgrad x + P_0 $, where $ \Harz := ( \Rey \Reym )^{1/2} B'_z $, $ \pgrad $, and $ P_0 $ are respectively a Hartmann number defined in terms of the flow-normal component of the applied magnetic field, a streamwise pressure gradient, and an unimportant constant. Equations~\eqrefab{eq:baseGovEqs} then become
\begin{equation}
\label{eq:baseX2}
\DD^2 U + \Harz^2 \DD B + \Rey \pgrad = 0, \quad \DD^2 B + \DD U = 0,
\end{equation}
the general solution of which can be expressed as
\begin{subequations}
\label{eq:generalUB}
\begin{align}
\taga
U( z ) & = C_0 + C_1 \frac{ \sinh( \Harz z ) }{ \Harz } + C_2 \frac{ \cosh( \Harz ) - \cosh( \Harz z ) }{ \cosh( \Harz ) - 1 }, \\
\tagb
B( z ) & = K_0 + K_1 z + C_1 \frac{ 1 - \cosh( \Harz z ) }{ \Harz ^ 2 } + C_2 \frac{ \sinh( \Harz z ) - \sinh( \Harz ) z }{ \Harz( \cosh( \Harz ) - 1 ) },
\end{align}
\end{subequations}
where $ C_0 $, $ C_1 $, $ C_2 $, $ K_0 $, and $ K_1 $ are constants such that
\begin{equation}
\label{eq:Pi}
\Pi = ( C_2 \Harz \coth( \Harz / 2 ) - \Harz ^2 K_1 ) / \Rey.
\end{equation}
We remark that the terms proportional to $ C_1 $ correspond to antisymmetric velocity and symmetric magnetic field solutions with respect to $ z $, whereas the constant $ C_2 $ gives rise to symmetric velocity and antisymmetric magnetic field profiles. Moreover, the term proportional to $ K_1 $ in~\eqrefb{eq:generalUB} can be interpreted as the magnetic field due to a uniform current of magnitude $ K_1 / \Reym $, driven in the spanwise direction. For values of the Hartmann number approaching zero~\eqref{eq:generalUB}, becomes
\begin{subequations}
\label{eq:generalUBHydro}
\begin{align}
\taga
U( z ) &= C_0 + C_1 z + C_2 ( 1 - z^2 ) + \ord( \Harz ^ 2 ), \\
\tagb
B( z ) &= K_0 + K_1 z - C_1 z^2 / 2  - C_2 z ( 1 - z^2 ) / 3 + \ord( \Harz ^ 2 ), 
\end{align}
\end{subequations}
and~\eqref{eq:Pi} reduces to $ \Pi = 2 C_2 / \Rey $. As expected, in this non-MHD limit $ U $ is a quadratic function of $ z $ and, even though $ B( z ) $ is nonzero, the streamwise induced field $ I_x = \Prm^{1/2} \Harz B $ vanishes.

In both of the free-surface and channel problems considered here, the choice of velocity normalisation ($U(0) = 1$) and boundary conditions (equations~\eqrefa{eq:noSlip} and \eqrefb{eq:stressBCBase}) implies that $ C_0 = C_1 = 0 $, and $ C_2 = 1 $. Moreover, if the walls are insulating the constants $ K_0 $ and $ K_1 $ vanish due to~\eqrefa{eq:insulatingWallBase} and~\eqrefa{eq:insulatingSurfBase}. In conducting-wall problems $ K_0 $ is again set to zero, either because of~\eqrefa{eq:insulatingSurfBase}, or, in channel problems, by convention (a nonzero $ K_0 $ can be absorbed in the applied magnetic field $ B'_x $). However, $ K_1 = ( \sinh( \Harz ) - \Harz \cosh( \Harz ) ) / ( \Harz ( \cosh( \Harz ) - 1 ) ) $ is in this case non-vanishing, due to~\eqrefa{eq:conductingWallBase}. 

\begin{figure}
\begin{center}
\includegraphics{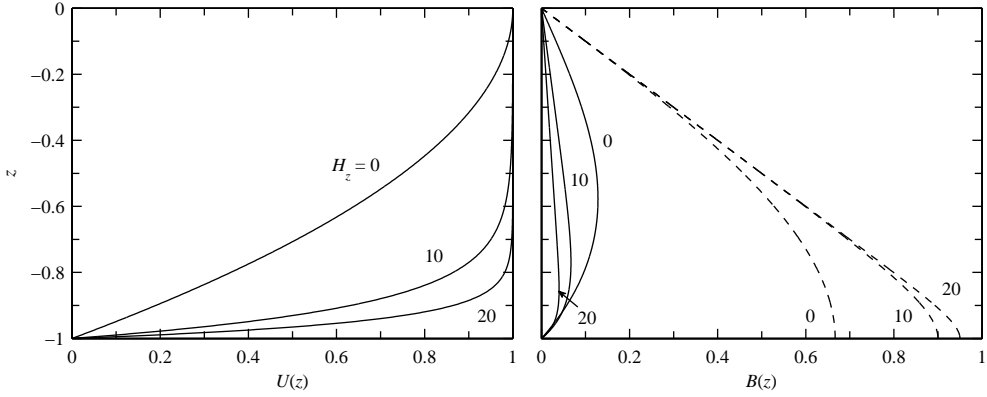}
\end{center}
\caption{\label{fig:baseHartmann}Steady-state velocity (left) and magnetic field profile (right) for Hartmann flow with $ \Harz \in \{0, 10, 20\} $. In the right-hand panel, the solid (dashed) lines correspond to insulating (conducting) boundary conditions at $ z = -1 $.}
\end{figure}

Figure~\ref{fig:baseHartmann} illustrates the functional form of these two classes of velocity and magnetic field profiles for Hartmann numbers in the range 0--20. Compared to the parabolic (Poiseuille) profile in non-MHD flows, Hartmann velocity profiles are characterised by a flat core region and exponential boundary layers of thickness $ \ord( 1 / \Harz )$ near the no-slip walls. Moreover, the mean steady-state velocity, given by
\begin{equation}
\label{eq:baseUAverage}
\langle U \rangle := \int_{-1}^0 \dd z \,  U( z ) = \frac{ \cosh( \Harz ) - \sinh( \Harz ) / \Harz }{ \cosh( \Harz ) - 1 },
\end{equation}
increases monotonically from its non-MHD ($ \Harz = 0 $) value of $ 2 / 3 $ to unity as $ \Harz \tti $. In problems with conducting walls, $ | B( z ) | $ peaks at $ | z | = 1 $, and its gradient, which is proportional to the spanwise induced current $ J_y = \Reym^{-1} \DD I_x = \Harz ( \Rey \Reym )^{-1/2 } \DD B$, attains its maximum magnitude at $ z = 0 $, where $ | \DD B | = 1 $. Also, for large $ \Harz $ the current is close to its maximal value throughout the core. On the other hand, if the wall is insulating, the current distribution becomes concentrated over the Hartmann layer as $ \Harz $ grows, with the magnetic-profile gradient reaching its maximum absolute value $ | \DD B( \pm 1 ) | =  ( \cosh( \Harz ) - \sinh( \Harz ) / \Harz ) / ( \cosh( \Harz ) -1 ) = \ord( 1 ) $ at the walls, while at the core, where $ | \DD B | = \ord( 1 / \Harz ) $, the current tends to a $ \Harz $-independent value.   

Before proceeding, it is worthwhile to note a prominent qualitative difference between MHD and non-MHD velocity profiles, which concerns the existence of inflection points. Because the non-MHD velocity profile~\eqrefa{eq:generalUBHydro} is quadratic in $ z $, it has constant second derivative. On the other hand, it is possible to show that for suitable choices of the constants $ C_1 $ and $  C_2 $, and for sufficiently large $ \Harz $, the MHD velocity profile \eqrefa{eq:generalUB} possesses an inflection point, so that an inviscid instability can potentially exist. In all of the flows studied here, the choice of boundary conditions completely suppresses the antisymmetric component of $ U $ ($ C_1 = 0 $), and eliminates the possibility of inflectional instabilities. However, one can imagine situations (e.g.~a sheared free surface) where the conditions for the inflection point to exist are satisfied. Whether the non-ideal MHD flow develops in practice an instability mode of inviscid origin would be an interesting topic to investigate in the future.

\subsection{\label{sec:3dNormalModes} Three-dimensional normal modes and the associated Squire transformation}

In the three-dimensional temporal-normal-mode analysis we work with the Ansatz
\begin{equation}
\label{eq:3dNormalModes}
\begin{gathered}
\pertu = \Real( ( \modeux( z ), \modeuy( z ), \modeuz( z ) ) \ee^{ \ii ( \alpha x + \beta y ) + \gamma t } ), \quad \pertp = \Real( ( \modep( z ) \ee^{ \ii ( \alpha x + \beta y ) + \gamma t } ), \\
\pertb = \Real( ( \modebx( z ), \modeby( z ), \modebz( z ) ) \ee^{ \ii ( \alpha x + \beta y ) + \gamma t } ), \quad \psi = \Real( \modepsi( z ) \ee^{ \ii ( \alpha x + \beta y ) + \gamma t } ),
\end{gathered}
\end{equation}
where $ p:= p' + \dotp{ \baseb }{ \pertb } $ is the linear perturbation of the pressure field $ \mathcal{ P } $, $ \modeu_i $, $ \modep $, $ \modeb_i $, and $ \modepsi $ are complex functions of $ z $, $ \alpha \geq 0 $ and $ \beta \geq 0 $ are respectively the streamwise and spanwise wavenumbers, and $ \gamma \in \mathbb{ C } $ is the complex growth rate. In channel problems with conducting walls $ \modepsi $ is omitted, while in free-surface problems we also set
\begin{equation}
\label{eq:3dNormalModesA}
a = \Real( \modea \ee^{ \ii ( \alpha x + \beta y ) + \gamma t } ),
\end{equation} 
where $ \modea \in \mathbb{ C } $ is the complex free-surface amplitude. In~\eqref{eq:3dNormalModes} and~\eqref{eq:3dNormalModesA} we have adopted the convention used by \cite{Ho89}, under which $ \Real( \gamma ) =: \Gamma $ gives the modal growth rate, whereas $ C := - \Imag( \gamma ) / ( \alpha^2 + \beta^2 )^{1/2} $ is the phase velocity. The complex phase velocity $ c := \ii \gamma / ( \alpha^2 + \beta^2 )^{1/2} $, where $ \Real( c ) = C $ and $ \Imag( c ) ( \alpha^2 + \beta^2 )^{1/2} = \Gamma $ is frequently employed in the literature \cite[e.g.][]{Yih63,Yih69,Takashima96} in place of $ \gamma $. 

Substituting~\eqref{eq:3dNormalModes} into~\eqref{eq:linearizedGovEqs} and~\eqrefb{eq:goveqsExt} leads to the set of coupled ordinary differential equations
\begin{subequations}
\label{eq:3dNormalModesEqs}
\begin{align}
\taga
\mathrm{\Lambda} \modeux & = \ii \alpha \modep - \ii ( \alpha \basebx + \beta \baseby ) \modebx - \basebz \DD \modebx - \DD ( \basebx ) \modebz + ( \DD U ) \modeuz, \\
\tagb
\mathrm{\Lambda} \modeuy & = \ii \beta \modep - \ii ( \alpha \basebx + \beta \baseby ) \modeby - \basebz \DD \modeby - \DD( \baseby ) \modebz, \\
\tagc
\mathrm{\Lambda} \modeuz & = \DD \modep - \ii ( \alpha \basebx + \beta \baseby ) \modebz - \basebz \DD \modebz, \\
\tagd
0 & = \ii ( \alpha \modeux + \beta \modeuy ) + \DD \modeuz, \\
\tage
\mathrm{\Lambda}_{m} \modebx & = - \ii ( \alpha \basebx + \beta \baseby ) \modeux - \basebz \DD \modeux + ( \DD \basebx ) \modeuz - ( \DD U ) \modebz, \\
\tagf
\mathrm{\Lambda}_{m } \modeby & = - \ii ( \alpha \basebx + \beta \baseby ) \modeuy - \basebz \DD \modeuy + ( \DD \baseby ) \modeuz, \\
\tagg
\mathrm{\Lambda}_{m } \modebz & = - \ii ( \alpha \basebx + \beta \baseby ) \modeuz - \basebz \DD \modeuz, \\
\tagh
0 & = \ii ( \alpha \modebx + \beta \modeby ) + \DD \modeuz, \\
\tagi
0 & = \DD^2 \modepsi - ( \alpha^2 + \beta^2 ) \modepsi,
\end{align}
\end{subequations}
where $ \mathrm{\Lambda} := ( \DD^2 - ( \alpha^2 + \beta^2 ) ) \Rey^{-1} - ( \gamma + \ii \alpha U )  $ and $ \mathrm{\Lambda}_{ m } := ( \DD^2 - ( \alpha^2 + \beta^2 ) ) \Reym^{-1} - ( \gamma + \ii \alpha U ) $. Here, the velocity eigenfunctions are subject to the homogeneous boundary conditions
\begin{equation}
\label{eq:noSlip3DNormalModes}  
\modeux( \zw ) = \modeuy( \zw ) = \modeuz( \zw ) = 0
\end{equation}
at the no-slip boundaries, which follow from~\eqrefb{eq:noSlip}. Moreover, if the walls are insulating, \eqref{eq:insulatingWallPert} leads to
\begin{equation}
\label{eq:insulatingWall3DNormalModes}
\begin{gathered}
\modebx( \zw ) + \ii \alpha \modepsi( \zw ) = 0, \quad  \modeby( \zw ) + \ii \beta \modepsi( \zw ) = 0, \quad \\
\modebz( \zw ) + \DD \modepsi( \zw ) = 0, \quad \alpha \modeby( \zw ) - \beta \modebx( \zw ) = 0,
\end{gathered}
\end{equation}
while boundary conditions for the magnetic field eigenfunctions in conducting-wall problems are, in accordance with \eqref{eq:conductingWallPert},
\begin{equation}
\label{eq:conductingWall3DNormalModes}
\DD \modebx( \zw ) = \DD \modeby( \zw ) = \modebz( \zw ) = 0.
\end{equation}
At the free surface, the kinematic and stress conditions, respectively~\eqref{eq:kinematic} and~\eqref{eq:stressBCPert}, yield
\begin{subequations}
\label{eq:3dNormalModesBCSurf1}
\begin{gather}
\taga
\modeuz( 0 ) - ( \gamma + \ii \alpha U( 0 ) ) \modea = 0, \\
\tagbc
\DD \modeux( 0 ) + \ii \alpha \modeuz( 0 ) + \DD^2 U( 0 ) a = 0, \quad \DD \modeuy( 0 ) + \ii \beta \modeuz( 0 ) = 0, \\
\tagd
\begin{aligned}
0 & = \modep( 0 ) - 2 \Reyinv \DD \modeuz( 0 ) - ( \basebx( 0 ) \modebx( 0 ) + \baseby( 0 ) \modeby( 0 ) + \basebz( 0 ) \modebz( 0 ) ) \\
& \quad - \left( \frac{ \cos\theta }{ \Fro^2 } + \frac{ \alpha^2 + \beta^2 }{ \Web^2 } + \basebx( 0 )  \DD B_x( 0 ) + \baseby( 0 ) \DD B_y( 0 )
- \frac{ 2 \ii \alpha}{ \Rey } \DD U( 0 ) \right) \modea,
\end{aligned}
\end{gather}
\end{subequations}
while the insulating boundary conditions~\eqref{eq:insulatingSurfPert} become
\begin{subequations}
\label{eq:3dNormalModesBCSurf2}
\begin{gather}
\tagab
\modebx( 0 ) + \ii \alpha \modepsi( 0 ) + \DD B_x( 0 ) \modea = 0, \quad  \modeby( 0 ) + \ii \beta \modepsi( 0 ) + \DD B_y( 0 ) \modea = 0, \\
\tagcd
 \modebz( 0 ) + \DD \modepsi( 0 ) = 0, \quad \alpha \modeby( 0 ) - \beta \modebx( 0 ) + ( \alpha \DD B_y( 0 ) - \beta   \DD B_x( 0 ) ) \modea = 0.
\end{gather}
\end{subequations}

Equations~\eqref{eq:3dNormalModesEqs}, in conjunction with the prescribed boundary conditions (in channel problems these are~\eqref{eq:noSlip3DNormalModes}, and either \eqref{eq:insulatingWall3DNormalModes} or~\eqref{eq:conductingWall3DNormalModes}, while in free-surface problems the boundary conditions are~\eqref{eq:noSlip3DNormalModes}, \eqref{eq:3dNormalModesBCSurf1}, \eqref{eq:3dNormalModesBCSurf2}, and either \eqref{eq:insulatingWall3DNormalModes} or~\eqref{eq:conductingWall3DNormalModes}), constitute a differential eigenvalue problem, which must be solved for the eigenvalue $ \gamma $, the eigenfunctions $ \modeu_i $, $ \modep $, $ \modeb_i $ and, where appropriate, $ \modepsi $ and/or $ \modea $. As with several other hydrodynamic stability problems, it is possible to derive a Squire transformation \cite[][]{Squire33}, mapping each three-dimensional normal mode to a two-dimensional one, with $ \baseby = \beta = \modeuy = \modeby = 0 $, and smaller or equal growth rate $ \Real( \gamma ) $. In the free-surface MHD flows studied here the Squire-transformed variables are
\begin{subequations}
\label{eq:squire}
\begin{gather}
\taga
\squirealpha := (\alpha^2 + \beta^2)^{1/2}, \quad \squiregamma := \squirealpha \gamma / \alpha, \\
\tagb
\squirerey := \alpha \Rey / \squirealpha , \quad \squireprm = \Prm, \quad \squiregal := \Gal, \quad \squirecapil := \Capil, \\
\tagc
\quad \squireU := U, \quad \squireBx := ( \alpha \basebx + \beta \baseby ) / \alpha, \quad \squireBz := \squirealpha \basebz / \alpha, \\
\squireux := ( \alpha \modeux + \beta \modeuy ) / \squirealpha , \quad \squireuz := \modeuz, \quad \squirebx := ( \alpha \modebx + \beta \modeby ) / \squirealpha, \quad \squirebz = \modebz, \\
\squirep := \squirealpha \modep / \alpha , \quad \squirea := \alpha \modea / \squirealpha, \quad \squirepsi := \modepsi,
\end{gather}
\end{subequations}
where, for reasons that will become clear below, we have opted to express the transformations for $ \Reym $, $ \Fro $, and $ \Web $ implicitly through the corresponding ones for $ \Prm $, $ \Gal $, and $ \Capil $. It is well known \cite[e.g.][]{Stuart54,Lock55,BetchovCriminale67} that $ \squireui $, $ \squirebi $, $ \squirep $, and $ \squirepsi $ satisfy~(\ref{eq:3dNormalModesEqs}\,{\it a,\,c--e,\,g--i}) with $ \modeui \mapsto \squireui $, $ \modebi \mapsto \squirebi $, $ \modep \mapsto \squirep $,  $ \modepsi \mapsto \squirepsi $, $ U \mapsto \squireU $, $ B_i \mapsto \squireBi $, $ \alpha \mapsto \squirealpha $, $ \gamma \mapsto \squiregamma $, $ \Rey \mapsto \squirerey $, $ \Reym \mapsto \squirereym := \squireprm \squirerey $ and, importantly, $ \beta \mapsto 0 $. Here we verify that  the transformation is also compatible with the non-trivial boundary conditions in free-surface MHD, but only if $ U $ meets the shear-free boundary condition~\eqrefb{eq:stressBCBase}. Specifically, making suitable linear combinations of~\eqref{eq:3dNormalModesBCSurf1}, and using~\eqrefd{eq:3dNormalModesBCSurf2}, it is possible to derive the relations
\begin{subequations}
\label{eq:squireBC1}
\begin{gather}
\tagab
\squireuz( 0 ) - ( \squiregamma + \ii \squirealpha \squireU( 0 ) ) \squirea = 0, \quad \DD \squireux( 0 ) + \ii \squirealpha \squireuz( 0 ) + \DD^2 \squireU( 0 )  \squirea = 0, \\
\tagc
\begin{aligned}
0 & = \squirep( 0 ) - 2 \squirerey^{-1} \DD \squireuz( 0 ) - ( \squireBx( 0 ) \squirebx( 0 ) + \squireBz( 0 ) \squirebz( 0 ) ) \\
& \quad - \left( \cos(\squiretheta) \squirefro^{-2} + \squirealpha^2 \squireweb^{-1} + \squireBx( 0 ) \DD \squireBx( 0 ) -  2 \ii (\squirerey \squirealpha)^{-1} \DD\squireU( 0 ) \right) \squirea,
\end{aligned}
\end{gather}
\end{subequations}
where $ \squirefro := \squirerey \squiregal^{-1/2} $ and $ \squireweb := \squirecapil / \squirerey $, while linear combinations of~\eqref{eq:3dNormalModesBCSurf2} lead to
\begin{subequations}
\label{eq:squireInsulatingBCSurf}
\begin{equation}
\tagab
\squirebx( 0 ) + \ii \squirealpha \squirepsi( 0 ) + \DD \squireBx( 0 ) \squirea = 0, \quad \squirebz( 0 ) + \DD \squirepsi( 0 ) = 0.
\end{equation}
\end{subequations}
Equations~\eqrefa{eq:squireBC1}, \eqrefb{eq:squireBC1}, \eqrefa{eq:squireInsulatingBCSurf}, and~\eqrefb{eq:squireInsulatingBCSurf} are structurally similar to \eqrefa{eq:3dNormalModesBCSurf1}, \eqrefc{eq:3dNormalModesBCSurf1}, \eqrefa{eq:3dNormalModesBCSurf2}, and~\eqrefc{eq:3dNormalModesBCSurf2}, respectively, and it is also straightforward to check that $ \squireui $ and $ \squirebi $ satisfy Squire-transformed versions of \eqref{eq:noSlip3DNormalModes}--\eqref{eq:conductingWall3DNormalModes}.  On the other hand, a comparison immediately reveals that~\eqrefc{eq:squireBC1} and~\eqrefd{eq:3dNormalModesBCSurf1} are compatible only if $ \DD U( 0 ) = 0 $. Thus, unlike channel problems, the validity of the transformation in free-surface flows depends on the functional form of the steady-state velocity.

In plane Poiseuille flow, the correspondence between three and two-dimensional modes implies that for the purpose of determining the minimum (critical) Reynolds number for instability it suffices to restrict attention to two-dimensional normal modes. That is, it follows from the inequalities $ \Real( \squiregamma ) \geq \Real( \gamma ) $ and $ \squirerey \leq \Rey $, which are a consequence of~\eqrefa{eq:squire} and~\eqrefc{eq:squire}, that to each unstable three-dimensional mode there corresponds a less or equally stable two-dimensional one at smaller or equal Reynolds number. However, in the multiple-parameter problems studied here the question of whether or not the critical mode is two-dimensional depends on the path followed in parameter space as $ \Rey $ is increased. In particular, if the process of increasing the Reynolds number modifies any of the flow parameters that are unchanged by the Squire transformation, the critical two-dimensional mode is not guaranteed to be of vanishing spanwise wavenumber. 

The latter observation has been made by \cite{Hunt66} for channel MHD flows under purely streamwise applied magnetic fields ($B_y = B_z = 0$), where, according to~\eqrefc{eq:squire}, $ B_x $ and, equivalently, the Alfv\'en number  $ \Alf = 1 / B_x $ are unchanged by the transformation. Hence, if $ \Alf $ and $ \Prm $ are held fixed as the Reynolds number of the three-dimensional problem is increased, then indeed the first mode that becomes unstable has $ \beta = 0 $. However, if one requires the channel width, the fluid's material properties, and the external magnetic field strength all to remain fixed while the flow speed is increased (as was assumed by Hunt), then as $ \Rey $ grows $ \Alf $ necessarily increases as well, and three-dimensional modes may become unstable first. 

In the problems with flow-normal external magnetic field ($ B_x' = B_y = 0$) studied here, \eqrefb{eq:squire} and \eqrefc{eq:squire} imply that the Hartmann number $ \squirehar := \squireprm^{1/2} \squirerey \squireBz = \Har $ and the induced magnetic field profile $ \tilde B := \squireBx / ( \squireprm^{1/2} \squirehar ) = B $ of the two-dimensional flow are the same as in the three-dimensional case. Since the Hartmann number is independent of the characteristic velocity $ U_* $, this in turn indicates that when channel Hartmann flow is driven at progressively higher speeds with all geometrical and material parameters held fixed (as can be accomplished by means of a pump generating the streamwise pressure gradient $ \Pi $~\eqref{eq:Pi}), the critical mode has purely streamwise wavevector. In free-surface problems, \eqrefb{eq:squire} necessitates that $ \Gal $ and $ \Capil $ are additionally constrained, but this cannot be accomplished simply by varying $ U_* $ with all other aspects of the problem fixed. This is because (i) the capillary number $ \Capil $ directly depends on $ U_* $, and (ii) the boundary condition~\eqrefa{eq:stressBCBase}, in conjunction with~\eqref{eq:Pi}, introduces the parameter interdependence 
\begin{equation}
\label{eq:reTheta}
\Rey = \Gal \tan(\theta) \tanh( \Har / 2 ) / \Har \quad \text{or} \quad \Rey = \Gal \tan( \theta ) ( \sech( \Har ) - 1  ) / \Har^2, 
\end{equation}
valid respectively for problems with insulating and conducting walls. In the expressions above only $ \Rey $ depends on $ U_* $, indicating that the steady-state velocity cannot be changed without modifying at least some of the remaining properties of the flow. An option compatible with~\eqref{eq:squire} would be to fix $ U_* $, and vary $ \Rey $ by changing the fluid thickness $ l $, the inclination angle $ \theta $, and the external magnetic field $ \basebz $ in a manner that $ \Gal $ and $ \Har $ remain constant.     

\subsection{\label{sec:2dNormalModes}Two-dimensional normal modes}

In the two-dimensional normal-mode formulation, where $ \beta $, $ \modeuy $, $ \modeby $, and $ \baseby $ are all set to zero, the divergence-free conditions~(\ref{eq:3dNormalModesEqs}\,{\it d,\,h}) can be used to eliminate the streamwise velocity and magnetic field eigenfunctions, giving
\begin{equation}
\label{eq:2dNormalModesUB}
\pertu( \vpos, t ) = \Real( ( \ii  \DD \modeu( z ) / \alpha, 0, \modeu( z ) ) e^{ \ii \alpha x + \gamma t } ), \quad \pertb( \vpos, t ) = \Real( ( \ii \DD \modeb( z ) / \alpha, 0, \modeb( z ) ) e^{ \ii \alpha x + \gamma t } )
\end{equation}
for the perturbed velocity and magnetic fields, where we have dropped the $ z $ subscript from $ \modeuz $ and $ \modebz $ for notational clarity. Substituting~\eqref{eq:2dNormalModesUB} into~(\ref{eq:3dNormalModesEqs}\,{\it a,\,c,\,g}), and eliminating the pressure eigenfunction then leads to the coupled OS and induction equations \cite[e.g.][]{BetchovCriminale67,MullerBuhler01},
\begin{subequations}
\label{eq:orrSommerfeldInd}
\begin{multline}
\taga
\Rey^{-1} ( \DD^2 - \alpha^2 )^2 \modeu - ( \gamma + \ii \alpha U ) ( \DD^2 - \alpha^2 ) \modeu + \ii \alpha ( \DD^2 U ) \modeu \\
+ ( \ii \alpha \basebx + \basebz \DD ) ( \DD^2 - \alpha^2 ) \modeb - \ii \alpha ( \DD^2 \basebx ) \modeb = 0, 
\end{multline}
and
\begin{equation}
\tagb
\Reym^{-1} ( \DD^2 - \alpha^2 ) \modeb - ( \gamma + \ii \alpha U ) \modeb + ( \ii \alpha \basebx + \basebz \DD ) \modeu = 0,
\end{equation}
\end{subequations}
respectively. 

Whenever applicable, we write 
\begin{equation}
\label{eq:2dNormalModesBExt}
\pertbext = \Real( ( \ii \DD \modebext( z ) / \alpha, 0,\modebext( z ) ) e^{ \ii \alpha x + \gamma t } ) = - \Real( \ii \alpha \modepsi( z ), 0, \DD \modepsi( z )  e^{ \ii \alpha x + \gamma t } ) ).
\end{equation}
for the external magnetic field perturbations. Laplace's equation~\eqrefi{eq:3dNormalModesEqs} for the magnetic potential then becomes $( \DD^2 - \alpha^2 ) \modepsi = 0 $, which, in conjunction with the condition that $ \modepsi $ vanishes at infinity, has the closed-form solutions
\begin{subequations}
\label{eq:psiAnalytic}
\begin{equation}
\tagab
\modepsi( z ) = 
\begin{cases}
\modepsi( 0 ) e^{ - \alpha z }, & z > 0, \\
\modepsi( -1 ) e^{ \alpha( z + 1 ) }, & z < -1,
\end{cases}
\quad
\modepsi( z ) = 
\begin{cases}
\modepsi( 1 ) e^{ - \alpha( z - 1 ) }, & z > 1, \\
\modepsi( - 1 ) e^{ \alpha ( z + 1 ) }, & z < -1,
\end{cases}
\end{equation} 
\end{subequations}
respectively for free-surface and channel problems with insulating walls. If the walls are conducting, only the expression valid for $ z > 0 $ is retained in free-surface problems, whereas in channel problems $ \modepsi $ is dropped altogether from the formulation.

In inductionless problems, \eqref{eq:zeroPm} substituted into~\eqrefa{eq:linearizedGovEqs} leads to the modified OS equation \cite[][]{Stuart54,Lock55}
\begin{equation}
\label{eq:orrSommerfeldZeroPm}
( \DD^2 - \alpha^2 )^2 \modeu - ( \ii \alpha \Harx + \Harz \DD )^2 \modeu - \Rey ( \gamma + \ii \alpha U ) ( \DD^2 - \alpha^2 ) \modeu + \ii \alpha \Rey ( \DD^2 U ) \modeu = 0, 
\end{equation} 
where $ \Harx := ( \Rey \Reym )^{1/2} \basebx $. Equation~\eqref{eq:orrSommerfeldZeroPm}, which replaces the coupled system~\eqref{eq:orrSommerfeldInd}, can also be obtained by letting $ \Prm \searrow 0 $ with $ \Harx $ and $ \Harz $ fixed. In that limit, the induced magnetic field $ I_x $ vanishes (see \S\ref{sec:steadyState}), and~\eqrefb{eq:orrSommerfeldInd} reduces to 
\begin{equation}
\label{eq:zeroPmNormalModes}
( \ii \alpha \basebx + \basebz \DD ) ( \DD^2 - \alpha^2 ) \modeb =  -\Reym ( \ii \alpha \basebx + \basebz \DD ) \modeu, 
\end{equation}
leading to~\eqref{eq:orrSommerfeldZeroPm} upon substitution in~\eqrefa{eq:orrSommerfeldInd}.

As for the boundary conditions, substituting for $ \modeux $, $ \modebx $, and $ \modep $ in the no-slip, kinematic, and stress conditions~\eqref{eq:noSlip3DNormalModes} and~\eqref{eq:3dNormalModesBCSurf1} by means of~(\ref{eq:3dNormalModesEqs}\,{\it c,\,d,\,h}), leads to
\begin{subequations}
\label{eq:2dNormalModesBC1}
\begin{gather}
\taga
\modeu( \zw ) = \DD \modeu( \zw ) = 0, \\
\tagbc
\modeu( 0 ) - ( \gamma + \ii \alpha \bux( 0 ) ) \modea = 0, \quad
\DD^2 \modeu( 0 ) + \alpha^2 \modeu( 0 ) - \ii \alpha \DD^2 U( 0 ) \modea = 0,
\end{gather}
and
\begin{multline}
\tagd
( ( ( \DD^2 - 3 \alpha ^ 2 ) \DD - \Rey ( \gamma + \ii \alpha \bux ) \DD + \ii \alpha \Rey ( \DD U ) ) \modeu )|_{z=0} + \Rey ( \basebz ( \DD^2 - \alpha^2 ) - \ii \alpha ( \DD \basebx ) ) \modeb |_{z=0} \\
- \alpha^2 \left( \Gal \Rey^{-1} + \alpha^2 \Capil^{-1} + \Rey \basebx( 0 ) \DD \basebx( 0 ) - 2 \ii \alpha  \DD U( 0 ) \right) \modea = 0.
\end{multline}
\end{subequations}
Moreover, using~\eqref{eq:psiAnalytic} to eliminate $ \modepsi $, the magnetic field boundary conditions at insulating boundaries, \eqref{eq:insulatingWall3DNormalModes} and~\eqref{eq:3dNormalModesBCSurf2}, yield
\begin{equation}
\label{eq:2dNormalModesInsulatingChannel}
\DD \modeb( \pm 1 ) \pm \alpha \modeb( \pm 1 ) = 0
\end{equation}
and
\begin{subequations}
\label{eq:2dNormalModesInsulatingFreeSurf}
\begin{equation}
\tagab 
\DD \modeb( -1 ) - \alpha \modeb( - 1 ) = 0, \quad \DD \modeb( 0 ) + \alpha \modeb( 0 ) - \ii \alpha \DD \basebx( 0 ) \modea = 0,
\end{equation}
\end{subequations}
respectively for channel and free-surface problems. If, on the other hand, the walls are
conducting~\eqref{eq:conductingWall3DNormalModes} leads to
\begin{equation}
\label{eq:conductingWall2DNormalModes}
\modeb( \zw ) = 0.
\end{equation}
In inductionless problems, the boundary conditions for $ \modeb $ are not required, and~\eqrefd{eq:2dNormalModesBC1} becomes
\begin{multline}
\label{eq:normalStressZeroPm} 
( ( \DD^2 - 3 \alpha^2 ) \DD - \Rey( \gamma + \ii \alpha U )\DD + \ii \alpha \Rey ( \DD U ) - \Harz ( \ii \alpha \Harx + \Harz \DD ) ) \modeu|_{z=0} \\
- \alpha^2 \left( \Gal \Rey^{-1} + \alpha^2 \Capil^{-1} - 2 \ii \alpha \DD U( 0 ) \right) \modea = 0.
\end{multline} 

To summarise, in both of the free-surface and channel MHD stability problems studied here, the steady-state configuration and the governing differential equations are respectively \eqref{eq:generalUB} (with $ z $ restricted to the appropriate interval, and the integration constants set according to the wall conductivity) and~\eqref{eq:orrSommerfeldInd}. The boundary conditions for free-surface problems with insulating wall are~\eqref{eq:2dNormalModesBC1} and~\eqref{eq:2dNormalModesInsulatingFreeSurf}, while if the wall is conducting~\eqref{eq:conductingWall2DNormalModes} is enforced in place of~\eqrefa{eq:2dNormalModesInsulatingFreeSurf}. In channel problems the boundary conditions are~\eqrefa{eq:2dNormalModesBC1}, and either~\eqref{eq:2dNormalModesInsulatingChannel} or~\eqref{eq:conductingWall2DNormalModes}. In inductionless problems, the governing equations are replaced by~\eqref{eq:orrSommerfeldZeroPm}, the magnetic field boundary conditions~\eqref{eq:2dNormalModesInsulatingChannel}--\eqref{eq:conductingWall2DNormalModes} are omitted, and~\eqrefd{eq:2dNormalModesBC1} is replaced by~\eqref{eq:normalStressZeroPm}.  

\section{\label{sec:energyBalance}Energy balance}

\subsection{\label{sec:energyBalanceGeneral}Formulation for two-dimensional perturbations}

Following the analysis by \cite{Stuart54} and \cite{SmithDavis82}, respectively for channel flows with homogeneous boundary conditions and non-MHD free-surface flows, we now derive energy-balance relations for normal modes in free-surface MHD with insulating boundary conditions. The resulting formulation will contribute towards a physical interpretation of the results presented in \S\ref{sec:discussion}, and can also provide consistency checks for numerical schemes \cite[][]{SmithDavis82,GiannakisFischerRosner08}. In order to keep complexity at a minimum, we restrict attention to two-dimensional normal modes, setting the spanwise wavenumber $ \beta $ equal to zero and assigning an arbitrary length $ L_y $ to the size of the domain in the $ y $ direction. Moreover, we assume that the steady-state velocity and magnetic field profiles satisfy~\eqref{eq:baseX2} subject to the boundary conditions~\eqrefa{eq:insulatingWallBase}, \eqrefa{eq:insulatingSurfBase}, \eqrefa{eq:noSlip}, and \eqrefb{eq:stressBCBase}. Similar energy equations apply for the other types of stability problems studied here, but in the interests of brevity, we do not explicitly consider their derivation. In this section we work in Cartesian tensor notation, using $ u_i $, $ b_i $, and $ b'_i $ to respectively denote the components of $ \pertu $, $ \pertb $, and $ \pertbext $ in the $(x,y,z) $ coordinate system, and $ \epsilon_{ijk} $ to denote the Levi-Civita symbol. Summation is assumed over repeated tensorial indices.

First, we define the kinetic energy density and the internal magnetic energy density of the perturbations as $ \mathcal{E}_u := u_i u_i / 2 $ and $ \mathcal{E}_b := b_i b_i / 2 $, respectively. The integrals of these quantities over the unperturbed fluid domain $ \dmn := ( 0, L_x ) \times ( 0, L_y ) \times ( -1, 0 ) $, where $ L_x := 2 \upi / \alpha $ is the modal wavelength, then yield the total kinetic and internal magnetic energies
\begin{subequations}
\label{eq:EuEb}
\begin{equation}
\tagab
E_u := \int_\dmn \dd V \, \mathcal{E}_u, \quad E_b := \int_\dmn \dd V \, \mathcal{ E }_b,
\end{equation}
\end{subequations}
where $ \dd V := \dd x \, \dd y \, \dd z $. Similarly, the magnetic energy density in the exterior of the fluid, given by $ \mathcal{E}_{ b' } := b'_i b'_i / 2 $, leads to the external magnetic energies
\begin{equation}
\label{eq:EbExt}
E_{ b'-} := \int_\dmnminus \dd V \, \mathcal{ E }_{ b' }, \quad \quad E_{ b'+ } := \int_\dmnplus \dd V \, \mathcal{ E }_{ b' }, \quad E_{b'} := E_{b'-} + E_{b'+},
\end{equation}
where $ \dmnminus := ( 0, L_x ) \times ( 0, L_y ) \times ( -\infty, -1 ) $ and $ \dmnplus := ( 0, L_x ) \times ( 0, L_y ) \times( 0, \infty ) $. The two forms of energy associated with the free surface are the  potential energy $ E_p $ and the surface-tension energy $ E_\sigma $, defined in terms of the corresponding densities (per unit surface) as
\begin{equation}
\label{eq:EpEsigma}
E_p := \int_\dmns \dd S \, \mathcal{ E }_p , \quad E_\sigma := \int_\dmns \dd S \, \mathcal{ E }_\sigma,
\end{equation}
where $ \dmns := ( 0, L_x ) \times ( 0, L_y ) $, $ \dd S := \dd x \, \dd y $, and 
\begin{equation}
\label{eq:epesigma}
\mathcal{ E }_p := \left( \cos(\theta) \Fro^{-2}  + \basebx( 0 ) \DD \basebx( 0 ) \right) a^2 / 2, \quad \mathcal{ E }_\sigma := ( \dx a )^2/ (2 \Web ).  
\end{equation}
We remark that $ E_p $ is equal to the work needed to displace the free surface from $ z = 0 $ to $ z = a( x, y, t ) $ in the presence of the gravitational field and the flow-normal gradient of the steady-state magnetic pressure. Also, noting that $  ( 1 + ( \dx a )^2 )^{1/2} \, \dd x \, \dd y = ( 1 + ( \dx a )^2 + \ord( a^4 ) ) \, \dd x \, \dd y $ is the area element on the free surface, $ \mathcal{ E }_\sigma $ is equal to the work done against surface tension in order to increase the free-surface area from its unperturbed value to that corresponding to the amplitude $ a( x, y, t ) $. The potential and surface-tension energies make up the total free-surface energy
\begin{equation}
\label{eq:Ea}
E_a := E_p + E_\sigma.
\end{equation}

The time-evolution of the energy in the fluid domain follows from the linearised equations~\eqref{eq:linearizedGovEqs}, which, upon elimination of the Lorentz force $ \pertfl $ in~\eqrefa{eq:linearizedGovEqs} by means of~\eqrefb{eq:linearizedGovEqs}, read 
\begin{subequations}
\label{eq:linearizedGovEqsTensor}
\begin{gather}
\taga
\dt u_i = - U_j \, \partial_j u_i - u_j \, \partial_j U_i +  b_j \, \partial_j B_i + \Rey^{-1} \lapl u_i  + B_j \, \partial_j b_i - \partial_i p, \\
\tagb
j_i = \Reym^{-1} \epsilon_{ i j k } \partial_j b_k = e_i + \epsilon_{ijk} U_j b_k + \epsilon_{ijk} u_j B_k, \\
\tagc
\dt b_i = - U_j \, \partial_j  b_i +  b_j \, \partial_j U_i - u_j \, \partial_j B_i + \Reym^{-1}  \lapl b_i + B_j \, \partial_j u_i, \\
\tagde
\partial_i u_i = 0, \quad \partial_i b_i = 0. 
\end{gather}
\end{subequations}
In particular, forming the contraction of~\eqrefa{eq:linearizedGovEqsTensor} with $ u_i $, the contraction of~\eqrefb{eq:linearizedGovEqsTensor} with $ b_i $, and adding the results together, leads to the energy equation   
\begin{equation}
\label{eq:energyDensityBalance}
\dt ( \mathcal{ E }_u + \mathcal{ E }_b ) =  g_R  + g_M + g_J + g_\nu + g_\eta + \partial_i ( - q^{(\mathcal{E})}_i + q^{(em)}_i +  q^{(mech)}_i ),
\end{equation}
where
\begin{subequations}
\label{eq:powerDensity}
\begin{gather}
\tagac
g_R := - u_i u_j \partial_j U_i, \quad
g_M := b_i b_j \partial_j U_i, \quad
g_J :=  u_i u_j ( \partial_j B_i - \partial_i B_j ), \\
\tagde
g_\nu := - s_{ i j }  s_{ i j }/ (2 \Rey), \quad
g_\eta := - \Reym  j_i j_i,
\end{gather}
\end{subequations}
with $ s_{ i j } := \partial_i u_j + \partial_j u_i $, and
\begin{subequations}
\label{eq:powerFlux}
\begin{gather}
\tagab
q^{(\mathcal{E})}_i :=  U_i( \mathcal{ E }_u + \mathcal{ E }_b ),\quad
q^{(em)}_i := \epsilon_{ ijk } \, b_j \, ( e_k + \epsilon_{ klm } U_l \, b_m ), \\
\tagcd
q^{(mech)}_i := u_j ( - ( p - p_b ) \delta_{ i j } + \Rey^{-1} s_{ i j } ), \quad p_b := B_i b_i.
\end{gather}
\end{subequations}
We remark that in deriving~\eqref{eq:energyDensityBalance} we have used the divergence-free conditions~\eqrefef{eq:linearizedGovEqs} (as well as the corresponding ones for $ U_i $ and $ B_i $) and the relations
\begin{equation}
\label{eq:laplUB}
u_i \lapl u_i = \partial_i (  u_j s_{ i j } ) - s_{ i j } s_{ i j } / 2, \quad
\Reym^{-1} b_i  \lapl b_i = \partial_i( \epsilon_{ i j k } b_j  j_k ) - \Reym  j_i  j_i.
\end{equation} 
In a similar manner, Faraday's law $ \dt b'_i = - \epsilon_{ i j k } \partial_j e'_k $, governing $ b'_i $ and the external electric field $ e'_i $, in conjunction with the curl-free property $ \epsilon_{ i j k } \partial_j b'_k = 0 $ (this holds due to the insulating nature of the medium surrounding the fluid), yields the energy equation
\begin{equation}
\label{eq:energyDensityBalanceExt}
\dt \mathcal{ E }_{ b' } = \epsilon_{ i j k } \partial_i ( b'_j e'_k )
\end{equation}
for the external magnetic energy density. As for the surface energies~\eqref{eq:epesigma}, multiplying the kinematic boundary condition~\eqref{eq:kinematic} by the free-surface amplitude $ a $ and suitable constants leads to the rate equations  
\begin{subequations}
\label{eq:energyDensityBalanceSurface}
\begin{align}
\taga
\dt \mathcal{ E }_g & = - U( 0 ) \, \dx \mathcal{ E }_g + \left( \cos( \theta )\Fro^{-2} + \basebx( 0 ) \DD \basebx( 0 ) \right) a \pertuz|_{z=0}, \\
\tagb
\dt \mathcal{ E }_\sigma & = - U( 0 ) \, \dx \mathcal{ E }_\sigma + ( \dx( \pertuz|_{ z = 0 } \, \dx a ) - \pertuz |_{ z = 0 } \, \dx^2 a ) / \Web.
\end{align} 
\end{subequations}

Each of the terms on the right-hand side of~\eqref{eq:energyDensityBalance} has a physical interpretation. First, the source terms $ g_ R $ and $ g_M$ in are respectively the Reynolds and Maxwell stresses, i.e.~the energy transfer rate between the basic flow and the velocity and magnetic field perturbations. Physically, the Reynolds stress is an outcome of the mechanical exchange of energy occurring as the velocity perturbations transport mass within the non-uniform steady-state velocity field. In contrast, the energy transfer mechanism corresponding to $ g_M $ is electromagnetic in nature. Its origin lies in the stretching/shrinking  of the perturbed magnetic field $ \pertb $ by the basic flow. Also, noting that $ \partial_j B_i - \partial_i B_j = \Reym \epsilon_{ j i k } J_k $, where $ J_k := \Reym^{-1} \epsilon_{klm}\partial_l B_m $ is the steady-state current, $ g_J $ is interpreted as the energy transferred from the basic current to the perturbations, the so-called current interaction, while $ g_\nu $ and $ g_\eta $ are the viscous and resistive dissipation terms. Among  the fluxes $ \{ q^{(\mathcal{E})}_i, q^{(em)}_i, q^{(mech)}_i \} $, $ q^{(\mathcal{E})}_i $ is the perturbation energy transported by the basic flow, $ q^{(em)}_i $ is the electromagnetic energy flux, evaluated in the rest frame of the unperturbed fluid (recall that in the non-relativistic limit the electric-field perturbation in the rest frame of the fluid is $ \perte + \crossp{ \baseu }{ \pertb } $), and $ q^{(mech)}_i $ is the momentum flux due to mechanical stresses acting on the fluid. 

We derive a global version of the above energy equations by integrating~\eqref{eq:energyDensityBalance} over $ \dmn $, and making use of the divergence theorem to reduce the volume integrals of $ q^{(\mathcal{E})}_i $, $ q^{(em)}_i $, and $ q^{(mech)}_i $ to surface integrals. To begin, a consequence of the assumed periodicity in $ x $ is that $ \int_\dmn \dd V \, \partial_i q^{(\mathcal{E})}_i  $ vanishes, and that the surface energies~\eqref{eq:EpEsigma} obey 
\begin{subequations}
\label{eq:energyBalanceSurface}
\begin{align}
\taga
\dt E_p &= \int_\dmns \dd S \, \left( \cos( \theta )\Fro^{-2} + \basebx( 0 ) \DD \basebx( 0 ) \right )a \pertuz|_{ z = 0 }, \\
\tagb
\dt E_\sigma &= - \int_\dmns \dd S \, \Web^{-1} \dx^2 a \, \pertuz|_{ z=0 }.
\end{align}
\end{subequations}
Also, integrating~\eqref{eq:energyDensityBalanceExt} over $ \dmnminus \cup \dmnplus $, and using the magnetic field boundary conditions \eqref{eq:insulatingWallPert} and~\eqref{eq:insulatingSurfPert}, leads to
\begin{equation}
\label{eq:EbPrime}
\dt E_{ b' } = - \int_\dmn \dd V \, \partial_i q_i^{(em)} - \DD \basebx( 0 ) \int_\dmns \dd S \, a e_y |_{ z= 0 } - U( 0 ) \int_\dmns \dd S \, ( \pertbx \pertbz )|_{ z = 0 },
\end{equation}
while the relation
\begin{equation}
\label{eq:EpEsigma2}
\dt ( E_p + E_\sigma ) = - \int_\dmn \dd V \, \partial_i q_i^{(mech)} - \frac{ \DD^2U( 0 ) }{ \Rey } \int_\dmns  \dd S \, a \pertux |_{ z = 0 } 
\end{equation} 
follows from the stress boundary conditions~\eqref{eq:stressBCPert}. Then, integrating~\eqref{eq:energyDensityBalance}, and eliminating $ q_i^{(em)} $ and $ q_i^{(mech)} $ by means of~\eqref{eq:EbPrime} and~\eqref{eq:EpEsigma2}, we arrive at the conservation equation
\begin{equation}
\label{eq:energyBalance}
\dt E = G_R + G_M + G_J + G_\nu + G_\eta + G_{a \nu } + G_{a J},
\end{equation}
for the total energy $ E := E_u + E_b + E_{b'} + E_p + E_\sigma $. Here the volume terms
\begin{equation}
\label{eq:energyBalanceTerms}
\begin{gathered}
G_R := \int_\dmn \dd V \, g_R, \quad G_M := \int_\dmn \dd V \, g_M, \quad G_J := \int_\dmn \dd V \, g_J, \\
G_\nu := \int_\dmn\dd V \, g_\nu, \quad G_\eta := \int_\dmn \dd V \, g_\eta
\end{gathered}
\end{equation}
are respectively the total Reynolds stress, Maxwell stress, current interaction, viscous dissipation, and resistive dissipation. Moreover, the surface terms 
\begin{subequations}
\label{eq:surfaceEnergyBalanceTerms}
\begin{align}
\taga
G_{ a U} & :=  - \frac{ \DD^2U( 0 ) }{ \Rey }  \int_\dmns \dd S \, a \pertux|_{ z = 0 }, \\
\tagb
G_{a J } & := - \Reym J_y( 0 ) \int_\dmns \dd S \, a ( j_y - ( \basebx \pertuz - \basebz \pertux) )  |_{ z = 0 }
\end{align}
\end{subequations}
represent the energy transferred to the free surface by viscous and electromagnetic forces, respectively. In particular, noting that  $ \basebx \pertuz - \basebz \pertux $ is the current induced by the velocity-field perturbations within the steady-state magnetic field, the quantity $ j_y - ( \basebx \pertuz - \basebz \pertux ) $ in~\eqrefb{eq:surfaceEnergyBalanceTerms} can be interpreted as the current induced by the steady-state fluid motion within the perturbed magnetic field \cite[][]{Hunt66}.

\subsection{\label{sec:energyBalanceNormalModes}Energy balance for two-dimensional normal modes}

The results of the preceding section can be applied to the special case of the two-dimensional normal mode solutions. First, the expressions 
\begin{equation}
\label{eq:modeeueb}
E_u = L  e^{ 2 \Gamma t } \int_{-1}^0 \dd z \, \modeeu( z ), \quad E_b = L e^{ 2 \Gamma t } \int_{-1}^0 \dd z \, \modeeb( z ),
\end{equation}
follow by substituting for $ u_i $ and $ b_i $ in~\eqref{eq:EuEb} using~\eqref{eq:2dNormalModesUB}, where $ L := L_x L_y / 4 \alpha^2 $ is a constant, and
\begin{equation}
\label{eq:modeeu}
\modeeu := | \DD \modeu |^2 + \alpha^2 | \modeu |^2 , \quad \modeeb := | \DD \modeb |^2 + \alpha^2 | \modeb |^2
\end{equation}
are the modal kinetic and magnetic energy densities, averaged over the streamwise and spanwise directions. The external magnetic energy $ E_b' $~\eqref{eq:EbExt} can be computed in terms of the internal magnetic field using the solution~\eqrefa{eq:psiAnalytic} for $ \modepsi $ to evaluate the integral over $ z $, and the insulating boundary conditions~\eqref{eq:insulatingWall3DNormalModes} and~\eqref{eq:3dNormalModesBCSurf2} to substitute for the magnetic potential at the fluid domain boundaries. Specifically, we have
\begin{equation}
\label{eq:modeebext}
E_b' = L e^{ 2 \Gamma t } \left( \int_{ -\infty }^{-1} \dd z + \int_{0}^\infty \dd z \right) ( | \DD \modeb '( z ) |^2 + \alpha^2 | \modeb'( z ) |^2 ) = L e^{ 2 \Gamma t } \alpha ( | \modeb( -1 ) |^2 + | \modeb( 0 ) |^2 ).
\end{equation}
In addition, the normal-mode solution~\eqref{eq:3dNormalModesA} (with $ \beta = 0 $) for the free-surface displacement in conjunction with~\eqref{eq:EpEsigma} yields
\begin{equation}
\label{eq:modeEpEsigma}
E_p = L e^{2 \Gamma t } \alpha^2 \left( \cos(\theta ) \Fro^{-2} + \basebx( 0 ) \DD \basebx( 0 ) \right) | \modea |^2, \quad E_\sigma = L  e^{ 2 \Gamma t} \alpha^4 \Web^{-1} | \modea |^2.
\end{equation}

We treat the source terms on the right-hand side of~\eqref{eq:energyBalance} in a similar manner. The volume terms~\eqref{eq:energyBalanceTerms} become
\begin{equation}
\label{eq:modeG}
\begin{gathered}
G_R = L e^{2 \Gamma t } \int_{-1}^{0} \dd z \modegr( z ), \quad G_M = L e^{2 \Gamma t } \int_{-1}^0 \dd z \, \modegm( z ), \\
\quad G_J = L e^{2 \Gamma t } \int_{-1}^0 \dd z \, \modegj( z ), \quad G_\nu = L e^{2 \Gamma t } \int_{-1}^{0} \dd z \, \modegnu( z ), \quad G_\eta = L e^{2 \Gamma t } \int_{-1}^0 \dd z \, \modegeta( z ),
\end{gathered}
\end{equation}
where
\begin{subequations}
\label{eq:modeGVolume}
\begin{gather}
\tagab
\modegr :=  2 \alpha ( \DD U ) \Imag( \modeu^* \DD \modeu ), \quad
\modegm := - 2 \alpha ( \DD U )  \Imag( \modeb^* \DD \modeb ), \\
\tagc
\modegj :=  2 \alpha  \DD \basebx \, \Imag( \modeu^* \DD \modeb - \modeb^* \DD \modeu ), \\
\tagd
\modegnu :=  -  2 \Rey^{-1} ( | \DD^2 \modeu |^2 - 2 \alpha^2 \Real( \modeu^* \DD^2 \modeu ) + \alpha^4 | \modeu |^2 ), \\
\tage
\modegeta := -  2 \Reym^{-1} ( | \DD^2 \modeb |^2 - 2 \alpha^2 \Real( \modeb^* \DD^2 \modeb ) + \alpha^4 | \modeb |^2 ).
\end{gather}
\end{subequations}
Moreover, the surface terms~\eqref{eq:surfaceEnergyBalanceTerms} evaluate to
\begin{subequations}
\label{eq:modeGSurface}
\begin{align}
\taga
G_{ a U } & = - L e^{2 \Gamma t } 2 \alpha \Rey^{-1} \DD^2U( 0 ) \, \Imag( \modea \DD \modeu^*( 0 ) ), \\
\tagb
G_{ a J } & = L e^{ 2 \Gamma t } 2 \alpha J_y( 0 ) ( \Imag( ( \DD^2 \modeb( 0 ) - \alpha^2 \modeb( 0 ) ) \modea^* ) \\
& \quad - 2 \alpha \basebz \DD \basebx( 0 ) \Imag( \modea \DD \modeu^*( 0 ) ) + 2 \alpha^2 \basebx( 0 ) \DD \basebx( 0 ) \Real( \modea \modeu^*( 0 ) ) ), 
\end{align}
\end{subequations}
Finally, noting that the energy growth rate of the normal-mode solutions~\eqref{eq:2dNormalModesUB} and~\eqref{eq:3dNormalModesA} satisfies $ \dt E = 2 \Gamma E $, and inserting~\eqref{eq:modeGVolume} and~\eqref{eq:modeGSurface} into~\eqref{eq:energyBalance}, we obtain
\begin{align}
\label{eq:gammaEnergyBalance}
\Gamma = \Gamma_\mathrm{ R } + \Gamma_M + \Gamma_J + \Gamma_\eta + \Gamma_\nu + \Gamma_{ a U } + \Gamma_{ a J },
\end{align}
where $ \Gamma_{[\cdot]} := G_{[\cdot]} / 2 E $. 

Equation~\eqref{eq:gammaEnergyBalance} expresses the modal growth rate as a sum of contributions from the various MHD energy generation and dissipation mechanisms. In the ensuing discussion, we oftentimes aggregate the terms on its right-hand side, writing $ \Gamma = \Gamma_{mech} + \Gamma_{em} $, where
\begin{subequations}
\label{eq:gammaMechEm}
\begin{equation}
\tagab
\Gamma_{mech} := \Gamma_{R} + \Gamma_{aU} + \Gamma_J + \Gamma_{\nu}, 
\quad
\Gamma_{em}  := \Gamma_{M} + \Gamma_{aJ} + \Gamma_\eta
\end{equation}
\end{subequations}
represent the net mechanical and electromagnetic contributions to $ \Gamma $, respectively. Moreover, we introduce normalised versions of the energy transfer densities~\eqref{eq:modeGVolume}, writing $ \skew3\hat\Gamma_{[\cdot]}( z ) := L \skew3\hat g_{[\cdot]}( z ) / ( 2 E )|_{t=0} $ for each of the source terms in~\eqref{eq:modeGVolume}, so that $ \int_{-1}^0 \dd z \, \skew3\hat\Gamma_{[\cdot]}( z ) = \Gamma_{[\cdot]} $. We also note that he corresponding energy transfer decomposition for inductionless problems can be obtained from~\eqref{eq:gammaEnergyBalance}  by formally setting the magnetic field energies, $ E_b $ and $ E_b' $, the Maxwell stress $ \Gamma_M $, as well as all terms dependent on the induced magnetic field to zero and, as follows from~\eqref{eq:zeroPmNormalModes}, replacing~\eqrefe{eq:modeGVolume} with
\begin{equation}
\label{eq:modeGEtaZeroPm}
\modegeta = - 2 \Rey^{-1} ( \Harz^2 | \DD \modeu |^2 + 2 \alpha \Harx \Harz \Imag( \modeu^* \DD \modeu ) + \alpha^2 \Harx^2 | \modeu |^2 ). 
\end{equation}

\section{\label{sec:discussion}Results and discussion}

We now investigate the stability properties of the models established in \S\ref{sec:problemFormulation}, restricting attention to problems with purely flow-normal external magnetic field (\ie $ \Harx = 0 $ and $ \Harz = \Har $) and magnetic Prandtl number no greater than $ 10^{-4} $ (including the inductionless limit $ \Prm \searrow 0 $). Albeit small, the chosen upper boundary for $ \Prm $ encompasses all known laboratory and industrial fluids. Our focus will be on the stability of travelling gravity and Alfv\'en waves, neither of which are present in channel Hartmann flow. In addition, the behaviour of the hard instability mode, which is the free-surface analogue of the even unstable mode in channel problems \cite[][]{Takashima96}, will be examined.  All numerical work was carried out using a spectral Galerkin method for the coupled OS and induction equations for free-surface MHD \cite[][]{GiannakisFischerRosner08}.

Following a review of free surface-Poiseuille flow in \S\ref{sec:hydroProblems}, we consider in \S\ref{sec:zeroPmProblems} inductionless problems, and then, in \S\ref{sec:mhdProblems}, flows at nonzero $ \Prm $. In the interest of commonality with the literature for channel Hartmann flow, we frequently use the complex phase velocity $ c := \ii \gamma / \alpha = C + \ii \Gamma / \alpha $, where $ C $ and $ \Gamma $ are respectively the modal phase velocity and growth rate, in place of the complex growth rate $ \gamma $. The complex phase velocity will also be employed whenever reference is made to neutral-stability curves in the $ ( \Rey, \alpha ) $ plane. In particular, we consider these curves to be the loci $ \Imag( c( \Rey, \alpha ) ) = 0 $; a definition that does not necessarily agree on the $ \alpha = 0 $ axis with the equivalent one, $ \Real( \gamma( \Rey, \alpha ) ) = 0 $, in terms of $ \gamma $ (see \S\ref{sec:mhdTopEnd}). We denote throughout the critical Reynolds number for the onset of instability and the wavenumber and phase velocity of the critical mode by $ \Reyc $, $ \alphac $, and $ C_c $, respectively.  

Motivated by the discussion of the Squire transformation in \S\ref{sec:3dNormalModes}, we have opted to perform our analysis  using the parameter set $ \{ \alpha, \Rey, \Prm, \Har, \Gal, \Capil \} $, rather than, say, parameterising gravity by means of the Froude number $ \Fro $ and inclination angle $ \theta $, and surface tension by means of the Weber number $ \Web $. In particular, all ensuing calculations where $ \Rey $ is varied (namely eigenvalue contours in the $ ( \Rey, \alpha ) $ plane, and critical-Reynolds-number calculations) are performed at constant $ \{ \Prm, \Har, \Gal, \Capil \} $, as under that condition the onset of instability is governed by two-dimensional ($ \beta = \modeuy = \modeby = 0 $) modes. Computing eigenvalue contours at constant $ \Har $ and, where applicable, $ \Prm $ is also the conventional choice in the literature for channel Hartmann flow \cite[e.g.][]{Lock55,PotterKutchey73,Takashima96}. However, in free-surface Poiseuille flow these calculations are typically performed at fixed inclination angle $ \theta $ \cite[][]{Yih63,Lin67,Yih69,DeBruin74,LamBayazitoglu86,FloryanDavisKelly87}, rather than fixed $ \Gal $. Yet, as follows from~\eqref{eq:reTheta}, as long as $ \Rey \Har / \tanh( \Har / 2 ) \ll \Gal $ or $ \Rey \Har ^ 2 / ( \sech( \Har ) - 1 ) \ll \Gal $, respectively for insulating and conducting lower wall, $ \theta $ remains small, which is the case for several of the calculations presented here. Among the various dimensionless groups associated with surface tension, our capillary number $ \Capil $ is equivalent to the parameter $ S' = 3 / ( 2 \Capil ) $ employed by \cite{Yih63}, whereas, for instance, \cite{SmithDavis82}, \cite{LamBayazitoglu86}, and \cite{FloryanDavisKelly87}  use $ S = \Rey / \Capil $, $ S_t = 1 / \Web = 1 / ( \Rey \Capil ) $, and $ \zeta = 3^{1/3} ( \Gal / \cos( \theta ) )^{2/3} \sin( \theta ) / ( 2 \Capil ) $, respectively. 

Unless otherwise stated, we set $ \Gal = 8.3 \times 10^7 $, which, for a typical liquid metal with kinematic viscosity $ \nu = 3 \times 10^{-7} \metre^2 \second^{-1} $ in a $ g = 9.81 \metre\second^{-2} $ gravitational field, corresponds to $ l^3 \cos( \theta ) = ( 0.00913 \metre )^3 $. Capillary effects are of minor importance for the instabilities we wish to explore, but we nominally work at $ \Capil = 0.07 $, which is the capillary number computed for dynamic viscosity $ \mu = 1.5 \times 10^{-3} \newton \second \metre^{-2} $ and surface-tension coefficient $ \sigma = 0.1 \newton\metre^{-1} $ (both of which are typical liquid-metal values), assuming a velocity scale $ U_* = 4.7 \metre\second^{-1} $.  

\subsection{\label{sec:hydroProblems}Free-surface Poiseuille flow}

\begin{figure}
\begin{center}
\includegraphics{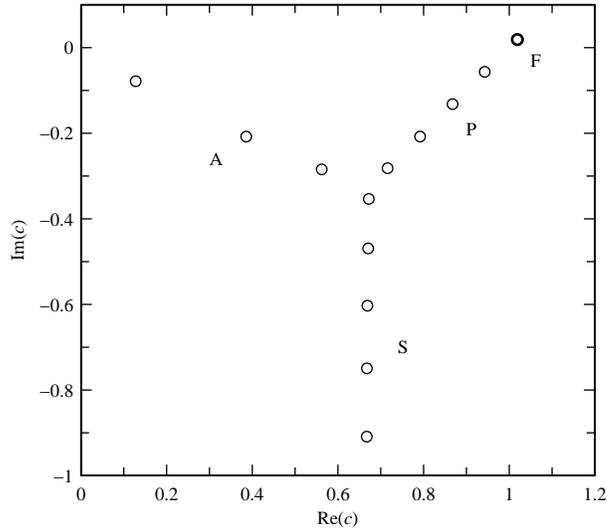}
\caption{\label{fig:spectrumAHydro}Eigenvalues for free-surface Poiseuille flow at $ \Rey = 7 \times 10^5 $, $ \alpha = 2 \times 10^{-3 } $, $ \Gal = 8.3 \times 10^7 $, and $ \Capil = 0.07 $, showing the A, P, and S branches on the complex-$ c $ plane. Mode~F, represented by a boldface marker, is unstable, and has growth-rate $ \Gamma = \alpha \Imag( c ) = 3.7447 \times 10^{-5} $ and energy transfer rates (the terms on the right-hand side of~\eqref{eq:gammaEnergyBalance} with $ \Gamma_M $, $ \Gamma_J $, $ \Gamma_\eta $, and $ \Gamma_{a J } $ set to zero) $ \Gamma_R = -7.7875 \times 10^{-6} $, $ \Gamma_\nu = -1.1048 \times 10^{-4} $, and $ \Gamma_{ a U } = 1.5572 \times 10^{-4}$.}
\end{center}
\end{figure}

Our baseline scenario is non-MHD free-surface flow with the Poiseuille velocity profile $ U( z ) = 1 - z^2 $. As shown in the spectrum in figure~\ref{fig:spectrumAHydro} and table~\ref{table:spectrumAHydro}, evaluated at $ \Rey = 7 \times 10^5 $ and $ \alpha = 2 \times 10^{-3} $, the eigenvalues form the characteristic three-branch structure in the complex plane encountered in plane Poiseuille flow \cite[][]{Mack76,DongarraStraughanWalker96,Kirchner00,MelenkKirchnerSchwab00}, with the difference that, because they lack reflection symmetry with respect to $ z $, the modes do not arise as symmetric--antisymmetric pairs. Following standard nomenclature \cite[][]{Mack76}, we label the branches A, P, and S, where $ 0 < \Real( c ) < \langle U \rangle = 2/ 3 $ and $ \langle U \rangle < \Real( c ) < 1 $, respectively for modes in the A and P families, while S modes have $ C = \langle U \rangle $, asymptotically as $ \Real( c ) \to -\infty $ \cite[][]{GroschSalwen64}. Among the A, P, and S modes, only the ones at the top end of the spectrum carry appreciable surface energy. For instance, in table~\ref{table:spectrumAHydro}, $ E_a / E  $ drops from 0.00652 for mode P$_1 $ to $ \ord( 10^{-7} ) $ for mode S$_2$. 

In addition to the above shear modes, the depicted free-surface spectrum contains an unstable surface mode, denoted by F, which propagates downstream with phase velocity greater than the steady-state velocity at the free surface (\ie $ \Real( c ) > 1 $). This so-called \emph{soft instability} is driven by viscous stresses acting on the free surface \cite[][]{Yih63}. In the notation of \S\ref{sec:energyBalance}, the corresponding energy transfer rate $ \Gamma_{aU} > | \Gamma_\nu + \Gamma_R | $ to mode~F exceeds the net rate of energy dissipated by viscous and Reynolds stresses \cite[see also][]{SmithDavis82}, resulting in a positive growth rate $ \Gamma = \Gamma_{aU} + \Gamma_\nu + \Gamma_R $.

\begin{table}
\begin{tabular*}{\linewidth}{@{\extracolsep{\fill}}llrl}
& & \multicolumn{1}{c}{$c$} & \multicolumn{1}{c}{$E_a/E$}\\
\cline{3-4}
1 &  $ \mbox{F} $ &  $ 1.019322365642126\mathrm{E}+00 + 1.872373565774912\mathrm{E}-02 \, \ii $ & $ 3.81670\mathrm{E}-02 $ \\ 
2 &  $ \mbox{P}_{ 1 } $ &  $ 9.430932551528158\mathrm{E}-01 - 5.660800610069469\mathrm{E}-02 \, \ii $ & $ 6.52344\mathrm{E}-03 $ \\ 
3 &  $ \mbox{A}_{ 1 } $ &  $ 1.280197187921976\mathrm{E}-01 - 7.844228782612407\mathrm{E}-02 \, \ii $ & $ 1.61245\mathrm{E}-04 $ \\ 
4 &  $ \mbox{P}_{ 2 } $ &  $ 8.676593690519436\mathrm{E}-01 - 1.322655312947944\mathrm{E}-01 \, \ii $ & $ 6.09230\mathrm{E}-04 $ \\ 
5 &  $ \mbox{P}_{ 3 } $ &  $ 7.920814474574365\mathrm{E}-01 - 2.078761371940656\mathrm{E}-01 \, \ii $ & $ 1.14211\mathrm{E}-04 $ \\ 
6 &  $ \mbox{A}_{ 2 } $ &  $ 3.862388506852979\mathrm{E}-01 - 2.079139567980078\mathrm{E}-01 \, \ii $ & $ 1.41771\mathrm{E}-04 $ \\ 
7 &  $ \mbox{P}_{ 4 } $ &  $ 7.161372374734705\mathrm{E}-01 - 2.819337921185398\mathrm{E}-01 \, \ii $ & $ 2.31032\mathrm{E}-05 $ \\ 
8 &  $ \mbox{A}_{ 3 } $ &  $ 5.621911688341056\mathrm{E}-01 - 2.845782600055865\mathrm{E}-01 \, \ii $ & $ 3.99770\mathrm{E}-05 $ \\ 
9 &  $ \mbox{S}_{ 1 } $ &  $ 6.724206086971892\mathrm{E}-01 - 3.536832210862633\mathrm{E}-01 \, \ii $ & $ 3.85934\mathrm{E}-06 $ \\ 
10 &  $ \mbox{S}_{ 2 } $ &  $ 6.706989371744410\mathrm{E}-01 - 4.693930705398125\mathrm{E}-01 \, \ii $ & $ 1.01726\mathrm{E}-07 $ \\ 

\end{tabular*}
\caption{\label{table:spectrumAHydro}Complex phase velocity $ c $ and free-surface energy $ E_a $ (normalised by the total modal energy $ E = E_u + E_a $) of the 10 least stable modes of the spectrum in figure~\ref{fig:spectrumAHydro}. The modes are tabulated in order of decreasing $ \Imag( c ) $, and labelled $ \mathrm{ A }_i $, $ \mathrm{ P }_i $, $ \mathrm{ S }_i $, or F according to their family, where $ i $ denotes the rank, again in order of decreasing $ \Imag( c ) $, within a given family.}
\end{table}

Besides mode~F, whenever the speed of propagation of surface waves in the absence of a basic flow is large compared to the steady-state velocity, the spectrum also contains an upstream-propagating ($ \Real( c ) < 0 $) surface mode \cite[\eg figure~4 in][]{GiannakisFischerRosner08}. As $ \Rey $ grows, that mode joins the A branch, and eventually becomes unstable. The latter instability, oftentimes referred to as the \emph{hard instability} \cite[][]{Lin67,DeBruin74,FloryanDavisKelly87}, is the free-surface analogue of the Tollmien-Schlichting wave in plane Poiseuille flow \cite[][]{Lin44}, \ie it is caused by positive Reynolds stress associated with a critical layer that develops for suitable values of the mode's phase velocity. 

As shown in \figrefa{fig:reAlphaZeroPm}, the growth-rate contours of the hard mode in the $ ( \Rey, \alpha ) $ plane are qualitatively similar to those of the unstable mode in channel flow \cite[][]{Shen54}. In fact, if gravitational and surface-tension forces are decreased to zero the critical parameters of the hard mode approach the $ ( \Reyc, \alphac, C_c ) = (5772.2, 1.021, 0.264) $ values computed for plane Poiseuille flow \cite[][]{Orszag71}, revealing their common nature. Increasing $ \Gal $ results in the instability region extending to progressively larger wavenumbers (see \eg the $ \Har = 0 $ results in table~\ref{table:criticalHardZeroPm}), where the upper and lower branches of the neutral-stability curve $ \Imag( c ) = 0 $ intersect in a cusp-like manner. However, the hard mode's critical Reynolds number does not vary monotonically with the strength of the flow-normal gravitational force \cite[][]{DeBruin74,FloryanDavisKelly87}. In particular, for the $ \Capil = 0.07 $ capillary number used here, $ \Reyc $ decreases with $ \Gal \lesssim 10^{7} $ (\eg in table~\ref{table:criticalHardZeroPm}, $ \Reyc $ drops to $ 3711.3 $ for $ \Gal = 8.3 \times 10^{6} $), but becomes an increasing function of the Galilei number for sufficiently strong gravitational fields, eventually exceeding the corresponding critical Reynolds number for channel flow. For instance, in \figrefa{fig:reAlphaZeroPm} the hard mode's critical parameters are $ ( \Reyc, \alphac, C_c ) = (7361.0, 2.815, 0.184 ) $.

As for the soft mode, it is evident from the structure of the $ \Imag( c ) = 0 $ contour in \figrefa{fig:reAlphaZeroPm}, which runs parallel to the $ \log( \alpha ) $ axis for $ \alpha \ll 1 $, that its region of instability in the $ (\Rey, \alpha ) $ plane extends to arbitrarily small wavenumbers. This makes the soft mode amenable to study using regular perturbation theory for large wavelengths ($ \alpha \searrow 0 $), when, in contrast, an analytic treatment of the hard mode would require the full machinery of singular asymptotic expansions \cite[e.g.][]{DrazinReid81}. In particular, \cite{Yih63} has established that the lower branch of the soft mode's neutral-stability curve is the $ \alpha = 0 $ axis and further, that its upper branch, shown in \figrefa{fig:reAlphaZeroPm}, emanates from a bifurcation point located at $ ( \Reyb, 0 ) = ( ( 5 \Gal / 8 )^{1/2}, 0 ) $, with corresponding phase velocity $ C_b = 2 $ (see also appendix~\ref{app:pertZeroPm}).

As a check on the proximity of the bifurcation point $ ( \Reyb, 0 ) $ to the critical point $ ( \Reyc, \alphac ) $ of the soft mode for the problem in \figrefa{fig:reAlphaZeroPm}, which is strongly suggested by the direction of the $ \Imag( c ) = 0 $ contour for $ \alpha \ll 1 $, we have numerically computed the minimum Reynolds number $ \Rey_m $ for instability at fixed $ \alpha = 10^{-5} $. The $ \Rey_m = 7202.4298 $ numerical result is very close to the analytically determined value $ \Reyb = 7202.4301 $ for the Reynolds number at the bifurcation point, as was also observed in a number of calculations with $ \Gal \in [ 10^3, 10^9 ] $. Still, for our reference problem with $ ( \Gal, \Capil ) = ( 8.3 \times 10^7, 0.07 ) $, $ \Rey_m $ is smaller than the corresponding $ \Reyb $ by an amount of order $ \ord( 10^{-8} ) $, which in all likelihood is not due to numerics (\eg the discrepancy did not disappear by increasing the polynomial degree of the discretisation scheme). This observation is consistent with a fourth-order asymptotic result that for any capillary number $ \Capil $ there exists a lower bound in $ \Gal $ above which $ \dd \Rey / \dd \alpha $ is negative on the $ \Imag( c ) = 0 $ contour, in the neighbourhood of the bifurcation point \cite[][]{GiannakisRosnerFischer08}. For  $ \Capil = 0.07 $ that lower bound amounts to $ \Gal \approx 3.13 \times 10^5 $, indicating that for the problem in \figrefa{fig:reAlphaZeroPm}, $ \alphac = 0 $ is not an exact statement. However, we expect the smallness of $ \alphac $ to render any unstable modes with $ \Rey < \Reyb $ irrelevant in a laboratory context, even if the true critical Reynolds number were to deviate significantly from $ \Reyb $. 

\subsection{\label{sec:zeroPmProblems}Inductionless free-surface Hartmann flow}

\begin{figure}
\begin{center}
\includegraphics{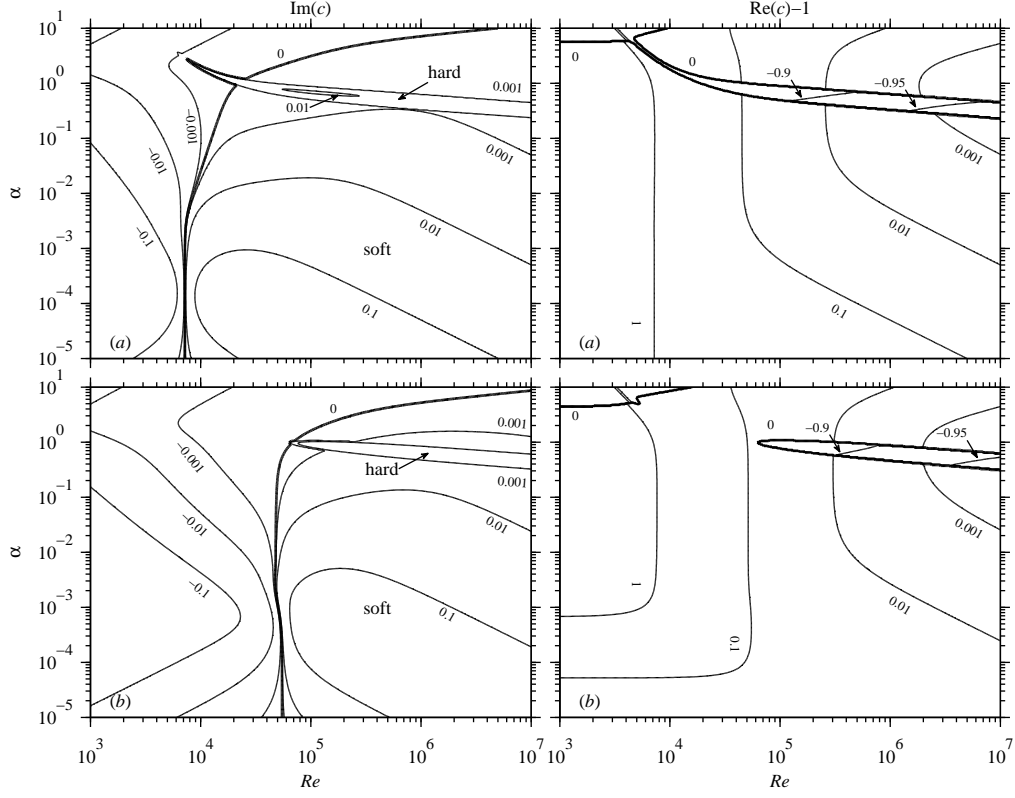}
\end{center}
\caption{\label{fig:reAlphaZeroPm}Imaginary (left-hand panels) and real (right-hand panels) parts of the complex phase velocity $ c $ of the least stable mode in the $ ( \Rey, \alpha ) $ plane, computed at constant Galilei and capillary numbers $ \Gal = 8.3 \times 10^7 $ and $ \Capil = 0.07 $, for free-surface Poiseuille flow ({\it a}), and inductionless free-surface Hartmann flow with $ \Har = 3 $ ({\it b}). The regions of instability for the soft and hard modes, indicated in the left-hand panels, are distinguishable by the corresponding phase velocity $ \Real( c ) $, which exceeds unity in the case of the soft mode, but is less than the mean steady-state speed $ \langle U \rangle $ for the hard mode. In panels ({\it a}) and ({\it b}), $ \langle U \rangle $ is $2/3$ and $0.742$, respectively.}
\end{figure}

In inductionless Hartmann flow, the external magnetic field influences normal-mode stability on one hand by modifying the steady-state velocity profile~\eqrefa{eq:generalUB}, therefore altering the energy transfer to the perturbations mediated by viscosity (represented by the terms $ \Gamma_R $ and $ \Gamma_{aU} $ in~\eqref{eq:gammaEnergyBalance}) and, on the other hand, by means of the Lorentz force \eqrefb{eq:linearizedGovEqs}, whose only nonzero component $ f_x := - \Har^2 \Rey^{-1} u_x $ is streamwise. The latter has a direct dissipative effect associated with resistivity (the energy transfer term $ \Gamma_\eta $), but may also indirectly affect $ \Gamma_R $, $ \Gamma_{ a U } $, and the viscous-dissipation rate $ \Gamma_\nu $ by modifying the perturbed velocity field. In free-surface problems, as is the case with their fixed-boundary counterparts \cite[][]{Lock55,PotterKutchey73,Takashima96}, the combined outcome of the external magnetic field is to suppress instabilities. In fact, even moderate Hartmann numbers ($ \Har \sim 3$) are sufficient to shift the onset of the soft and hard instabilities to Reynolds numbers significantly higher than in non-MHD flows, in the manner illustrated by the eigenvalue contour plots in \figrefb{fig:reAlphaZeroPm}. 

\begin{figure}
\begin{center}
\includegraphics{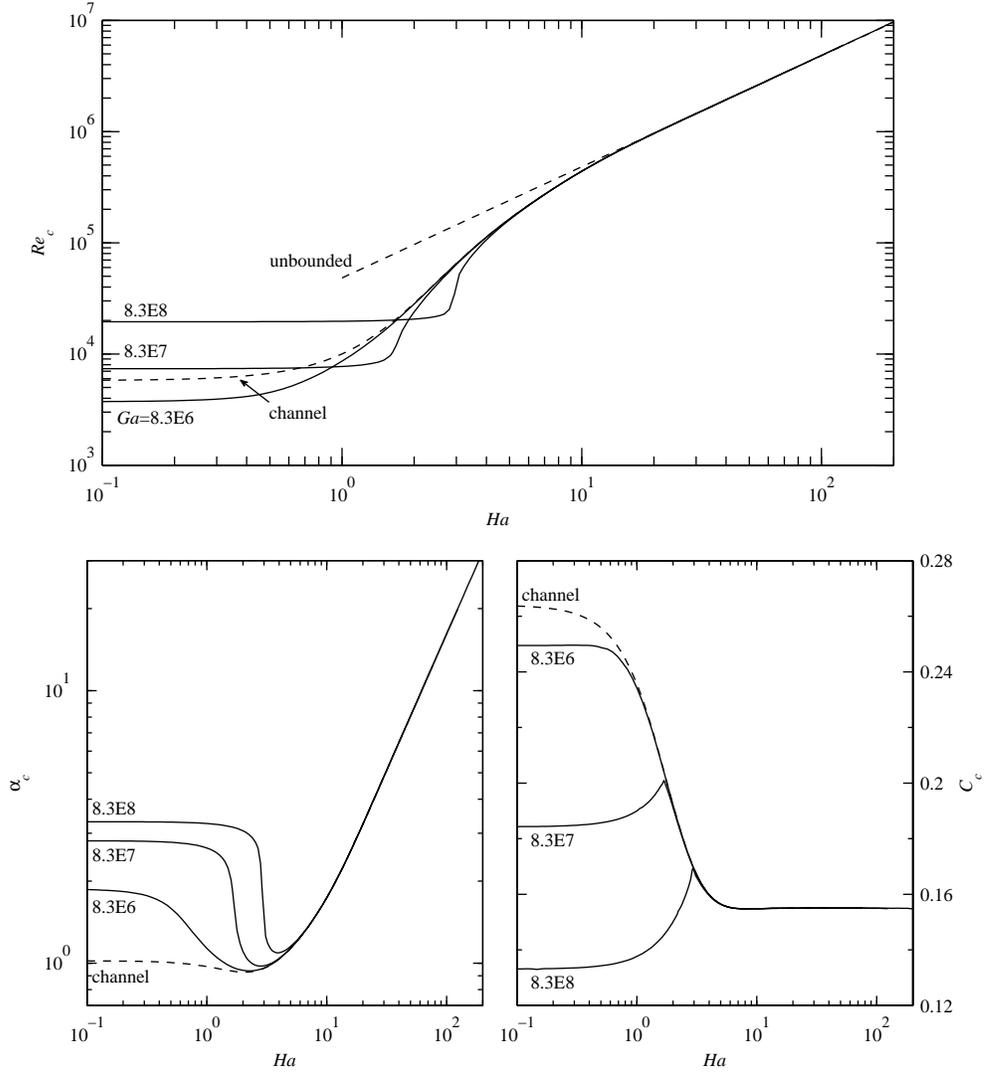}
\end{center}
\caption{\label{fig:criticalHardZeroPm}Critical Reynolds number $ \Reyc $, wavenumber $ \alphac $, and phase velocity $ C_c $ of the hard mode in inductionless free-surface Hartmann flow with $ \Gal / ( 8.3 \times 10^7 ) \in \{ 0.1, 1, 10 \} $, and values of the Hartmann number $ \Har $ logarithmically spaced on the interval $ [ 0.1, 200 ] $. The capillary number is $ \Capil = 0.07 $ throughout. The critical parameters for channel Hartmann flow \cite[][]{Takashima96}, as well as the result $ \Reyc = \text{48,250}\,\Har $ for the unbounded Hartmann layer \cite[][]{LingwoodAlboussiere99}, are plotted in dashed lines for reference.}
\end{figure}

The critical Reynolds number, wavenumber, and phase velocity of the hard mode, computed numerically in~figure~\ref{fig:criticalHardZeroPm} as a function of the Hartmann number $ \Har \in [ 0.1, 200 ] $ for representative values of the Galilei number $ \Gal / ( 8.3 \times 10^7 ) \in \{ 0.1, 1, 10 \} $, exhibit two distinct types of behaviour, depending on the relative strength of the gravitational and Lorentz forces. The first of these occurs when the Lorentz force is weak compared to gravity (\eg the $ \Gal = 8.3 \times 10^8 $ example in figure~3 for $ \Har \lesssim 2 $), and is characterised by small variation of the critical parameters with $ \Har $. As observed in the $ ( \Rey, \alpha ) $ plane, the main influence of the applied field in this regime is to reduce the wavenumber bandwidth of the cusp-like tip of the hard mode's instability region, with little change in the position $ ( \Reyc, \alphac ) $ of the intersection point between the upper and lower branches of the neutral-stability curve. Eventually, however, the tip collapses, and $ \alphac $  rapidly decreases towards the corresponding channel-flow result. For Hartmann numbers larger than that threshold the behaviour of the hard mode's critical parameters changes character and, as can be deduced by comparing table~\ref{table:criticalHardZeroPm} to the calculations in table~1 of \cite{Takashima96}, it becomes nearly identical to that of the unstable mode in inductionless channel Hartmann flow. In particular, the wavelength of the critical mode becomes shorter, as expected from the decreasing thickness of the Hartmann layer \cite[][]{Lock55}, and for sufficiently strong fields the critical Reynolds number as a function of $ \Har $ is well described by the $ \Reyc = \text{48,250}\,\Har $ linear increase computed by \cite{LingwoodAlboussiere99} for the unbounded Hartmann layer. In separate test calculations, where the Lorentz-force terms in~\eqref{eq:orrSommerfeldZeroPm} were set to zero but the Hartmann velocity profile was retained, we have observed that the critical parameters of the hard mode remain close to the results in table~\ref{table:criticalHardZeroPm}, in agreement with the observation by \cite{Lock55} that the principal contribution to the behaviour of $ ( \Reyc, \alphac, C_c ) $ comes from the modification of basic flow, rather than electromagnetic forces acting on the perturbed velocity field.

\begin{table}
\begin{tabular*}{\linewidth}{@{\extracolsep{\fill}}llll}
$ \Har $ & $ \Reyc $ & $ \alphac $ & $ C_c $ \\
\hline
\multicolumn{4}{l}{$ \Gal = 8.3 \times 10^6 $}\\
 $ 0 $ &  $ 3.7113036\mathrm{E}+03 $ & $ 1.87198\mathrm{E}+00 $ & $ 2.49413\mathrm{E}-01 $ \\ 
 $ 0.5 $ &  $ 4.5019913\mathrm{E}+03 $ & $ 1.59323\mathrm{E}+00 $ & $ 2.48917\mathrm{E}-01 $ \\ 
 $ 1 $ &  $ 8.5959446\mathrm{E}+03 $ & $ 1.13443\mathrm{E}+00 $ & $ 2.34231\mathrm{E}-01 $ \\ 
 $ 2 $ &  $ 2.8176098\mathrm{E}+04 $ & $ 9.41259\mathrm{E}-01 $ & $ 1.92059\mathrm{E}-01 $ \\ 
 $ 5 $ &  $ 1.6404990\mathrm{E}+05 $ & $ 1.13450\mathrm{E}+00 $ & $ 1.56426\mathrm{E}-01 $ \\ 
 $ 10 $ &  $ 4.3981065\mathrm{E}+05 $ & $ 1.73916\mathrm{E}+00 $ & $ 1.54789\mathrm{E}-01 $ \\ 
 $ 20 $ &  $ 9.6176655\mathrm{E}+05 $ & $ 3.23761\mathrm{E}+00 $ & $ 1.55011\mathrm{E}-01 $ \\ 
 $ 50 $ &  $ 2.4155501\mathrm{E}+06 $ & $ 8.07657\mathrm{E}+00 $ & $ 1.55029\mathrm{E}-01 $ \\ 
 $ 100 $ &  $ 4.8311017\mathrm{E}+06 $ & $ 1.61537\mathrm{E}+01 $ & $ 1.55030\mathrm{E}-01 $ \\ 

\\
\multicolumn{4}{l}{$ \Gal = 8.3 \times 10^7 $}\\
 $ 0 $ &  $ 7.3610164\mathrm{E}+03 $ & $ 2.81462\mathrm{E}+00 $ & $ 1.84251\mathrm{E}-01 $ \\ 
 $ 0.5 $ &  $ 7.4343292\mathrm{E}+03 $ & $ 2.77817\mathrm{E}+00 $ & $ 1.85689\mathrm{E}-01 $ \\ 
 $ 1 $ &  $ 7.7154319\mathrm{E}+03 $ & $ 2.64647\mathrm{E}+00 $ & $ 1.89974\mathrm{E}-01 $ \\ 
 $ 2 $ &  $ 2.3929863\mathrm{E}+04 $ & $ 1.10416\mathrm{E}+00 $ & $ 1.91208\mathrm{E}-01 $ \\ 
 $ 5 $ &  $ 1.6378495\mathrm{E}+05 $ & $ 1.13615\mathrm{E}+00 $ & $ 1.56420\mathrm{E}-01 $ \\ 
 $ 10 $ &  $ 4.3979016\mathrm{E}+05 $ & $ 1.73922\mathrm{E}+00 $ & $ 1.54788\mathrm{E}-01 $ \\ 
 $ 20 $ &  $ 9.6176624\mathrm{E}+05 $ & $ 3.23764\mathrm{E}+00 $ & $ 1.55011\mathrm{E}-01 $ \\ 
 $ 50 $ &  $ 2.4155501\mathrm{E}+06 $ & $ 8.07657\mathrm{E}+00 $ & $ 1.55029\mathrm{E}-01 $ \\ 
  $ 100 $ &  $ 4.8311017\mathrm{E}+06 $ & $ 1.61537\mathrm{E}+01 $ & $ 1.55030\mathrm{E}-01 $ \\ 

\\
\multicolumn{4}{l}{$ \Gal = 8.3 \times 10^8 $}\\
 $ 0 $ &  $ 1.9476764\mathrm{E}+04 $ & $ 3.30597\mathrm{E}+00 $ & $ 1.33088\mathrm{E}-01 $ \\ 
 $ 0.5 $ &  $ 1.9531440\mathrm{E}+04 $ & $ 3.29373\mathrm{E}+00 $ & $ 1.34239\mathrm{E}-01 $ \\ 
 $ 1 $ &  $ 1.9704003\mathrm{E}+04 $ & $ 3.25532\mathrm{E}+00 $ & $ 1.37611\mathrm{E}-01 $ \\ 
 $ 2 $ &  $ 2.0605875\mathrm{E}+04 $ & $ 3.06079\mathrm{E}+00 $ & $ 1.50076\mathrm{E}-01 $ \\ 
 $ 5 $ &  $ 1.6107058\mathrm{E}+05 $ & $ 1.15339\mathrm{E}+00 $ & $ 1.56349\mathrm{E}-01 $ \\ 
 $ 10 $ &  $ 4.3958470\mathrm{E}+05 $ & $ 1.73985\mathrm{E}+00 $ & $ 1.54786\mathrm{E}-01 $ \\ 
 $ 20 $ &  $ 9.6176320\mathrm{E}+05 $ & $ 3.23764\mathrm{E}+00 $ & $ 1.55011\mathrm{E}-01 $ \\ 
 $ 50 $ &  $ 2.4155501\mathrm{E}+06 $ & $ 8.07657\mathrm{E}+00 $ & $ 1.55029\mathrm{E}-01 $ \\ 
 $ 100 $ &  $ 4.8311017\mathrm{E}+06 $ & $ 1.61537\mathrm{E}+01 $ & $ 1.55030\mathrm{E}-01 $ \\ 

\end{tabular*}
\caption{\label{table:criticalHardZeroPm}Critical Reynolds number $ \Reyc $, wavenumber $ \alphac $, and phase velocity $ C_c $ of the hard mode in inductionless free-surface Hartmann flow, computed for Galilei number $ \Gal / (8.3\times 10^7) \in \{ 0.1, 1, 10 \} $, capillary number $ \Capil = 0.07 $, and representative values of the Hartmann number $ \Har $ in the interval $ [ 0, 100 ] $.}
\end{table}

\begin{figure}
\begin{center}
\includegraphics{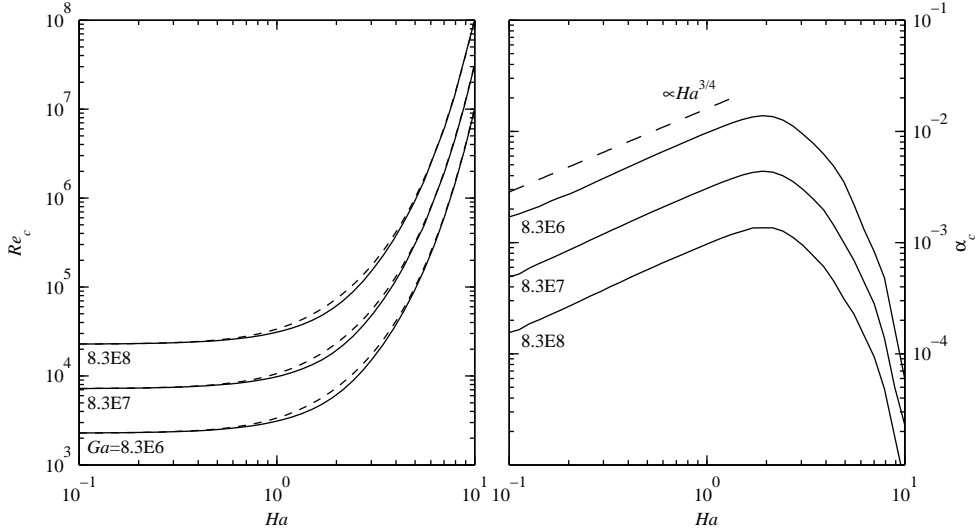}
\end{center}
\caption{\label{fig:criticalSoftZeroPm}Critical Reynolds number $ \Reyc $ and wavenumber $ \alphac $ as a function of the Hartmann number for the soft mode in inductionless free-surface Hartmann flow with $ \Gal / ( 8.3 \times 10^7 ) \in \{ 0.1, 1, 10 \} $. The capillary number is $ \Capil = 0.07 $ throughout. The Reynolds number at the bifurcation point $ \Reyb( \Har ) $, given by~\eqrefa{eq:criticalSoft}, is plotted in dotted lines. The critical phase velocity $ C_c $ for $ \Gal = 8.3 \times 10^7 $ is shown in figure~\ref{fig:criticalSoft}. The fluctuations present in the $ \Har \lesssim 0.2 $ and $ \Har \gtrsim 3 $ results for $ \alphac $ are a consequence of the ill-conditioning of the calculation described in the caption to table~\ref{table:criticalSoftZeroPm}.}
\end{figure}

As for the soft mode, it follows from the large-wavelength analysis in appendix~\ref{app:pertZeroPm} that when $ \Har $ is nonzero the $ \alpha = 0 $ axis remains part of its neutral-stability curve, and a bifurcation point $ ( \Reyb, 0 ) $, from which the upper part of the neutral-stability curve branches off, is again present in the $ ( \Rey, \alpha ) $ plane. In particular, the position of the bifurcation point on  the $ \alpha = 0 $ axis and the corresponding modal phase velocity $ C_b $, respectively determined from the coefficients $ \gamma_2 $ and $ \gamma_1 $ in the perturbative expansion $ \gamma = \gamma_1 \alpha + \gamma_2 \alpha^2 + \ord( \alpha )^3 $ for the complex growth rate $ \gamma $, are given by 
\begin{subequations}
\label{eq:criticalSoft}
\begin{align}
\taga
\Reyb  &= \frac{ (8 \Gal )^{1/2}  \sinh( \Har / 2 ) ( \Har - \tanh( \Har ) )^{1/2} }{ ( \Har \coth( \Har / 2 ) \sech^3( \Har ) ( 2 \Har ( 2 + \cosh( 2 \Har ) ) - 3 \sinh( 2 \Har ) ) )^{1/2} }, \\
\tagb
C_b &= 1 + \sech( \Har ),
\end{align}
\end{subequations}
where $ \Reyb \searrow ( 5 \Gal / 8 )^{1/2} $ and $ C_b \nearrow 2 $ tend to their non-MHD values when $ \Har $ is decreased to zero. Contrary to the non-MHD case examined in \S\ref{sec:hydroProblems}, however, for Hartmann numbers lying in a relatively narrow band the bifurcation point becomes clearly separated from the critical point $ (\Reyc, \alphac ) $. This is illustrated in figure~\ref{fig:criticalSoftZeroPm} and table~\ref{table:criticalSoftZeroPm}, where for the examined values of the Galilei number $ \Gal / 8.3 \times 10^7 \in \{ 0.1, 1, 10 \} $ the critical wavenumber follows an $ \alphac \propto \Har^{3/4} $ increase for $ \Har \lesssim 2 $ (\eg reaching $ \alpha \approx 0.0044 $ for $ \Har \approx 1.9 $ and $ \Gal = 8.3 \times 10^7 $), before rapidly diminishing again at larger Hartmann numbers. As is the case with the $ \Reyb \propto \Gal^{1/2} $ scaling in \eqrefa{eq:criticalSoft}, all the $ \Reyc $ results nearly collapse to a single curve when scaled by $ \Gal^{1/2} $. The calculations also suggest that an $ \alphac \propto \Gal^{-1/2} $ scaling applies for the critical wavenumber, but this cannot be firmly confirmed with the presently attainable level of numerical accuracy and precision. As expected, whenever $ \alphac $ is small, the deviation of the critical Reynolds number and phase velocity from~\eqref{eq:criticalSoft} is less significant, but even when $ \alphac $ is close to its maximum value the relative error is still acceptable. In table~\ref{table:criticalSoftZeroPm}, for instance, $ \Reyb $ overestimates $ \Reyc $ by approximately 20\% when $ \Har = 2 $, while $ C_b $ underestimates $ C_c $ by 4\%. The influence of $ \alphac > 0 $ on the critical Reynolds number can also be observed by examining the $ \Imag( c ) = 0 $ contour in \figrefb{fig:reAlphaZeroPm}, where $ \Rey $ decreases from $ 5.49 \times 10^4 $ when $ \alpha = 10^{-5} $ to $ 4.74 \times 10^4 \approx \Reyc $ when $ \alpha = 0.0032 \approx \alphac $.

\begin{table}
\begin{tabular*}{\linewidth}{@{\extracolsep{\fill}}llllll}
$ \Har $ & $ \Reyc / \Gal^{1/2} $ & $ \Reyb / \Gal^{1/2} $ & $ \alphac \Gal^{1/2} $ & $ C_c $ & $ C_b $ \\
\hline
\multicolumn{4}{l}{$ \Gal = 8.3 \times 10^6 $}\\
 $ 0.1 $ &  $ 7.9405\mathrm{E}-01 $ & $ 7.9417\mathrm{E}-01 $ & $ 4.850\mathrm{E}+00 $ & $ 1.9950\mathrm{E}+00 $ & $ 1.9950\mathrm{E}+00 $ \\ 
 $ 0.2 $ &  $ 8.0418\mathrm{E}-01 $ & $ 7.7498\mathrm{E}-01 $ & $ 7.854\mathrm{E}+00 $ & $ 1.9810\mathrm{E}+00 $ & $ 1.9803\mathrm{E}+00 $ \\ 
 $ 1 $ &  $ 1.0784\mathrm{E}+00 $ & $ 1.1687\mathrm{E}+00 $ & $ 2.795\mathrm{E}+01 $ & $ 1.7026\mathrm{E}+00 $ & $ 1.6481\mathrm{E}+00 $ \\ 
 $ 2 $ &  $ 2.1251\mathrm{E}+00 $ & $ 2.5619\mathrm{E}+00 $ & $ 3.978\mathrm{E}+01 $ & $ 1.3185\mathrm{E}+00 $ & $ 1.2658\mathrm{E}+00 $ \\ 
 $ 5 $ &  $ 3.2659\mathrm{E}+01 $ & $ 3.5000\mathrm{E}+01 $ & $ 8.500\mathrm{E}+00 $ & $ 1.0137\mathrm{E}+00 $ & $ 1.0135\mathrm{E}+00 $ \\ 
 $ 8 $ &  $ 5.2610\mathrm{E}+02 $ & $ 5.4649\mathrm{E}+02 $ & $ 1.448\mathrm{E}+00 $ & $ 1.0007\mathrm{E}+00 $ & $ 1.0007\mathrm{E}+00 $ \\ 

\\
\multicolumn{4}{l}{$ \Gal = 8.3 \times 10^7 $}\\
 $ 0.1 $ &  $ 7.9405\mathrm{E}-01 $ & $ 7.9417\mathrm{E}-01 $ & $ 4.483\mathrm{E}+00 $ & $ 1.9951\mathrm{E}+00 $ & $ 1.9950\mathrm{E}+00 $ \\ 
 $ 0.2 $ &  $ 8.0417\mathrm{E}-01 $ & $ 8.0498\mathrm{E}-01 $ & $ 7.779\mathrm{E}+00 $ & $ 1.9810\mathrm{E}+00 $ & $ 1.9803\mathrm{E}+00 $ \\ 
 $ 1 $ &  $ 1.0783\mathrm{E}+00 $ & $ 1.1687\mathrm{E}+00 $ & $ 2.776\mathrm{E}+01 $ & $ 1.7027\mathrm{E}+00 $ & $ 1.6481\mathrm{E}+00 $ \\ 
 $ 2 $ &  $ 2.1249\mathrm{E}+00 $ & $ 2.5619\mathrm{E}+00 $ & $ 3.984\mathrm{E}+01 $ & $ 1.3186\mathrm{E}+00 $ & $ 1.2658\mathrm{E}+00 $ \\ 
 $ 5 $ &  $ 3.2666\mathrm{E}+01 $ & $ 3.5000\mathrm{E}+01 $ & $ 8.254\mathrm{E}+00 $ & $ 1.0137\mathrm{E}+00 $ & $ 1.0135\mathrm{E}+00 $ \\ 
 $ 8 $ &  $ 5.2616\mathrm{E}+02 $ & $ 5.4649\mathrm{E}+02 $ & $ 1.342\mathrm{E}+00 $ & $ 1.0007\mathrm{E}+00 $ & $ 1.0007\mathrm{E}+00 $ \\ 

\\
\multicolumn{4}{l}{$ \Gal = 8.3 \times 10^8 $}\\
 $ 0.1 $ &  $ 7.9405\mathrm{E}-01 $ & $ 7.9417\mathrm{E}-01 $ & $ 4.453\mathrm{E}+00 $ & $ 1.9951\mathrm{E}+00 $ & $ 1.9950\mathrm{E}+00 $ \\ 
 $ 0.2 $ &  $ 8.0417\mathrm{E}-01 $ & $ 8.0498\mathrm{E}-01 $ & $ 7.702\mathrm{E}+00 $ & $ 1.9810\mathrm{E}+00 $ & $ 1.9803\mathrm{E}+00 $ \\ 
 $ 1 $ &  $ 1.0783\mathrm{E}+00 $ & $ 1.1687\mathrm{E}+00 $ & $ 2.799\mathrm{E}+01 $ & $ 1.7027\mathrm{E}+00 $ & $ 1.6481\mathrm{E}+00 $ \\ 
 $ 2 $ &  $ 2.1249\mathrm{E}+00 $ & $ 2.5619\mathrm{E}+00 $ & $ 3.942\mathrm{E}+01 $ & $ 1.3186\mathrm{E}+00 $ & $ 1.2658\mathrm{E}+00 $ \\ 
 $ 5 $ &  $ 3.2644\mathrm{E}+01 $ & $ 3.5000\mathrm{E}+01 $ & $ 9.348\mathrm{E}+00 $ & $ 1.0137\mathrm{E}+00 $ & $ 1.0135\mathrm{E}+00 $ \\ 
 $ 8 $ &  $ 5.2632\mathrm{E}+02 $ & $ 5.4649\mathrm{E}+02 $ & $ 1.225\mathrm{E}+00 $ & $ 1.0007\mathrm{E}+00 $ & $ 1.0007\mathrm{E}+00 $ \\ 

\end{tabular*}
\caption{\label{table:criticalSoftZeroPm}Critical Reynolds number $ \Reyc $, wavenumber $ \alphac $, and phase velocity $ C_c $ of the soft mode in inductionless free-surface Hartmann flow, computed for Galilei number $ \Gal / (8.3\times 10^7) \in \{ 0.1, 1, 10 \} $, capillary number $ \Capil = 0.07 $, and representative values of the Hartmann number $ \Har $ in the interval $ [ 0.1, 8 ] $. Also shown are the Reynolds number $ \Reyb $ at the bifurcation point and the corresponding phase velocity $ C_b $, determined by~\eqref{eq:criticalSoft}. In order to illustrate the dependence of the critical parameters on $ \Gal $, the results for $ \Reyc $ and $ \Reyb $ have been scaled by $ \Gal^{1/2} $, while $ \alphac $ has been scaled by $ \Gal^{-1/2} $. We remark that because the $ \Imag(c) $ contours for mode~F, which participates in the soft instability, are nearly parallel to the $ \log( \alpha ) $ axis when $ \alpha \ll 1 $ (see figure~\ref{fig:reAlphaZeroPm}\,{\it b}), and the gradient of $ \Real( \gamma( \Rey, \alpha ) ) $ becomes shallow as $ \Har $ grows (this is a consequence of the strong-field neutrality of mode~F discussed in the main text), critical-parameter calculations for the soft instability are significantly more poorly conditioned than the corresponding ones for the hard mode. As a result, the number of attained significant digits in the calculations is smaller than in table~\ref{table:criticalHardZeroPm}, especially so for $ \alphac $. In addition, there is an evidence of an $ \ord( 10^{-4} ) $ systematic drift in the results for $ \alphac $  when the optimisation solver used to compute $ ( \Reyc, \alphac, C_c ) $ is restarted with initial conditions determined from the output of preceding iterations, indicating that, with the current computational resources, some of our results have not yet reached an asymptotic limit. However, we do not expect this to impart significant changes to the shape of the curves in figure~\ref{fig:criticalSoftZeroPm}.}
\end{table}

The agreement between the analytical results for $ ( \Reyb, C_b ) $ and the numerically computed values for $ ( \Reyc, C_c ) $ steadily improves as $ \alphac( \Har ) $ enters the decreasing phase ($ \Har \gtrsim 2 $ in figure~\ref{fig:criticalSoftZeroPm}). As such, \eqrefa{eq:criticalSoft} can be used to deduce that at sufficiently large $ \Har $ the soft mode's critical Reynolds number $ \Reyc \sim (\Gal/ \Har)^{1/2} \exp( \Har )$ increases exponentially with the Hartmann number, and its critical phase velocity $ C_c \sim 1 + 2 \exp( - \Har ) $ decreases exponentially towards unity. Similar small-$\alpha$ calculations (see~\eqref{eq:strongFieldHydroHartmann} and~\eqref{eq:strongFieldZeroPmPoiseuille}) lead to the results that $ \Reyb $ also increases exponentially in (physically unrealistic) non-MHD problems with the Hartmann velocity profile, but only quadratically when the Lorentz-force terms in~\eqref{eq:orrSommerfeldZeroPm} are retained while the velocity profile keeps its non-MHD parabolic form. Therefore, the formation of the Hartmann velocity profile is the main driver of the critical-parameter behaviour of the soft instability as well, although it should be noted that the exponent in~\eqrefa{eq:strongFieldHydroHartmann} is smaller than the corresponding one derived from~\eqrefa{eq:criticalSoft} and, as a result, the critical Reynolds number of the full problem, including both Lorentz forces and the Hartmann velocity profile, outgrows that of the non-MHD test problem without bound.      

\begin{figure}
\begin{center}
\includegraphics{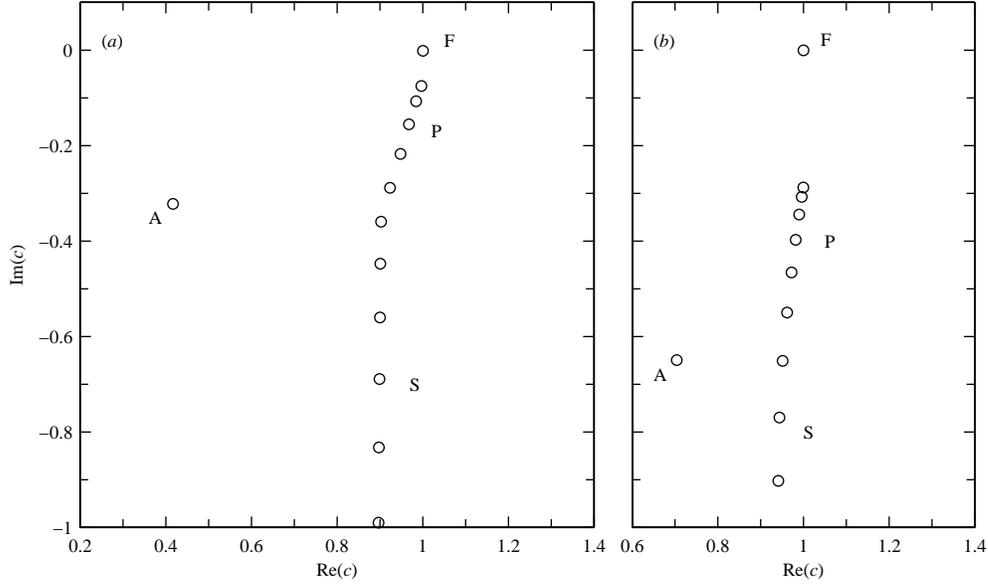}
\caption{\label{fig:spectrumAZeroPm}Eigenvalues of inductionless free-surface Hartmann flow  at $ \Rey = 7 \times 10^5 $, $ \alpha = 2 \times 10^{-3 } $, $ \Gal = 8.3 \times 10^7 $, and $ \Capil = 7 \times 10^{-2} $. The Hartmann number in panels ({\it a}) and ({\it b}) is $ \Har = 10 $ and $ 20 $, respectively. The evolution of this spectrum with $ \Har \in [ 0.1, 50 ] $ is shown in movie~1, available with the online version of the paper.}
\end{center}
\end{figure}

\begin{figure}
\begin{center}
\includegraphics{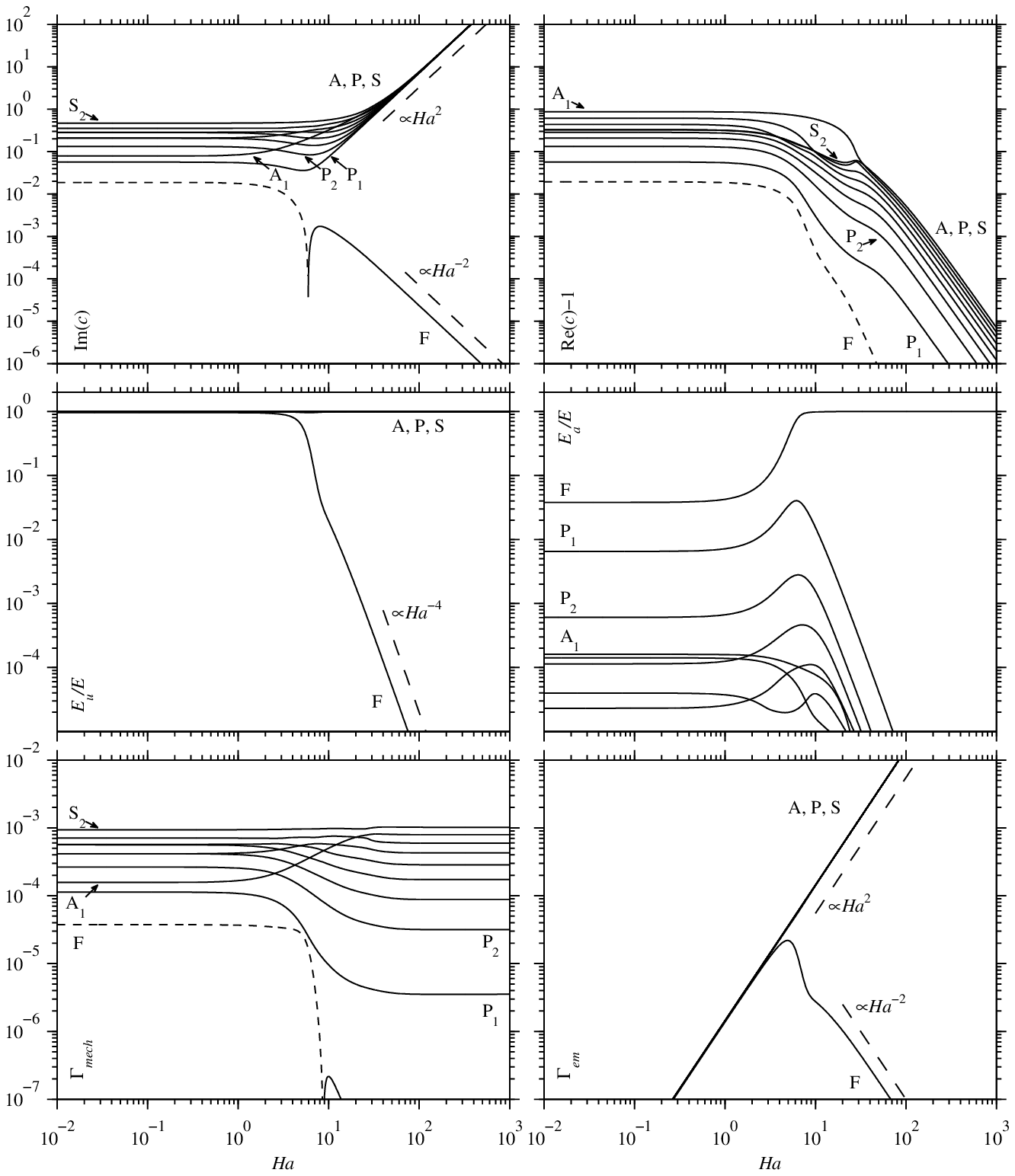}
\end{center}
\caption{\label{fig:harGammaAZeroPm}Complex phase velocity $ c $, kinetic and surface energies, $ E_u $ and $ E_a $ (normalised by the total energy $ E = E_u + E_a $), and mechanical and electromagnetic energy transfer rates, $ \Gamma_{mech } $ and $ \Gamma_{em} $, for the ten least stable modes of inductionless free-surface Hartmann flow at $ \Rey = 7 \times 10^5 $, $ \alpha = 2 \times 10^{-3} $, $ \Gal = 8.3 \times 10^7 $, and $ \Capil = 0.07 $, showing the qualitatively different dependence of the F and A, P, S modes on the Hartmann number $ \Har $. In the first and third-row panels, solid and dashed lines respectively correspond to negative and positive values. Besides mode F, the curves for modes A$_1$, P$_1$, P$_2$, and S$_2$ (the modes indexed according to their $ \Har = 0 $ values; see table~\ref{table:spectrumAHydro}) are indicated.}
\end{figure}

Turning now to the behaviour of the eigenvalues on the complex-$ c $ plane, a prominent feature of inductionless Hartmann flow, illustrated in figure~\ref{fig:spectrumAZeroPm} and movie~1 (see~table~\ref{table:spectrumAZeroPm} for corresponding numerical data), is that as $ \Har $ increases the P branch of the spectrum becomes aligned with the S branch, and the eigenvalues in the A branch collapse towards the P--S branch intersection point. In addition, with the exception of mode~F, which is seen to move along the $ \Real( c ) $-axis towards $ \Real( c ) = 1 $, the eigenvalues are translated to smaller values of $ \Imag( c ) $ (quadratically with $ \Har $, as shown in figure~\ref{fig:harGammaAZeroPm}). As in non-MHD problems, the phase velocity of the S-family modes is (asymptotically) equal to the average steady-state speed~\eqref{eq:baseUAverage}, which approaches unity as $ \Har $ grows, and  $ \Real( c ) $ lies in the interval $ ( \langle U \rangle, 1 ) $ for modes in the P branch. Moreover, the phase velocity of mode~F remains greater than unity, even for strong fields ($ \Har \sim 10^3) $. By performing suitable test calculations, we have verified that the branch-alignment and $ \Imag( c ) $-decrease observed for the A, P, and S modes are independently caused by the formation of the Hartmann velocity profile and the Lorentz force, respectively.

\begin{table}
\begin{tabular*}{\linewidth}{@{\extracolsep{\fill}}llrl}
& & \multicolumn{1}{c}{$c$} & \multicolumn{1}{c}{$E_a/E$}\\
\cline{3-4}
\multicolumn{4}{l}{$ \Har = 10 $}\\
1 &  $ \mbox{F} $ &  $ 1.000478880697718\mathrm{E}+00 - 1.492458547790363\mathrm{E}-03 \, \ii $ & $ 9.80634\mathrm{E}-01 $ \\ 
2 &  $ \mbox{P}_{ 1 } $ &  $ 9.969844758638947\mathrm{E}-01 - 7.510300218671238\mathrm{E}-02 \, \ii $ & $ 1.55113\mathrm{E}-02 $ \\ 
3 &  $ \mbox{P}_{ 2 } $ &  $ 9.844830185724478\mathrm{E}-01 - 1.070009578761852\mathrm{E}-01 \, \ii $ & $ 1.56260\mathrm{E}-03 $ \\ 
4 &  $ \mbox{P}_{ 3 } $ &  $ 9.679739162321193\mathrm{E}-01 - 1.555307117857878\mathrm{E}-01 \, \ii $ & $ 3.70125\mathrm{E}-04 $ \\ 
5 &  $ \mbox{P}_{ 4 } $ &  $ 9.480545293986525\mathrm{E}-01 - 2.173575052123932\mathrm{E}-01 \, \ii $ & $ 1.08068\mathrm{E}-04 $ \\ 
6 &  $ \mbox{P}_{ 5 } $ &  $ 9.235179572370349\mathrm{E}-01 - 2.886322049592563\mathrm{E}-01 \, \ii $ & $ 3.89742\mathrm{E}-05 $ \\ 
7 &  $ \mbox{A}_{ 1 } $ &  $ 4.169901573651648\mathrm{E}-01 - 3.220427370877234\mathrm{E}-01 \, \ii $ & $ 7.72718\mathrm{E}-05 $ \\ 
8 &  $ \mbox{P}_{ 6 } $ &  $ 9.026456030169425\mathrm{E}-01 - 3.595290667757633\mathrm{E}-01 \, \ii $ & $ 1.62104\mathrm{E}-05 $ \\ 
9 &  $ \mbox{S}_{ 1 } $ &  $ 9.009193226866442\mathrm{E}-01 - 4.475126687270328\mathrm{E}-01 \, \ii $ & $ 4.35020\mathrm{E}-06 $ \\ 
10 &  $ \mbox{S}_{ 2 } $ &  $ 9.003564254536676\mathrm{E}-01 - 5.601694489236487\mathrm{E}-01 \, \ii $ & $ 1.00325\mathrm{E}-06 $ \\ 

\\
\multicolumn{4}{l}{$ \Har = 20 $}\\
1 &  $ \mbox{F} $ &  $ 1.000045818436596\mathrm{E}+00 - 5.289119738767267\mathrm{E}-04 \, \ii $ & $ 9.98203\mathrm{E}-01 $ \\ 
2 &  $ \mbox{P}_{ 1 } $ &  $ 9.994955777811616\mathrm{E}-01 - 2.877934996693214\mathrm{E}-01 \, \ii $ & $ 1.41502\mathrm{E}-03 $ \\ 
3 &  $ \mbox{P}_{ 2 } $ &  $ 9.960450826999606\mathrm{E}-01 - 3.075429925109489\mathrm{E}-01 \, \ii $ & $ 1.54941\mathrm{E}-04 $ \\ 
4 &  $ \mbox{P}_{ 3 } $ &  $ 9.898991606163317\mathrm{E}-01 - 3.445398689126118\mathrm{E}-01 \, \ii $ & $ 5.56023\mathrm{E}-05 $ \\ 
5 &  $ \mbox{P}_{ 4 } $ &  $ 9.817920008849399\mathrm{E}-01 - 3.974550352653800\mathrm{E}-01 \, \ii $ & $ 2.52511\mathrm{E}-05 $ \\ 
6 &  $ \mbox{P}_{ 5 } $ &  $ 9.721963507775145\mathrm{E}-01 - 4.656744531110017\mathrm{E}-01 \, \ii $ & $ 1.23326\mathrm{E}-05 $ \\ 
7 &  $ \mbox{P}_{ 6 } $ &  $ 9.615513110612295\mathrm{E}-01 - 5.497038985606164\mathrm{E}-01 \, \ii $ & $ 6.15471\mathrm{E}-06 $ \\ 
8 &  $ \mbox{A}_{ 1 } $ &  $ 7.040044446493452\mathrm{E}-01 - 6.496646531013073\mathrm{E}-01 \, \ii $ & $ 2.32334\mathrm{E}-05 $ \\ 
9 &  $ \mbox{S}_{ 1 } $ &  $ 9.510925679889382\mathrm{E}-01 - 6.510431012818730\mathrm{E}-01 \, \ii $ & $ 3.02892\mathrm{E}-06 $ \\ 
10 &  $ \mbox{S}_{ 2 } $ &  $ 9.439623776920884\mathrm{E}-01 - 7.698041542899734\mathrm{E}-01 \, \ii $ & $ 1.41918\mathrm{E}-06 $ \\ 

\end{tabular*}
\caption{\label{table:spectrumAZeroPm}Complex phase velocity $ c $ and free-surface energy $ E_a $, normalised by the total energy $ E = E_u + E_a $, of the 10 least stable modes of the inductionless problems in figure~\ref{fig:spectrumAZeroPm}. The mean steady-state velocity $ \langle U \rangle $, given by \eqref{eq:baseUAverage}, is 0.9001 ($ \Har = 10$) and 0.9500 ($ \Har = 20 $). Due to the alignment of the P and S branches, there exists an ambiguity in distinguishing between the most stable P mode and the least stable S mode. Here we consider that the P branch comprises of the first six modes (in order of decreasing $ \Imag( c ) $) with $ \Real( c ) > \langle U \rangle $.}
\end{table}

The exponential $ \Reyb( \Har ) $ growth in~\eqrefa{eq:criticalSoft} is a somewhat misleading indicator for the magnetic field's stabilising effect on mode~F, which participates in the soft instability. The reason is that, unlike the remaining modes in the spectrum (as well as all channel modes), whose decay rate increases quadratically with $ \Har $ as a consequence of Lorentz damping, mode~F becomes asymptotically neutral for large magnetic field strengths. This behaviour is illustrated in figure~\ref{fig:harGammaAZeroPm}, where the complex phase velocity, as well as results for the energy components and energy transfer rates, are plotted as a function of $ \Har \in [ 10^{-2}, 10^3 ] $ for the ten least stable modes of the non-MHD problem in figure~\ref{fig:spectrumAHydro}. As the Hartmann number grows, the total mechanical energy transfer rate $ \Gamma_{mech} := \Gamma_R + \Gamma_\nu + \Gamma_{ a U } $ (the inductionless version of~\eqrefa{eq:gammaMechEm}) to mode~F,  experiences a sharp drop, caused by a decrease in $ \Gamma_{a U } $ associated with the flattening of the velocity profile. This, in conjunction with resistive dissipation $ \Gamma_\eta $, which in inductionless problems is the only component of the electromagnetic energy transfer rate $ \Gamma_{em} $, suffices to stabilise the mode for all $ \Har \gtrsim 6 $. However, instead of growing quadratically with $ \Har $, as it does for the A, P, and S families of modes, the decay rate $ - \Gamma $ of mode~F turns around, and approaches zero following a $ \Har^{-2} $ scaling. At the same time, the mode's energy content becomes almost entirely potential, with the kinetic energy following the power law $ E_u / E \propto \Har^{-4} $. In contrast, for sufficiently large Hartmann numbers, free-surface oscillations become negligible for the A, P, and S modes, as manifested by their decaying surface energy $E_a$.  

\begin{figure}
\begin{center}
\includegraphics{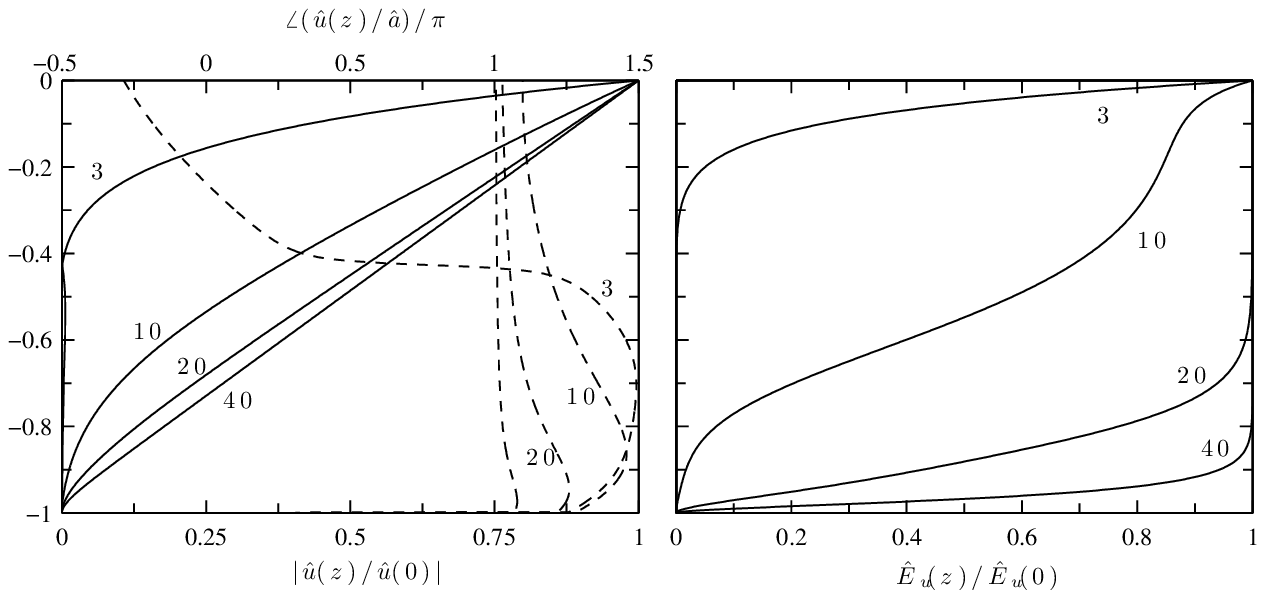}
\caption{\label{fig:modeFZeroPm}Modulus and phase of the velocity eigenfunction $ \modeu(z) $ (left; drawn with solid and dashed lines, respectively), and kinetic energy density $ \modeeu(z) $ (right) of mode~F for $ \Rey = 7 \times 10^{5} $, $ \alpha = 2 \times 10^{-3} $, $ \Gal = 8.3 \times 10^7 $, $ \Capil = 0.07 $, and Hartmann number $ \Har \in \{3, 10, 20, 40\} $. The phase of the eigenfunction is computed relative to the free-surface oscillation amplitude $ \modea $.}
\end{center}
\end{figure}

Figure~\ref{fig:modeFZeroPm} shows that as the Hartmann number is increased, the velocity eigenfunction $ \modeu( z ) $ corresponding to mode~F evolves from a typical surface-wave-like profile at small Hartmann numbers to a state of nearly constant shear, characterised by uniform distribution of the kinetic energy away from the wall. Moreover, in line with the kinematic boundary condition~\eqrefb{eq:2dNormalModesBC1} with $ C \approx 1 $ (\ie $ \gamma \approx - i \alpha $), at large Hartmann numbers $ \modeu( 0 ) $ exhibits a $ 180^\circ $ phase difference relative to the free-surface oscillation amplitude $ \modea $, which, in the real representation corresponds to the streamwise velocity perturbations $ u_x( 0 ) $ being $ 90^\circ $ out of phase with the free-surface oscillation amplitude $ a $ (recall that, in accordance with~\eqrefa{eq:2dNormalModesUB}, $ u_x = \Imag( \DD \modeu / \alpha \exp( \gamma t + \ii \alpha x ) ) $). 
      
The observed scaling of the kinetic energy of mode~F for strong magnetic fields is dimensionally consistent with a time-averaged equilibrium determined by the work done by Lorentz and gravitational stresses acting on the free surface. We approximate this balance by setting $ f \sim g $, where $ f \sim \Har^2 \Rey^{-1} | u_x( 0 ) | $ and $ g \sim \cos(\theta) \Fro^{ - 2 } | a | = \Gal \Rey^{-2} | a | $ are respectively estimates of the Lorentz and gravitational forces. Because at large $ \Har $ the velocity-eigenfunction gradient $ \DD \modeu $ is nearly constant over the inner part of the domain, and since at large wavelengths the ratio $  ( || u_x || / || u_z || )^2 \sim \alpha^{-2} $ is expected to be large, $ f \sim g $ leads to $ E_u / E_a \sim ( | u_x( 0 ) | \cos(\theta)/ ( \Fro^2 | a | ) )^2 \sim \Gal / \Har^4 $. The latter scaling is consistent with the $ E_u / E  $ results in figure~\ref{fig:harGammaAZeroPm} (note that $ E_u / E \approx E_u / E_a $ for $ E_u \ll E_a $), and in separate calculations we have checked that the $ E_u / E_a \propto \Gal $ scaling applies at fixed $ \Har $. Aside from figure~\ref{fig:harGammaAZeroPm}, a $ | \Gamma | \propto \Har^{-2 } $ strong-field behaviour for mode~F was also recorded in trial inductionless problems with $ U(z ) $ set to the Poiseuille profile, and is also expected on the basis of large-wavelength approximations (see~\eqrefc{eq:strongFieldZeroPmPoiseuille}). On the other hand, in non-MHD calculations with the Hartmann velocity profile, as well as in the corresponding small-$\alpha $ expansion~\eqrefc{eq:strongFieldHydroHartmann}, the modal growth rate $ \Gamma $ tends to a $ \Har $-independent negative value for $ \Har \gg 1 $. These observations, coupled with the dimensional argument for $ E_u / E_a $, suggest that the strong-field neutrality of mode~F is the outcome of a balance between gravitational and Lorentz forces, and does not rely on the form of the steady-state velocity profile.  

\subsection{\label{sec:mhdProblems}Free-surface Hartmann flow at $ \Prm \leq 10^{-4} $}

In low-$ \Prm $ channel Hartmann flow with insulating boundary conditions the critical Reynolds number, wavenumber, and phase velocity are known to be well-approximated by the inductionless scheme. In particular, the calculations by \cite{Takashima96} have established that for $ \Prm \leq 10^{-4 } $ the relative error incurred in $ \Reyc $ is less than $ 0.004 $, even at Hartmann numbers as high as 100, where $ \Reyc $ is of order $ 10^7 $. In free-surface flow (again with an insulating wall), however, we find that while the critical parameters of the hard mode are equally insensitive to $ \Prm \ll 1 $ as in channel problems, the soft mode's behaviour differs markedly between the small-$ \Prm $ and inductionless cases. Moreover, the boundary conditions now support a pair of travelling Alfv\'en waves, the upstream-propagating of which may become unstable at sufficiently high Alfv\'en numbers. When conducting boundary conditions are enforced, the Alfv\'en modes are removed from the spectrum, and the soft mode's critical Reynolds number becomes a decreasing function of $ \Har \gg 1 $. 

\subsubsection{\label{sec:mhdTopEnd}Properties of the least stable modes}

\begin{figure}
\begin{center}
\includegraphics{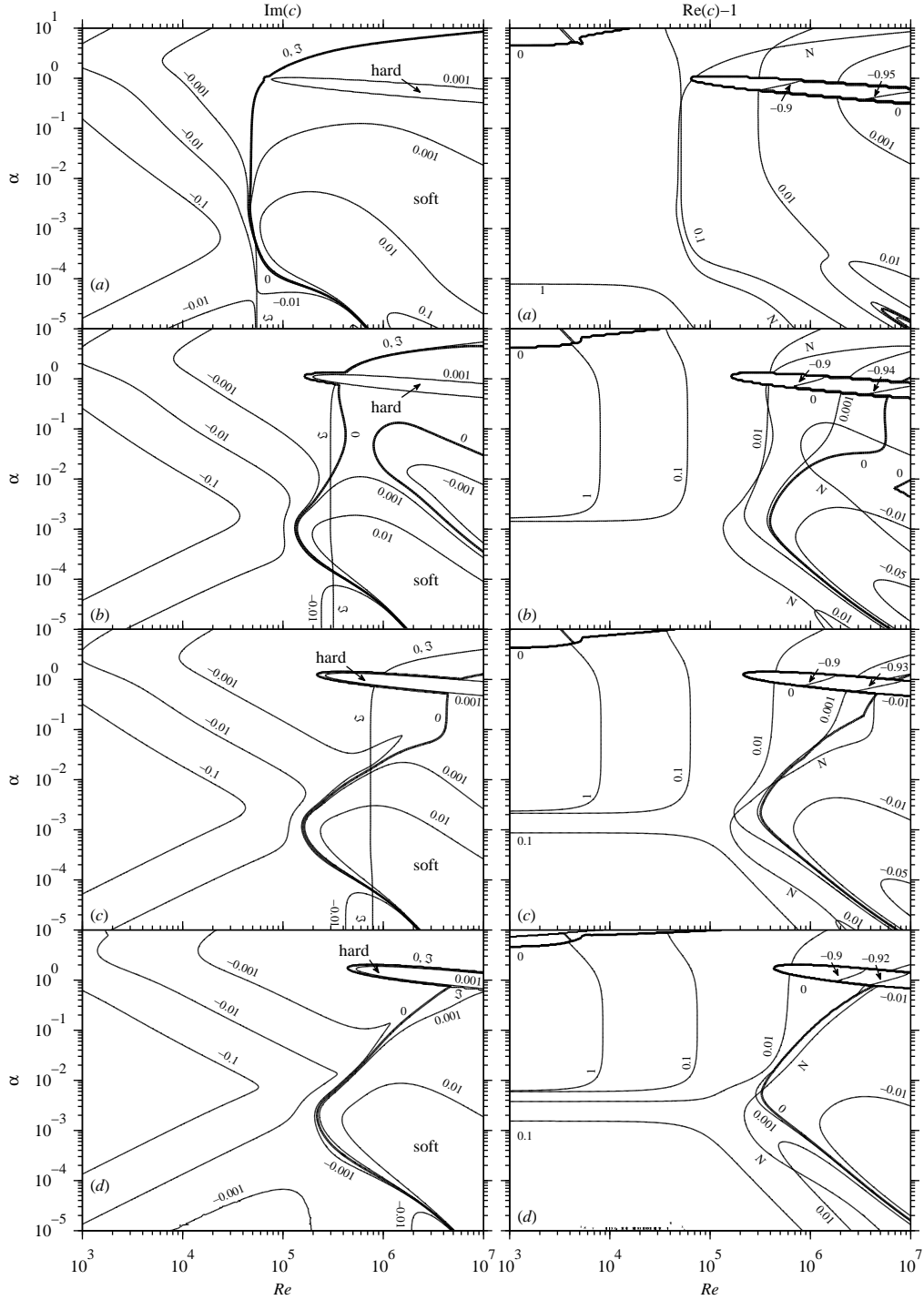}
\caption{\label{fig:reAlphaMhd}Contours of the complex phase velocity $ c $ in the $(\Rey, \alpha)$ plane for the least stable mode of free-surface Hartmann flow with an insulating wall at $ \Prm = 10^{-5} $, $ \Gal = 8.3 \times 10^7$, and $ \Capil = 0.07 $. The Hartmann number $ \Har $ in panels ({\it a--d}) is 3, 5, 6, and 10, respectively. The curves labelled $ \Im $ in the panels for $ \Imag(c ) $ are the $ \Imag( c ) = 0 $ contours of the corresponding inductionless problems. The neutral-stability curves for $ \Prm = 10^{-5} $, labelled~N, are drawn in the $ \Real( c ) $ panels for reference.}
\end{center}
\end{figure}

We illustrate the behaviour of the top end of the spectrum for problems with an insulating wall in figure~\ref{fig:reAlphaMhd}, where contours of the complex phase velocity in the $ ( \Rey, \alpha ) $ plane are plotted for the least stable mode at fixed $ \Prm = 10^{-5} $ (a value lying in the upper end of the $ \Prm $-regime for liquid metals) and moderately small Hartmann number $ \Har \in [ 3, 10 ]$. It is evident from the proximity of the portions of the inductionless and $ \Prm = 10^{-5} $ neutral-stability curves corresponding to the hard mode, as well as the close agreement between the critical-Reynolds-number calculations in tables~\ref{table:criticalHardZeroPm} and~\ref{table:criticalHardMhd}, that the influence of a small magnetic diffusivity on the hard instability is weak, with $ \Reyc $ being slightly smaller when $ \Prm $ is nonzero compared to its value in the inductionless limit. In the case of the soft mode, however, prominent differences in the stability properties exist even at small Hartmann numbers.

\begin{table}
\begin{tabular*}{\linewidth}{@{\extracolsep{\fill}}llll}
$ \Har $ & $ \Reyc $ & $ \alphac $ & $ C_c $ \\
\hline
\multicolumn{4}{l}{$ \Prm = 10^{-5} $}\\
 $ 0.5 $ &  $ 7.4343262\mathrm{E}+03 $ & $ 2.77817\mathrm{E}+00 $ & $ 1.85689\mathrm{E}-01 $ \\ 
 $ 1 $ &  $ 7.7154109\mathrm{E}+03 $ & $ 2.64640\mathrm{E}+00 $ & $ 1.89960\mathrm{E}-01 $ \\ 
 $ 2 $ &  $ 2.3923849\mathrm{E}+04 $ & $ 1.10426\mathrm{E}+00 $ & $ 1.91216\mathrm{E}-01 $ \\ 
 $ 5 $ &  $ 1.6371768\mathrm{E}+05 $ & $ 1.13613\mathrm{E}+00 $ & $ 1.56438\mathrm{E}-01 $ \\ 
 $ 10 $ &  $ 4.3954479\mathrm{E}+05 $ & $ 1.73924\mathrm{E}+00 $ & $ 1.54813\mathrm{E}-01 $ \\ 
 $ 20 $ &  $ 9.6100175\mathrm{E}+05 $ & $ 3.23776\mathrm{E}+00 $ & $ 1.55044\mathrm{E}-01 $ \\ 
 $ 50 $ &  $ 2.4134695\mathrm{E}+06 $ & $ 8.07606\mathrm{E}+00 $ & $ 1.55064\mathrm{E}-01 $ \\ 
 $ 100 $ &  $ 4.8268720\mathrm{E}+06 $ & $ 1.61512\mathrm{E}+01 $ & $ 1.55064\mathrm{E}-01 $ \\ 

\\
\multicolumn{4}{l}{$ \Prm  = 10^{-4} $}\\
 $ 0.5 $ &  $ 7.4342966\mathrm{E}+03 $ & $ 2.77815\mathrm{E}+00 $ & $ 1.85683\mathrm{E}-01 $ \\ 
 $ 1 $ &  $ 7.7152014\mathrm{E}+03 $ & $ 2.64657\mathrm{E}+00 $ & $ 1.89973\mathrm{E}-01 $ \\ 
 $ 2 $ &  $ 2.3879649\mathrm{E}+04 $ & $ 1.10492\mathrm{E}+00 $ & $ 1.91273\mathrm{E}-01 $ \\ 
 $ 5 $ &  $ 1.6340601\mathrm{E}+05 $ & $ 1.13611\mathrm{E}+00 $ & $ 1.56539\mathrm{E}-01 $ \\ 
 $ 10 $ &  $ 4.3859195\mathrm{E}+05 $ & $ 1.73911\mathrm{E}+00 $ & $ 1.54934\mathrm{E}-01 $ \\ 
 $ 20 $ &  $ 9.5885884\mathrm{E}+05 $ & $ 3.23738\mathrm{E}+00 $ & $ 1.55171\mathrm{E}-01 $ \\ 
 $ 50 $ &  $ 2.4078809\mathrm{E}+06 $ & $ 8.07604\mathrm{E}+00 $ & $ 1.55197\mathrm{E}-01 $ \\ 
 $ 100 $ &  $ 4.8155345\mathrm{E}+06 $ & $ 1.61542\mathrm{E}+01 $ & $ 1.55200\mathrm{E}-01 $ \\ 

\end{tabular*}
\caption{\label{table:criticalHardMhd}Critical Reynolds number $ \Reyc $, wavenumber $ \alphac $, and phase velocity $ C_c $ of the hard mode in free-surface Hartmann flow with insulating boundary conditions, computed at Galilei number $ \Gal = 8.3 \times 10^7 $, capillary number $ \Capil = 0.07 $, magnetic Prandtl number $ \Prm \in \{ 10^{-5}, 10^{-4} \} $, and Hartmann number $ \Har \in [ 0.5, 100 ] $.} 
\end{table}

First, the structure of the $ \Imag( c ) $ contours in figure~\ref{fig:reAlphaMhd} suggest that the $ \alpha = 0 $ axis is no longer part of the neutral-stability curve $ \Imag( c ) = 0 $, and this can be confirmed by means of large-wavelength perturbation theory. In particular, according to the discussion in appendix~\ref{app:pertMhd}, free-surface Hartmann flow with an insulating wall supports, besides mode~F, a second mode with vanishing complex growth rate $ \gamma $ in the limit $ \alpha \searrow 0 $, of magnetic origin. The zeroth-order degeneracy between mode F and this magnetic mode is broken at first order in $ \alpha $, where there exist two distinct solutions, respectively $ \gamma_1^{(F)} $ and $ \gamma_1^{(M)} $, for the coefficient $ \gamma_1 $ in the perturbative series $ \gamma = \gamma_0 + \gamma_1 \alpha + \gamma_2 \alpha^2 + \ord( \alpha^3 ) $. Both solutions have negative real part for all $ \Har > 0 $ and $ \Prm > 0 $ ($ \Real( \gamma_1^{(M)} ) $ is negative even when $ \Har $ equals zero), which implies that for any Reynolds number there exists an upper bound $ \alpha_m $ in $ \alpha $, below which mode~F is stable. That is, $ \Real( \gamma ) $ is negative for $ 0 < \alpha < \alpha_m $ or, as observed in figure~\ref{fig:reAlphaMhd}, $ \Imag( c ) < 0 $ for $ 0 \leq \alpha < \alpha_m $ (\cf inductionless flow). We remark that because $ \gamma $ vanishes in the limit $ \alpha \searrow 0 $, it is important to distinguish between the definitions $ \Imag( c ) = 0 $ and $ \Real( \gamma ) = 0 $ for the soft mode's neutral-stability curve, since with the latter definition the $ \alpha = 0 $ axis remains part of the curve even when $ \Real( \gamma_1 ) \neq 0 $. In separate numerical calculations we have checked that $ \alpha_m $ becomes smaller when $ \Prm $ is decreased at fixed $ \Har $, which is consistent with the large-wavelength result~\eqref{eq:gamma1Pm} that in the limit $ \Prm \searrow 0 $, $ \gamma_1^{(F)} $ reaches the inductionless value~\eqref{eq:gamma1ZeroPmHartmann}, while $ \gamma_1^{(M)} $ is singular. 

Performing a similar type of analysis establishes that (i) when the induced magnetic field $ B $ is set to zero, $ \Real( \gamma_1 ) $ is still negative for mode~F, as well as for the magnetic mode, (ii) when $ U $ and $ B $ are both set to zero, $ \gamma_1 $, now given by~\eqref{eq:gamma1FS}, is negative for the magnetic mode but vanishes for mode~F, and (iii) in Hartmann flow with a conducting wall, the $ \modeb( -1 ) = 0  $ constraint imposed on the magnetic field eigenfunction eliminates the magnetic mode, and $ \gamma_1^{(F) }$ becomes equal to the corresponding expansion coefficient in the inductionless limit. It therefore appears that the suppression of the soft instability for $ \alpha \searrow 0 $ is the combined outcome of the boundary conditions, which allow for the presence of the magnetic mode, and the coupling between mode F and the magnetic mode provided by the steady-state flow. 

The soft mode's departure from inductionless behaviour is also prominent at larger values of $ \alpha \Rey $. As can be seen in figure~\ref{fig:reAlphaMhd}, at moderate Hartmann numbers ($ \Har \sim 5$) regions of stability emerge in the $( \Rey, \alpha )$ plane that would contain unstable modes in the inductionless limit. Moreover, as $ \Har $ grows, a wedge-like instability region forms, extending to Reynolds numbers significantly smaller than in the corresponding inductionless problems. Unstable modes may now have phase velocity smaller than unity, but $ \Real( c ) > 1 $ is found to apply for the modes close to the tip of the wedge-like region (including the critical mode), at least for the $ \Har \leq 10^{3} $ interval covered in our calculations. 
 
\begin{figure}
\begin{center}
\includegraphics{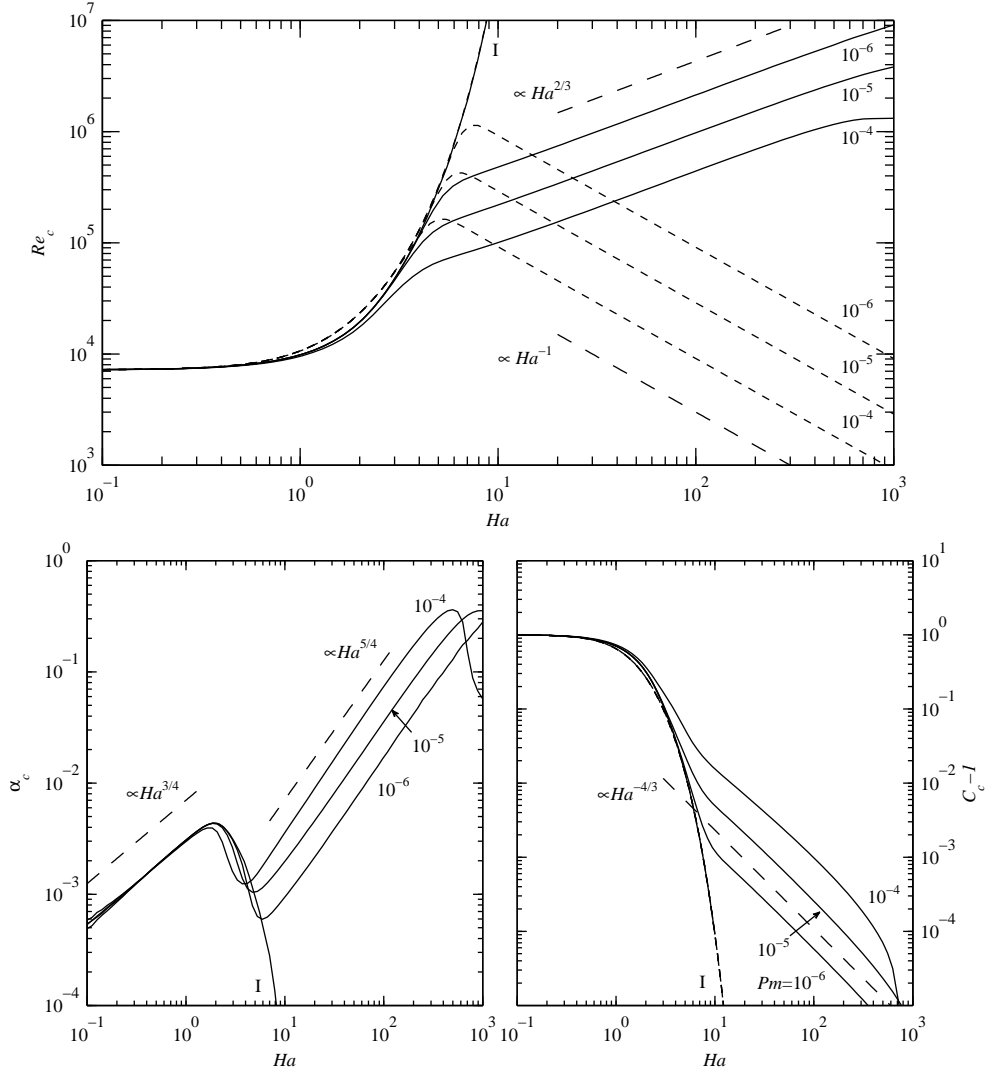}
\end{center}
\caption{\label{fig:criticalSoft}Critical Reynolds number $ \Reyc $, wavenumber $ \alphac $, and phase velocity $ C_c $ of the soft mode as a function of the Hartmann number $ \Har \in [ 10^{-1}, 10^3 ] $ for inductionless problems (curves labelled I), and insulating-wall problems with $ \Prm \in \{ 10^{-6}, 10^{-5}, 10^{-4 } \} $. Also shown, in dotted lines, are the Reynolds number $ \Reyb $, and the corresponding phase velocity $ C_b $ at the bifurcation point of the neutral-stability curve, evaluated for conducting-wall problems by means of the large-wavelength results~\eqref{eq:criticalReMhdConducting} and~\eqrefb{eq:criticalSoft}. The Galilei and capillary numbers are $ \Gal = 8.3 \times 10^7 $ and $ \Capil = 0.07 $ throughout. As in figure~\ref{fig:criticalSoftZeroPm}, the fluctuations in the $ \alphac( \Har ) $ curves are due to the ill-conditioning of critical-parameter calculations for the soft mode. Aside from these fluctuations, we do not expect significant effects on the shape of the curves due to ill-conditioning.}
\end{figure}

The critical parameters of the soft mode as a function of $ \Har \in [ 10^{-1}, 10^{3} ] $ are plotted in figure~\ref{fig:criticalSoft} for logarithmically spaced values of the magnetic Prandtl number $ \Prm \in \{ 10^{-6}, 10^{-5}, 10^{-4} \} $ (see also table~\ref{table:criticalSoftMhd}). As expected from the contour plots in figure~\ref{fig:reAlphaMhd}, when the Hartmann number is small, $ \Reyc  $ is close to its value in the inductionless limit. For instance, the relative difference between the $ \Har = 1 $ results in table~\ref{table:criticalSoftMhd} and the corresponding inductionless results in table~\ref{table:criticalSoftZeroPm} is less than 3.5\%. However, once the magnetic field strength exceeds a threshold, which decreases with $ \Prm $, $ \Reyc( \Har ) $ branches off from the exponential growth~\eqrefa{eq:criticalSoft}, and follows closely the power law $ \Reyc \propto \Har^{2/3} $. During that transition, the decrease of the wavenumber with $ \Har $ observed in the inductionless limit is reversed, switching over to an $ \alphac \propto \Har ^ { 5 / 4} $ scaling. Moreover, the exponential decrease~\eqrefb{eq:criticalSoft} of the critical phase velocity relative to the steady-state flow at the free surface becomes a $ C_c - 1 \propto \Har^{-4/3 } $ power law. In this intermediate Hartmann-number regime, the results for $ \Reyc( \Har ) $, $ \alphac( \Har ) $, and $ C_c( \Har ) - 1 $ collapse to nearly single curves if scaled by $ \Prm^{1/3} $, $ \Prm^{-1/3} $, and $ \Prm^{-2/3} $, respectively (though the agreement is not as good for the $ \Prm = 10^{-4} $ data for $ C_c - 1 $). The power-law behaviour of the critical parameters is only transient, however. Eventually, at sufficiently large Hartmann numbers, $ \Reyc( \Har ) $ levels off (for $ \Prm = 10^{-4} $ this occurs around $ \Har = 500 $), and $ \alphac( \Har ) $ becomes a decreasing function once again.

\begin{table}
\begin{tabular*}{\linewidth}{@{\extracolsep{\fill}}llll}
$ \Har $ & $ \Reyc $ & $ \alphac $ & $ C_c -1 $ \\
\hline
\multicolumn{4}{l}{$ \Prm = 10^{-6} $}\\
 $ 0.1 $ &  $ 7.2341\mathrm{E}+03 $ & $ 5.353\mathrm{E}-04 $ & $ 9.9504\mathrm{E}-01 $ \\ 
 $ 1 $ &  $ 9.8246\mathrm{E}+03 $ & $ 3.069\mathrm{E}-03 $ & $ 7.0268\mathrm{E}-01 $ \\ 
 $ 10 $ &  $ 4.7810\mathrm{E}+05 $ & $ 9.369\mathrm{E}-04 $ & $ 1.2076\mathrm{E}-03 $ \\ 
 $ 100 $ &  $ 2.1205\mathrm{E}+06 $ & $ 1.670\mathrm{E}-02 $ & $ 5.7707\mathrm{E}-05 $ \\ 
 $ 1000 $ &  $ 9.3869\mathrm{E}+06 $ & $ 2.113\mathrm{E}-01 $ & $ 1.8874\mathrm{E}-06 $ \\ 

\\
\multicolumn{4}{l}{$ \Prm  = 10^{-5} $}\\
 $ 0.1 $ &  $ 7.2343\mathrm{E}+03 $ & $ 5.355\mathrm{E}-04 $ & $ 9.9505\mathrm{E}-01 $ \\ 
 $ 1 $ &  $ 9.8253\mathrm{E}+03 $ & $ 3.072\mathrm{E}-03 $ & $ 7.0314\mathrm{E}-01 $ \\ 
 $ 10 $ &  $ 2.1891\mathrm{E}+05 $ & $ 1.978\mathrm{E}-03 $ & $ 4.9470\mathrm{E}-03 $ \\ 
 $ 100 $ &  $ 9.7469\mathrm{E}+05 $ & $ 3.552\mathrm{E}-02 $ & $ 2.5559\mathrm{E}-04 $ \\ 
 $ 1000 $ &  $ 3.8912\mathrm{E}+06 $ & $ 2.244\mathrm{E}-01 $ & $ 4.1975\mathrm{E}-06 $ \\ 

\\
\multicolumn{4}{l}{$ \Prm  = 10^{-4} $}\\
 $ 0.1 $ &  $ 7.2322\mathrm{E}+03 $ & $ 6.005\mathrm{E}-04 $ & $ 9.9563\mathrm{E}-01 $ \\ 
 $ 1 $ &  $ 9.5341\mathrm{E}+03 $ & $ 2.979\mathrm{E}-03 $ & $ 7.3307\mathrm{E}-01 $ \\ 
 $ 10 $ &  $ 1.0034\mathrm{E}+05 $ & $ 3.571\mathrm{E}-03 $ & $ 1.5546\mathrm{E}-02 $ \\ 
 $ 100 $ &  $ 4.4148\mathrm{E}+05 $ & $ 7.375\mathrm{E}-02 $ & $ 9.8053\mathrm{E}-04 $ \\ 
 $ 1000 $ &  $ 1.3161\mathrm{E}+06 $ & $ 6.202\mathrm{E}-02 $ & $ 1.4324\mathrm{E}-06 $ \\ 

\end{tabular*}
\caption{\label{table:criticalSoftMhd}Critical Reynolds number $ \Reyc $, wavenumber $ \alphac $, and phase velocity $ C_c $ of the soft mode in free-surface Hartmann flow with insulating boundary conditions, computed at Galilei number $ \Gal = 8.3 \times 10^7 $, capillary number $ \Capil = 0.07 $, magnetic Prandtl number $ \Prm \in \{ 10^{-6}, 10^{-5}, 10^{-4} \} $, and Hartmann number $ \Har \in [ 10^{-1}, 10^3 ] $. These calculations are affected by a similar ill-conditioning as the corresponding ones for inductionless problems in table~\ref{table:criticalSoftZeroPm}. In particular, we have observed an $ \ord( 10^{-4} ) $ systematic drift in some of the results for $ \alphac $, arising  when the optimisation solver used to compute $ ( \Reyc, \alphac, C_c ) $ is restarted using the output of previous calculations for initialisation. With the presently available computational resources we were not able to perform a sufficiently large number of iterations so as to establish convergence in $ \alphac $. However, we expect the results for $ \Reyc $ and $ C_c $ to be affected by this issue to a lesser extent.} 
\end{table}

The deviation of the critical Reynolds number of the soft instability from~\eqrefa{eq:criticalSoft} is even more pronounced in problems with a conducting wall. As outlined in appendix~\ref{app:pertMhd}, in this case the first-order coefficient in the perturbation series for $ \gamma $ is given by the same expression as in inductionless flows (\ie \eqref{eq:gamma1ZeroPmHartmann}), which has vanishing real part, and as a result the $ \alpha = 0 $ axis remains part of the neutral-stability curve. This enables the derivation of the closed-form expression~\eqref{eq:criticalReMhdConducting} for the Reynolds number $ \Reyb $ at the bifurcation point, which, in the manner shown in figure~\ref{fig:criticalSoft}, becomes a decreasing function of $ \Har $, varying like $ \Reyb \sim ( \Gal / \Prm )^{1/2} \Har^{-1} $ at large Hartmann numbers. Even though we have not explicitly computed the soft mode's critical parameters for conducting-wall problems, we have verified in eigenvalue contour plots that, as in inductionless flows, $ \Reyb $ is close to $ \Reyc $. In any case, since $ \Reyb $ is an upper bound for $ \Reyc $, the behaviour of $ \Reyb( \Har ) $ suffices to conclude that in free-surface Hartmann flow with a conducting wall the external magnetic field leads to a reduction of the critical Reynolds number for instability for all $ \Prm > 0 $.  

\subsubsection{\label{sec:roleOfJ}The role of the steady-state induced magnetic field}

Except for the large-wavelength instability suppression observed in problems with insulating boundary conditions, where, as stated in \S\ref{sec:mhdTopEnd}, the modal decay rate is nonzero to linear order in $ \alpha $ even when $ B $ vanishes, the discrepancy between inductionless and nonzero-$ \Prm $ behaviour for mode~F is mainly caused by the steady-state induced magnetic field~\eqrefb{eq:generalUB} and the associated contribution $ \pertfj := \Reym \crossp{ \basej }{ \pertb } \sim \Prm^{1/2} \Har || \DD B || ||\pertb|| $ to the linearised Lorentz force~\eqrefb{eq:linearizedGovEqs} from the spanwise current $ \basej := \Reym^{-1} \curl{ \baseb } = \Reym^{-1} \Har \Prm^{1/2} \DD B \vy $. As a demonstration, in figure~\ref{fig:modeFMhd} and table~\ref{table:modeFMhd} we have computed the complex phase velocity of mode~F, as well as certain energy components and energy transfer rates, as a function of $ \Har \in [ 10^{-2}, 10^{3} ] $  for (i) $ \Prm = 10^{-5} $ flows with insulating and conducting walls, (ii) the corresponding inductionless problems, and (iii) insulating and conducting-wall problems with $ \Prm = 10^{-5} $ and $ B $ set to zero. 

\begin{figure}
\begin{center}
\includegraphics{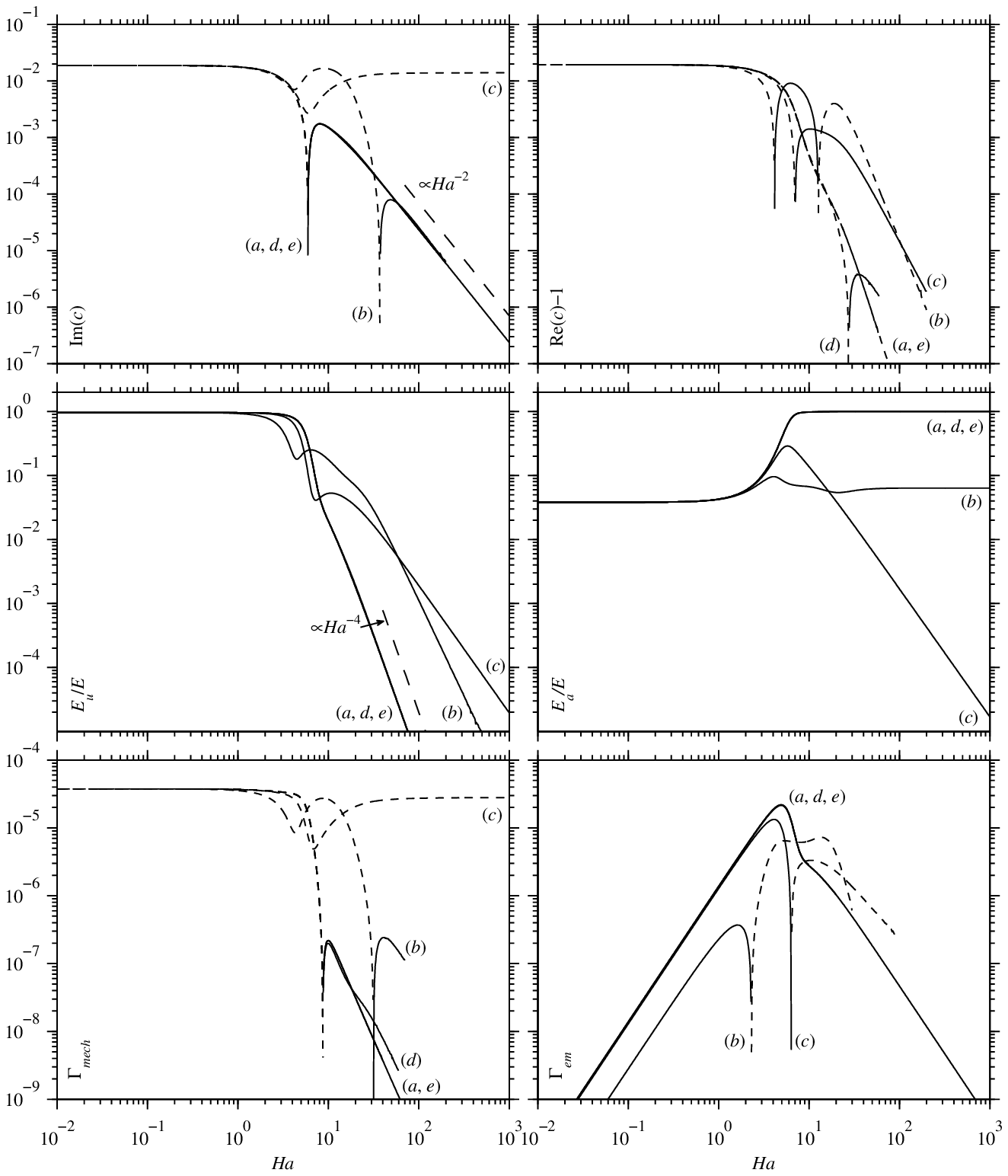}
\caption{\label{fig:harGammaCompareF}Complex phase velocity $ c $, kinetic and surface energies, $ E_u $ and $ E_a $ (normalised by the total energy $ E $), and mechanical and electromagnetic energy transfer rates, $ \Gamma_{mech} $ and $ \Gamma_{em} $, for mode~F in free-surface Hartmann flow at $ \Rey = 7 \times 10^{5} $, $ \alpha = 2 \times 10^{-3} $, $ \Gal = 8.3 \times 10^7 $, $ \Capil = 0.07 $, and $ \Har \in [ 10^{-2}, 10^3 ] $. The curves labelled ({\it a}) correspond to inductionless flow, while ({\it b}) and ({\it c}) were evaluated at $ \Prm = 10^{-5} $, respectively with insulating and conducting boundary conditions. Curves~({\it d}) and ({\it e}) are for the same problems as ({\it b}) and ({\it c}), respectively, but with the magnetic field profile $ B $ set to zero. In the first and third-row panels, solid and dashed lines respectively correspond to negative and positive values. As discussed in \cite{GiannakisFischerRosner08}, accurate numerical evaluation of some of the energy transfer rates can be problematic, especially when a large polynomial degree is required for the spectral solution to~\eqref{eq:orrSommerfeldInd} to converge. For this reason, we have not been able to evaluate $ \Gamma_{mech} $ and $ \Gamma_{em } $ for Hartmann numbers as large as $ 10^3 $. Roundoff errors also limit the convergence of the eigenvalue for mode~F to about five significant digits  when $ \Har $ becomes large in cases ({\it b--e}). Apart from the $ \Real( c ) $ graph for case ({\it c}), this level of numerical precision is not sufficient to continue the logarithmic plots in the first row to $ \Har \gtrsim 100 $, where both of $ | \Imag( c ) | $ and $ | \Real( c ) - 1 |$ become small compared to $ | c | \approx 1 $.}
\end{center}
\end{figure}

\begin{table}
\begin{tabular*}{\linewidth}{@{\extracolsep{\fill}}lrrrrr}
& Inductionless & Insulating & Conducting & Ins. ($B=0$) & Cond. ($B=0$) \\
\cline{2-6}
$\Gamma$ &$-1.0578\mathrm{E}-06$ &$9.2461\mathrm{E}-06$ &$2.2910\mathrm{E}-05$ &$-1.1177\mathrm{E}-06$ &$-1.0578\mathrm{E}-06$  \\ 
$C-1$ &$4.5818\mathrm{E}-05$ &$3.9237\mathrm{E}-03$ &$-9.1720\mathrm{E}-04$ &$3.3397\mathrm{E}-05$ &$4.5808\mathrm{E}-05$  \\
\\ 
$|\modeu(0)/\modea|$ &$1.0618\mathrm{E}-06$ &$1.2127\mathrm{E}-05$ &$2.2983\mathrm{E}-05$ &$1.1197\mathrm{E}-06$ &$1.0618\mathrm{E}-06$  \\ 
$\angle(\modeu(0)/\modea)$ &$-0.9725 \pi $ &$-0.2240 \pi $ &$0.0254 \pi $ &$-0.9810 \pi $ &$-0.9725 \pi $  \\ 
$|\modeb(0)/\modea|$ & &$1.6557\mathrm{E}-03$ &$1.2622\mathrm{E}-04$ &$1.7458\mathrm{E}-05$ &$3.6763\mathrm{E}-08$  \\ 
$\angle(\modeb(0)/\modea)$ & &$-0.4145 \pi $ &$-0.4980 \pi $ &$-0.9258 \pi $ &$-0.9928 \pi $  \\
\\ 
$\mathrm{E}_u/\mathrm{E}$ &$1.7966\mathrm{E}-03$ &$6.8769\mathrm{E}-02$ &$3.3217\mathrm{E}-02$ &$1.9035\mathrm{E}-03$ &$1.7965\mathrm{E}-03$  \\ 
$\mathrm{E}_b/\mathrm{E}$ & &$1.6250\mathrm{E}-03$ &$9.2557\mathrm{E}-01$ &$2.5348\mathrm{E}-06$ &$2.6228\mathrm{E}-06$  \\ 
$\mathrm{E}_{b+}/\mathrm{E}$ & &$4.3846\mathrm{E}-01$ &$1.8511\mathrm{E}-03$ &$8.9635\mathrm{E}-04$ &$3.9822\mathrm{E}-09$  \\ 
$\mathrm{E}_{b-}/\mathrm{E}$ & &$4.3696\mathrm{E}-01$ & &$8.9621\mathrm{E}-04$ &  \\ 
$\mathrm{E}_a/\mathrm{E}$ &$9.9820\mathrm{E}-01$ &$5.4185\mathrm{E}-02$ &$3.9363\mathrm{E}-02$ &$9.9630\mathrm{E}-01$ &$9.9820\mathrm{E}-01$  \\
\\ 
$\Gamma_R$ &$-9.2340\mathrm{E}-09$ &$6.8264\mathrm{E}-06$ &$-1.6850\mathrm{E}-07$ &$8.3862\mathrm{E}-09$ &$-9.2227\mathrm{E}-09$  \\ 
$\Gamma_M$ & &$1.1921\mathrm{E}-04$ &$2.9758\mathrm{E}-08$ &$3.0503\mathrm{E}-07$ &$9.7837\mathrm{E}-12$  \\ 
$\Gamma_J$ & &$4.0665\mathrm{E}-05$ &$2.1215\mathrm{E}-05$ &$0$ &$0$  \\ 
$\Gamma_\nu$ &$-2.1990\mathrm{E}-08$ &$-4.2321\mathrm{E}-05$ &$-3.9619\mathrm{E}-07$ &$-4.3489\mathrm{E}-08$ &$-2.2006\mathrm{E}-08$  \\ 
$\Gamma_\eta$ &$-1.0266\mathrm{E}-06$ &$-1.2435\mathrm{E}-04$ &$-1.9008\mathrm{E}-05$ &$-1.3873\mathrm{E}-06$ &$-1.0323\mathrm{E}-06$  \\ 
$\Gamma_{aU}$ &$2.4801\mathrm{E}-15$ &$-1.1324\mathrm{E}-12$ &$5.8983\mathrm{E}-14$ &$4.7636\mathrm{E}-15$ &$2.0486\mathrm{E}-15$  \\ 
$\Gamma_{aJ}$ & &$9.2090\mathrm{E}-06$ &$2.1239\mathrm{E}-05$ &$0$ &$0$  \\ 
$\Gamma_{mech}$ &$-3.1224\mathrm{E}-08$ &$5.1710\mathrm{E}-06$ &$2.0650\mathrm{E}-05$ &$-3.5102\mathrm{E}-08$ &$-3.1229\mathrm{E}-08$  \\ 
$\Gamma_{em}$ &$-1.0266\mathrm{E}-06$ &$4.0697\mathrm{E}-06$ &$2.2610\mathrm{E}-06$ &$-1.0823\mathrm{E}-06$ &$-1.0323\mathrm{E}-06$  \\ 

\end{tabular*}
\caption{\label{table:modeFMhd}Growth rate and phase velocity, velocity and magnetic field eigenfunctions at the free-surface, energy components, and energy transfer rates for mode~F at $ \Rey = 7 \times 10^{5} $, $ \alpha = 2 \times 10^{-3} $, $ \Gal = 8.3 \times 10^7 $, $ \Capil = 0.07 $, $ \Har = 20 $, and $ \Prm = 10^{-5} $. The data in column~1, counting from the left, are for the corresponding inductionless problem, and therefore all entries involving the magnetic field eigenfunction are omitted. The results in columns~2 and~4 are for insulating boundary conditions, while for those in columns~3 and~5 conducting boundary conditions have been imposed (\ie the external magnetic energy component $ E_{b-} $ is omitted). For the calculations in columns~4 and~5 the induced magnetic field $ B $ has been set to zero, and as a result $ \Gamma_J $ and $ \Gamma_{aJ} $ vanish.}
\end{table}

The results for the test problems with $ B = 0 $ agree fairly well with the corresponding ones in the inductionless limit. Here the main differences are that the total electromagnetic energy transfer rate $ \Gamma_{ em } := \Gamma_\eta + \Gamma_M $ now includes a small, positive Maxwell stress, and for $ \Har \gtrsim 30 $, that the modal phase velocity is less than the free-surface steady-state velocity. The error is particularly small for the conducting-wall problem, since in that case the energy transfer to the modes via Maxwell stress, which is not captured by the inductionless scheme, is suppressed due to the nature of the boundary conditions. In contrast, when $ B $ is included, the current-interaction term $ \Gamma_J $ causes the total mechanical energy transfer rate $ \Gamma_{mech} := \Gamma_\nu + \Gamma_R + \Gamma_{a U } + \Gamma_J $ to remain positive for Hartmann numbers larger than the $ \Har \approx 9 $ value above which it becomes negative in inductionless and $ B = 0 $ problems. The total electromagnetic energy transfer rate $ \Gamma_{em } := \Gamma_\eta + \Gamma_M + \Gamma_{aJ } $, where the surface term $ \Gamma_{a J } $ is appreciable and positive, also deviates markedly from its $ \Har $-dependence in the inductionless limit. 

In the insulating-wall example, the combined effect produces a more than two-fold increase of the Hartmann number required for instability suppression relative to the inductionless flow. Moreover, the modal energy in the strong-field limit, instead of becoming almost exclusively gravitational, is split into a $ \Har $-independent mix between magnetic and surface parts. Although we have observed a scaling of the form $ E_b / E_a \propto \Prm / \Gal $ for $ \Har \gg 1 $, we have not been able to account for it by invoking a work-balance argument (\cf \S\ref{sec:zeroPmProblems}). 

If now the wall is conducting, the current interaction term $ \Gamma_J $ becomes sufficiently large so as to cause the growth rate $ \Imag( c ) \alpha $ to asymptote to a positive value, rather than approach zero from below. In addition, the energy in the magnetic degrees of freedom dominates, with both surface and kinetic contributions decaying to zero. The greater influence of the steady-state current in problems with a conducting wall is consistent with the fact that $ | B( z ) | $ is of order unity throughout the core of the flow domain, while it is of order $  1 / \Har $ when the wall is insulating (see \S\ref{sec:steadyState}).

The data for $ \Gamma_R $ and $ \Gamma_\nu $ in table~\ref{table:modeFMhd} exhibit a particularly large discrepancy between the nonzero-$ B $ problem with an insulating wall and its inductionless counterpart, signalling that the Lorentz force associated with $ \basej $ causes significant changes in structure of the velocity eigenfunction $ \modeu( z ) $. Indeed, as illustrated in figure~\ref{fig:modeFMhd}, the perturbed velocity field of the full MHD problem bears little resemblance to the inductionless examples in figure~\ref{fig:modeFZeroPm}. In particular, instead of evolving towards a state of constant shear as $ \Har $ grows, $ | \modeu( z ) | $, as well as the kinetic energy density $ \modeeu( z) $, develop a maximum in the wall region, where $ | \basej | $ is concentrated (see figure~\ref{fig:baseHartmann}). The increased eigenfunction curvature leads in turn to higher viscous and resistive dissipation, but these are more than counter-balanced by positive current interaction, and by positive Reynolds and Maxwell stresses. As for the magnetic field eigenfunction $ \modeb(z) $, its modulus varies by less than $ 10^{-2} $ over the fluid domain, and since the modal wavenumber $ \alpha = 0.002 $ is small, the energy of the magnetic field penetrating into the exterior region, $ E_{b'} $~\eqref{eq:modeebext}, exceeds the internal magnetic energy $ E_{b} $~\eqref{eq:modeeueb} by two orders of magnitude (see table~\ref{table:modeFMhd}). We remark that in problems with a conducting wall, where the current distribution over the inner part of the domain is nearly uniform, $ \modeu( z ) $ is qualitatively similar to the inductionless case, and $ \modeb( z ) $ varies nearly linearly from zero at the wall to its free-surface value.

\begin{figure}
\begin{center}
\includegraphics{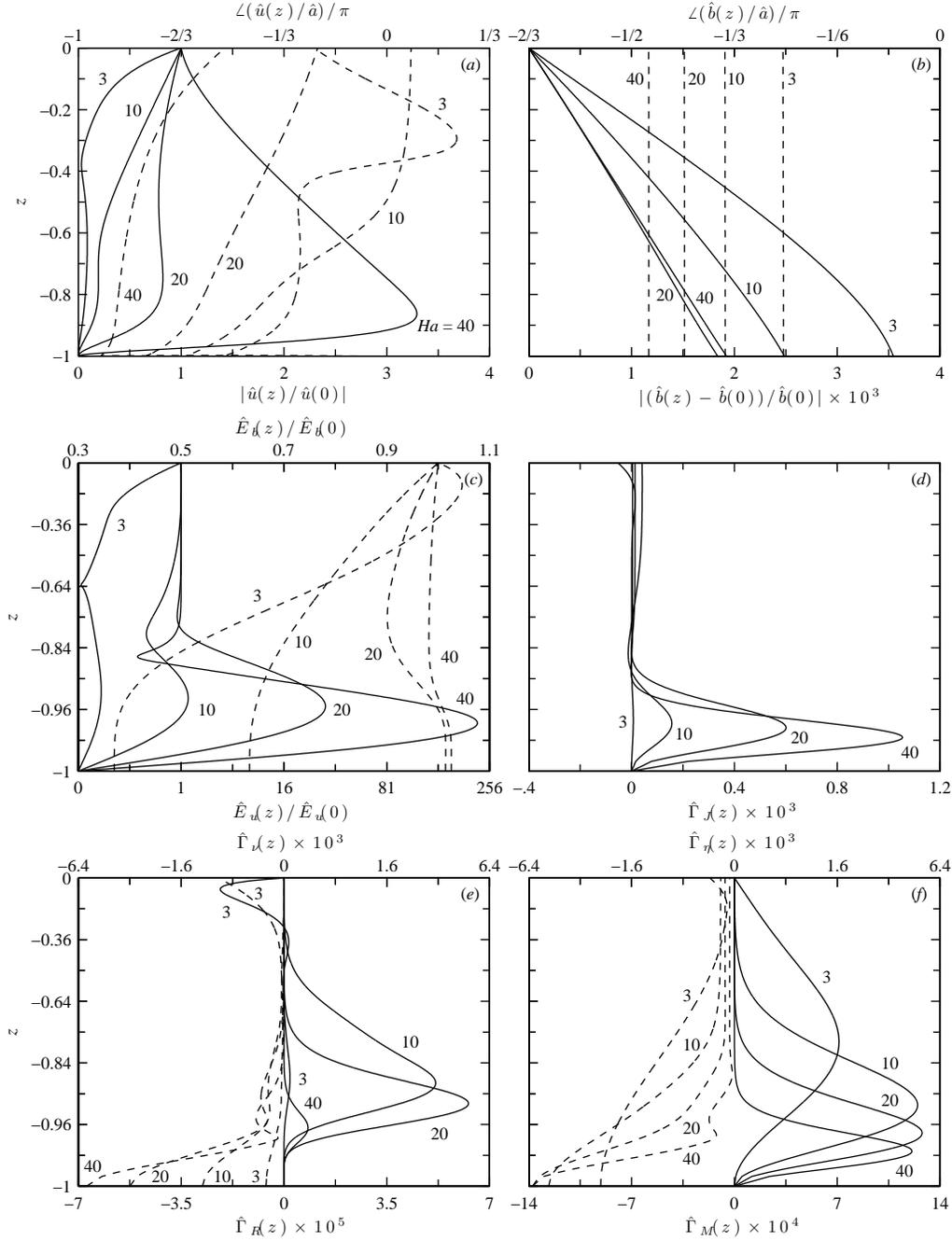}
\caption{\label{fig:modeFMhd}Velocity eigenfunction ({\it a}), magnetic field eigenfunction ({\it b}), kinetic and magnetic energy densities ({\it c}), current interaction ({\it d}), viscous dissipation and current-interaction densities ({\it e}), and resistive dissipation and Maxwell-stress densities ({\it f}) for mode~F in insulating-wall problems with $ \Prm = 10^{-5} $, $ \Har \in \{ 3, 10, 20, 40 \} $, and remaining flow-parameters equal to those in the inductionless problems in figure~\ref{fig:modeFZeroPm}. In panels~({\it a}), ({\it b}), ({\it e}), and ({\it f}) solid and dotted lines are associated with the lower and upper horizontal axes, respectively. In order to better depict the peaks in $ \modeeu( z ) $, the corresponding axis has quartic scaling. Likewise, quadratic scaling has been used for the $ z $ coordinate in panels ({\it c}--{\it f}), as well as the axes for $ \skew3\hat\Gamma_\nu(z) $ and $ \skew3\hat\Gamma_\eta(z) $.}
\end{center}
\end{figure}

\subsubsection{Travelling Alfv\'en modes}

\begin{figure}
\begin{center}
\includegraphics{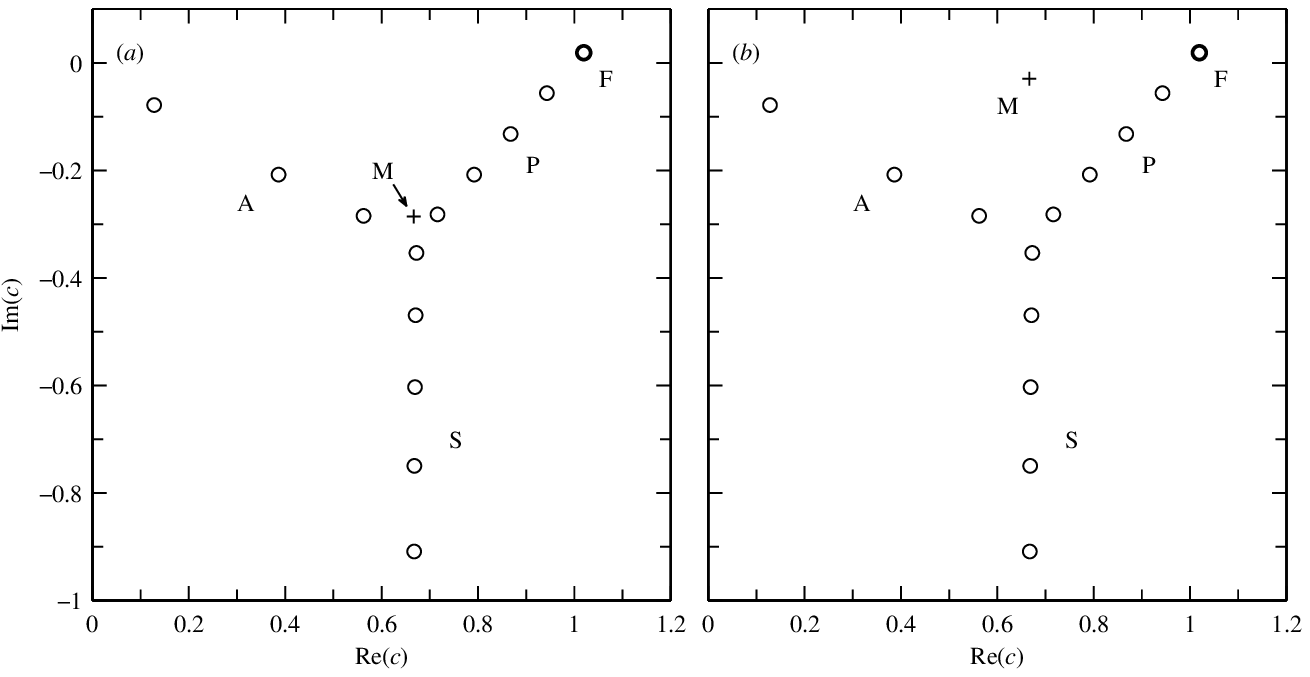}
\caption{\label{fig:spectrumAMhdPm}Eigenvalues of free-surface Hartmann flow with an insulating wall at $ \Rey = 7 \times 10^5 $, $ \alpha = 2 \times 10^{-3 } $, $ \Gal = 8.3 \times 10^7 $, $ \Capil = 0.07 $, and $ \Har = 0.1 $. The magnetic Prandtl number $ \Prm $ in panels~({\it a}) and ({\it b}) is $ 10^{-5} $ and $ 10^{-4} $, respectively. Mode~M, plotted using a + marker, is singular in the inductionless limit $ \Prm \searrow 0 $, while mode~F, plotted in boldface, is unstable in both panels. The dependence of this spectrum on $ \Prm \in [ 10^{-6}, 10^{-4} ] $ is shown in movie~2, available with the online version of the paper.} 
\end{center}
\end{figure}

Having studied the behaviour of the least stable modes in some detail, we now examine the influence of a small, but nonzero, magnetic Prandtl number on the more stable modes. As illustrated by the spectra in figure~\ref{fig:spectrumAMhdPm} and table~\ref{table:spectrumAMhdPm}, computed for $ \Har = 0.1 $ and $ \Prm \in \{ 10^{-5} , 10^{-4 } \} $, when the applied magnetic field is weak, free-surface flows with an insulating wall exhibit, in addition to the A, F, P, and S~modes, a mode labelled~M that is not present in inductionless flows. Mode~M is stable, and like S-family modes, its phase velocity is close to the mean steady-state speed. However, it stands out from the remaining modes because of (i) the strong $ \Prm $-dependence of its decay rate, which decreases by an order of magnitude between the $ \Prm = 10^{-5} $ and $ \Prm = 10^{-4} $ calculations in table~\ref{table:spectrumAMhdPm} (the corresponding relative variation for the A, F, P, and S modes is less than $ 10^{-4} $), and (ii) its mostly magnetic energy content  (\eg in table~\ref{table:spectrumAMhdPm}, $ E_b / E \approx 0.95 $ for mode~M, whereas $ E_b / E \lesssim 10^{-4} $ for modes in the A, F, P, and S families). Although we have not confirmed this analytically, numerical calculations strongly suggest that mode~M is singular in the inductionless limit $ \Prm \searrow 0 $. In particular, as shown in movie~2, when $ \Prm $ is increased from small values the complex phase velocity of mode~M is seen to approach the upper part of the spectrum from arbitrarily small values of $ \Imag( c ) $, moving along the S eigenvalue branch. We remark, however, that the complex phase velocity does not cross the $ \Imag( c ) = 0 $ axis. Instead, if $ \Prm $ is allowed to be of order unity, mode~M eventually participates in the formation of magnetic eigenvalue branches \cite[][]{GiannakisFischerRosner08}. 

\begin{table}
\begin{tabular*}{\linewidth}{@{\extracolsep{\fill}}llrll}
& & \multicolumn{1}{c}{$c$} & \multicolumn{1}{c}{$E_a/E$} & \multicolumn{1}{c}{$E_b/E$}\\
\cline{3-5}
\multicolumn{5}{l}{$ \Har = 0.1 $, $ \Prm = 10^{-5} $}\\
1 &  $ \mbox{F} $ &  $ 1.01931082772748\mathrm{E}+00 + 1.87122221676069\mathrm{E}-02 \, \ii $ & $ 3.82063\mathrm{E}-02 $ & $ 2.65603\mathrm{E}-04 $ \\ 
2 &  $ \mbox{P}_{ 1 } $ &  $ 9.43120008162654\mathrm{E}-01 - 5.65860384497908\mathrm{E}-02 \, \ii $ & $ 6.53040\mathrm{E}-03 $ & $ 4.87128\mathrm{E}-05 $ \\ 
3 &  $ \mbox{A}_{ 1 } $ &  $ 1.28101973089488\mathrm{E}-01 - 7.85027707446042\mathrm{E}-02 \, \ii $ & $ 1.61230\mathrm{E}-04 $ & $ 1.74331\mathrm{E}-06 $ \\ 
4 &  $ \mbox{P}_{ 2 } $ &  $ 8.67716150616648\mathrm{E}-01 - 1.32222546160657\mathrm{E}-01 \, \ii $ & $ 6.09816\mathrm{E}-04 $ & $ 6.65946\mathrm{E}-06 $ \\ 
5 &  $ \mbox{P}_{ 3 } $ &  $ 7.92170269201228\mathrm{E}-01 - 2.07817958769442\mathrm{E}-01 \, \ii $ & $ 1.14325\mathrm{E}-04 $ & $ 3.72558\mathrm{E}-06 $ \\ 
6 &  $ \mbox{A}_{ 2 } $ &  $ 3.86411583077811\mathrm{E}-01 - 2.07964549091180\mathrm{E}-01 \, \ii $ & $ 1.41756\mathrm{E}-04 $ & $ 2.47046\mathrm{E}-06 $ \\ 
7 &  $ \mbox{P}_{ 4 } $ &  $ 7.16273931735328\mathrm{E}-01 - 2.81864800130393\mathrm{E}-01 \, \ii $ & $ 2.31289\mathrm{E}-05 $ & $ 9.27739\mathrm{E}-06 $ \\ 
8 &  $ \mbox{A}_{ 3 } $ &  $ 5.62415456020230\mathrm{E}-01 - 2.84607742157634\mathrm{E}-01 \, \ii $ & $ 3.99484\mathrm{E}-05 $ & $ 4.07061\mathrm{E}-06 $ \\ 
9 &  $ \mbox{M} $ &  $ 6.66774912099173\mathrm{E}-01 - 2.86025163476446\mathrm{E}-01 \, \ii $ & $ 6.24179\mathrm{E}-07 $ & $ 9.74910\mathrm{E}-01 $ \\ 
10 &  $ \mbox{S}_{ 1 } $ &  $ 6.72566612117371\mathrm{E}-01 - 3.53685916601131\mathrm{E}-01 \, \ii $ & $ 3.85519\mathrm{E}-06 $ & $ 1.26283\mathrm{E}-06 $ \\ 

\\
\multicolumn{5}{l}{$ \Har = 0.1 $, $ \Prm = 10^{-4} $}\\
1 &  $ \mbox{F} $ &  $ 1.01931125301428\mathrm{E}+00 + 1.87093734887185\mathrm{E}-02 \, \ii $ & $ 3.82168\mathrm{E}-02 $ & $ 6.24640\mathrm{E}-05 $ \\ 
2 &  $ \mbox{M} $ &  $ 6.66800739433602\mathrm{E}-01 - 2.97757500101643\mathrm{E}-02 \, \ii $ & $ 1.47480\mathrm{E}-06 $ & $ 9.97323\mathrm{E}-01 $ \\ 
3 &  $ \mbox{P}_{ 1 } $ &  $ 9.43117677138117\mathrm{E}-01 - 5.65832503068096\mathrm{E}-02 \, \ii $ & $ 6.53056\mathrm{E}-03 $ & $ 1.04209\mathrm{E}-05 $ \\ 
4 &  $ \mbox{A}_{ 1 } $ &  $ 1.28101057320598\mathrm{E}-01 - 7.85034486458773\mathrm{E}-02 \, \ii $ & $ 1.61230\mathrm{E}-04 $ & $ 3.43338\mathrm{E}-07 $ \\ 
5 &  $ \mbox{P}_{ 2 } $ &  $ 8.67714460364094\mathrm{E}-01 - 1.32221228532799\mathrm{E}-01 \, \ii $ & $ 6.09793\mathrm{E}-04 $ & $ 1.03611\mathrm{E}-06 $ \\ 
6 &  $ \mbox{P}_{ 3 } $ &  $ 7.92167689972354\mathrm{E}-01 - 2.07816923033732\mathrm{E}-01 \, \ii $ & $ 1.14320\mathrm{E}-04 $ & $ 2.11886\mathrm{E}-07 $ \\ 
7 &  $ \mbox{A}_{ 2 } $ &  $ 3.86416377565185\mathrm{E}-01 - 2.07965044026100\mathrm{E}-01 \, \ii $ & $ 1.41753\mathrm{E}-04 $ & $ 2.75247\mathrm{E}-07 $ \\ 
8 &  $ \mbox{P}_{ 4 } $ &  $ 7.16262532911781\mathrm{E}-01 - 2.81865240487700\mathrm{E}-01 \, \ii $ & $ 2.31293\mathrm{E}-05 $ & $ 4.63160\mathrm{E}-08 $ \\ 
9 &  $ \mbox{A}_{ 3 } $ &  $ 5.62410360672946\mathrm{E}-01 - 2.84606171228795\mathrm{E}-01 \, \ii $ & $ 3.99490\mathrm{E}-05 $ & $ 7.99622\mathrm{E}-08 $ \\ 
10 &  $ \mbox{S}_{ 1 } $ &  $ 6.72561149596221\mathrm{E}-01 - 3.53688962352034\mathrm{E}-01 \, \ii $ & $ 3.85565\mathrm{E}-06 $ & $ 1.00596\mathrm{E}-08 $ \\ 

\end{tabular*}
\caption{\label{table:spectrumAMhdPm}Complex phase velocity $ c $, free-surface energy $ E_a $, and magnetic energy $ E_b $ of the 10 least stable modes of the spectra in figure~\ref{fig:spectrumAMhdPm}. The energies $ E_a $ and $ E_b $ have been normalised by the total energy $ E = E_u + E_b + E_a $.}
\end{table}

The existence of modes of magnetic origin (hence the designation M) in the spectrum of the coupled OS and induction equations~\eqref{eq:orrSommerfeldInd} is consistent with the fact that the limit $ \Prm \searrow 0 $ at fixed $ \Har $, which, as discussed in~\S\ref{sec:2dNormalModes}, leads to the approximate stability equation~\eqref{eq:orrSommerfeldZeroPm}, is singular (effectively, the differential order of the stability problem is reduced from six to four). Modes of this type are also present in channel Hartmann flows, as well as in test problems with $ B = 0 $. However, we have found no evidence of mode~M in numerical calculations with conducting boundary conditions, which correlates with the absence of the $ \gamma_1^{(M)} $ magnetic solution in the large-wavelength approximations for conducting-wall problems in appendix~\ref{app:pertMhd}. 

\begin{figure}
\begin{center}
\includegraphics{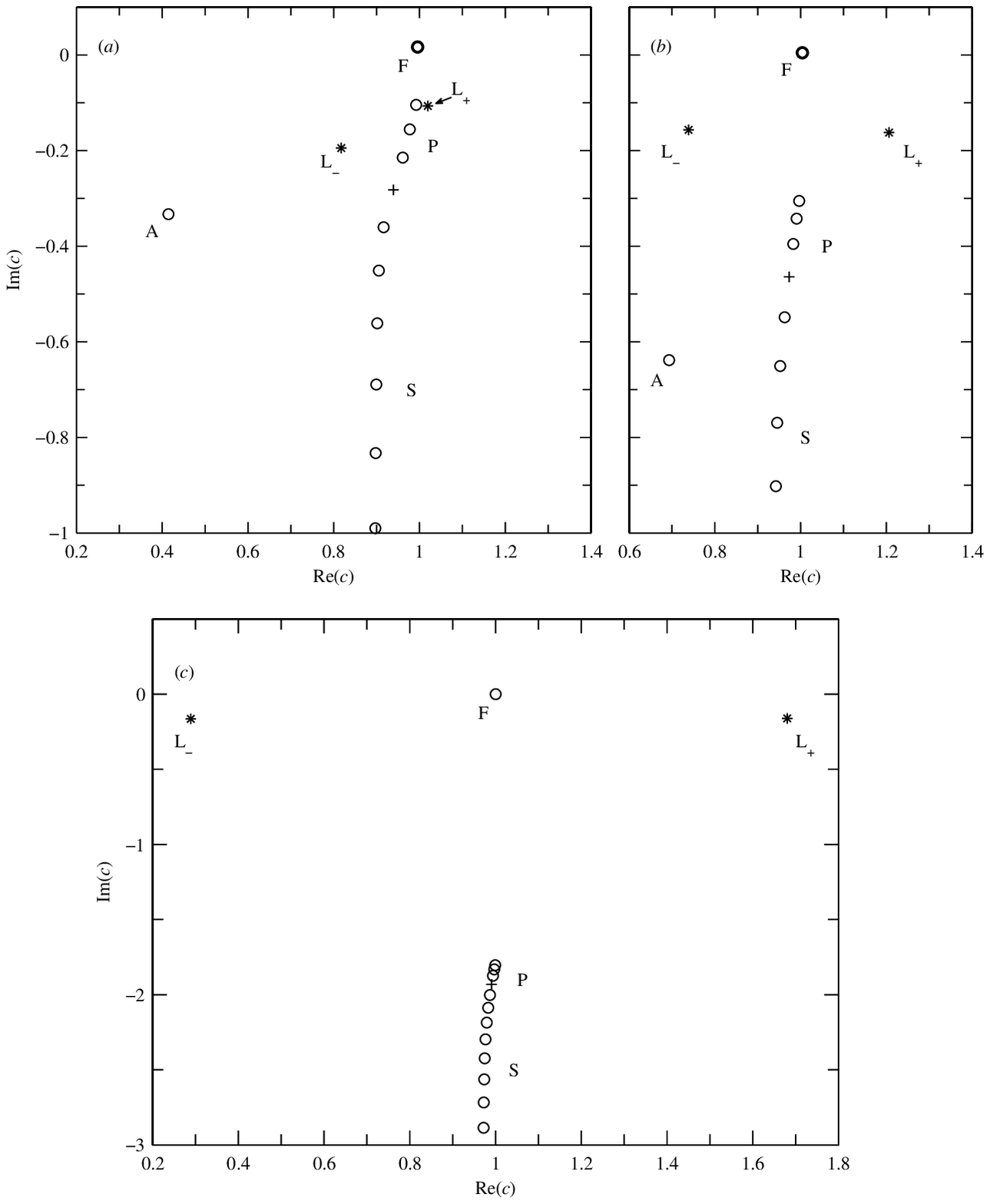}
\caption{\label{fig:spectrumAMhd}Eigenvalues of free-surface Hartmann flow with insulating wall at $ \Rey = 7 \times 10^5 $, $ \alpha = 2 \times 10^{-3 } $, $ \Prm = 10^{-5} $, $ \Gal = 8.3 \times 10^7 $, $ \Capil = 0.07 $, and Hartmann number $ \Har = 10 $ ({\it a}), $ 20 $ ({\it b}), and 50 ({\it c}). The modes labelled L$_-$ and L$_+$, and plotted using * markers, are travelling Alfv\'en waves, and can be continuously traced to modes A$_2$ and P$_1$, respectively, as $ \Har \searrow 0 $. The mode plotted using a + marker originates from mode~M in figure~\ref{fig:spectrumAMhdPm}({\it a}). In panels ({\it a}) and ({\it b}), mode~F, plotted in boldface, is unstable. The evolution of this spectrum with $ \Har \in [ 0.1, 50 ] $ is shown in movie~3, available with the online version of the paper.} 
\end{center}
\end{figure}

As shown in the spectra in figure~\ref{fig:spectrumAMhd}, evaluated for insulating-wall problems with $ \Prm = 10^{-5} $, $ \Har \in \{ 10, 20, 50 \} $, and otherwise the same parameters as the inductionless calculations in figure~\ref{fig:spectrumAZeroPm}, when the external magnetic field is of appreciable strength the collapse of the A eigenvalue branch and the alignment of the P and S branches observed in the inductionless limit (see \S\ref{sec:zeroPmProblems}) are also present in $ \Prm > 0$ problems. In addition, according to the data in table~\ref{table:spectrumAMhd}, the magnetic energy for all but the first handful of modes is relatively small ($ E_b/E \lesssim 10^{-1} $), and in those cases the accuracy of the inductionless approximation is very acceptable (for instance, a comparison with table~\ref{table:spectrumAZeroPm} shows that the relative error for mode P$_4$ at $ \Har = 10 $ is at the 1\% level). Still, as already discussed in \S\ref{sec:mhdTopEnd} and \S\ref{sec:roleOfJ}, nonzero-$ \Prm $ spectra deviate substantially from the corresponding inductionless ones in the behaviour of mode~F, whose main distinguishing features in figure~\ref{fig:spectrumAMhd} and table~\ref{table:spectrumAMhd} in comparison with figure~\ref{fig:spectrumAZeroPm} and table~\ref{table:spectrumAZeroPm} are (i) sub-unity phase velocity $ C \approx 0.9959 $ when $ \Har = 10 $, (ii) positive growth rate in both of the $ \Har = 10 $ and $ \Har = 20 $ calculations, and (iii) magnetic-energy predominance for strong background fields  (\eg $ E_b / E \approx 0.93 $ for $ \Har = 50$).

\begin{table}
\begin{tabular*}{\linewidth}{@{\extracolsep{\fill}}llrll}
& & \multicolumn{1}{c}{$c$} & \multicolumn{1}{c}{$E_a/E$} & \multicolumn{1}{c}{$E_b/E$}\\
\cline{3-5}
\multicolumn{5}{l}{$ \Har = 10 $, $ \Prm = 10^{-5} $}\\
1 &  $ \mbox{F} $ &  $ 9.95865078137089\mathrm{E}-01 + 1.63947948356842\mathrm{E}-02 \, \ii $ & $ 6.67566\mathrm{E}-02 $ & $ 7.56599\mathrm{E}-01 $ \\ 
2 &  $ \mbox{P}_{ 1 } $ &  $ 9.92557588155474\mathrm{E}-01 - 1.04442847126442\mathrm{E}-01 \, \ii $ & $ 1.82412\mathrm{E}-03 $ & $ 1.03289\mathrm{E}-01 $ \\ 
3 &  $ \mbox{L}_{ + } $ &  $ 1.01939554477557\mathrm{E}+00 - 1.06858631729338\mathrm{E}-01 \, \ii $ & $ 5.76706\mathrm{E}-03 $ & $ 3.85449\mathrm{E}-01 $ \\ 
4 &  $ \mbox{P}_{ 2 } $ &  $ 9.77461544779890\mathrm{E}-01 - 1.55881324364227\mathrm{E}-01 \, \ii $ & $ 6.76741\mathrm{E}-04 $ & $ 1.08003\mathrm{E}-01 $ \\ 
5 &  $ \mbox{L}_{-} $ &  $ 8.17555458976949\mathrm{E}-01 - 1.94975997718683\mathrm{E}-01 \, \ii $ & $ 7.43663\mathrm{E}-04 $ & $ 2.66480\mathrm{E}-01 $ \\ 
6 &  $ \mbox{P}_{ 3 } $ &  $ 9.61162906229426\mathrm{E}-01 - 2.14936006245129\mathrm{E}-01 \, \ii $ & $ 1.26044\mathrm{E}-04 $ & $ 8.66194\mathrm{E}-02 $ \\ 
\slshape7 & \slshape $ \mbox{P}_{ 4 } $ &  $ \text{\slshape9.39093462022514}\text{\slshape{E}}-\text{\slshape01} - \text{\slshape2.82501396482664}\text{\slshape{E}}-\text{\slshape{01}} \, \text{\slshape{i}} $ & $ \text{\slshape5.08031}\text{\slshape{E}}-\text{\slshape{06}} $ & $ \text{\slshape3.25970}\text{\slshape{E}}-\text{\slshape{02}} $ \\ 
8 &  $ \mbox{A}_{ 1 } $ &  $ 4.14592505991546\mathrm{E}-01 - 3.33142243705258\mathrm{E}-01 \, \ii $ & $ 7.03352\mathrm{E}-05 $ & $ 1.24080\mathrm{E}-02 $ \\ 
9 &  $ \mbox{P}_{ 5 } $ &  $ 9.16439057608324\mathrm{E}-01 - 3.60617428043375\mathrm{E}-01 \, \ii $ & $ 2.76783\mathrm{E}-06 $ & $ 7.44498\mathrm{E}-03 $ \\ 
10 &  $ \mbox{S}_{ 1 } $ &  $ 9.05320512052477\mathrm{E}-01 - 4.51161148416211\mathrm{E}-01 \, \ii $ & $ 1.69292\mathrm{E}-06 $ & $ 1.52332\mathrm{E}-03 $ \\ 

\\
\multicolumn{5}{l}{$ \Har = 20 $, $ \Prm = 10^{-5} $}\\
1 &  $ \mbox{F} $ &  $ 1.00392370142334\mathrm{E}+00 + 4.62308174479437\mathrm{E}-03 \, \ii $ & $ 5.41852\mathrm{E}-02 $ & $ 8.77046\mathrm{E}-01 $ \\ 
2 &  $ \mbox{L}_{-} $ &  $ 7.38547682046289\mathrm{E}-01 - 1.56955992195965\mathrm{E}-01 \, \ii $ & $ 8.21253\mathrm{E}-04 $ & $ 4.47778\mathrm{E}-01 $ \\ 
3 &  $ \mbox{L}_{ + } $ &  $ 1.20624039578785\mathrm{E}+00 - 1.62495095508186\mathrm{E}-01 \, \ii $ & $ 1.00742\mathrm{E}-03 $ & $ 5.53021\mathrm{E}-01 $ \\ 
4 &  $ \mbox{P}_{ 1 } $ &  $ 9.96643310724506\mathrm{E}-01 - 3.05478418809574\mathrm{E}-01 \, \ii $ & $ 2.85312\mathrm{E}-08 $ & $ 7.90718\mathrm{E}-04 $ \\ 
5 &  $ \mbox{P}_{ 2 } $ &  $ 9.90773359402505\mathrm{E}-01 - 3.42351571140559\mathrm{E}-01 \, \ii $ & $ 1.12321\mathrm{E}-07 $ & $ 1.41202\mathrm{E}-03 $ \\ 
6 &  $ \mbox{P}_{ 3 } $ &  $ 9.82893615854751\mathrm{E}-01 - 3.95487920055681\mathrm{E}-01 \, \ii $ & $ 3.08386\mathrm{E}-07 $ & $ 1.84037\mathrm{E}-03 $ \\ 
\slshape7 & \slshape $ \mbox{P}_{ 4 } $ &  $ \text{\slshape9.73387208655838}\text{\slshape{E}}-\text{\slshape01} - \text{\slshape4.64261831294372}\text{\slshape{E}}-\text{\slshape01} \, \text{\slshape{i}} $ & $ \text{\slshape5.33617}\text{\slshape{E}}-\text{\slshape07} $ & $ \text{\slshape1.74777}\text{\slshape{E}}-\text{\slshape{03}} $ \\ 
8 &  $ \mbox{P}_{ 5 } $ &  $ 9.62772461311780\mathrm{E}-01 - 5.48990582273123\mathrm{E}-01 \, \ii $ & $ 6.23475\mathrm{E}-07 $ & $ 1.32879\mathrm{E}-03 $ \\ 
9 &  $ \mbox{A}_{ 1 } $ &  $ 6.93299438967882\mathrm{E}-01 - 6.38530573016983\mathrm{E}-01 \, \ii $ & $ 8.03501\mathrm{E}-06 $ & $ 9.71982\mathrm{E}-03 $ \\ 
10 &  $ \mbox{S}_{ 1 } $ &  $ 9.52461437608548\mathrm{E}-01 - 6.50819443090838\mathrm{E}-01 \, \ii $ & $ 5.46661\mathrm{E}-07 $ & $ 8.54110\mathrm{E}-04 $ \\ 

\\
\multicolumn{5}{l}{$ \Har = 50 $, $ \Prm = 10^{-5} $}\\
1 &  $ \mbox{F} $ &  $ 1.00024009887062\mathrm{E}+00 - 7.84769189348587\mathrm{E}-05 \, \ii $ & $ 6.17833\mathrm{E}-02 $ & $ 9.29735\mathrm{E}-01 $ \\ 
2 &  $ \mbox{L}_{ + } $ &  $ 1.68062858014303\mathrm{E}+00 - 1.61493055991922\mathrm{E}-01 \, \ii $ & $ 1.67082\mathrm{E}-04 $ & $ 5.14154\mathrm{E}-01 $ \\ 
3 &  $ \mbox{L}_{ - } $ &  $ 2.89175320132634\mathrm{E}-01 - 1.64256263206813\mathrm{E}-01 \, \ii $ & $ 1.57691\mathrm{E}-04 $ & $ 4.87064\mathrm{E}-01 $ \\ 
4 &  $ \mbox{P}_{ 1 } $ &  $ 9.99199429127421\mathrm{E}-01 - 1.80434774096787\mathrm{E}+00 \, \ii $ & $ 1.82872\mathrm{E}-10 $ & $ 1.94749\mathrm{E}-03 $\\ 
5 &  $ \mbox{P}_{ 2 } $ &  $ 9.97091976769849\mathrm{E}-01 - 1.83142407634326\mathrm{E}+00 \, \ii $ & $ 7.93664\mathrm{E}-10 $ & $ 7.16630\mathrm{E}-04 $ \\ 
6 &  $ \mbox{P}_{ 3 } $ &  $ 9.94227660233984\mathrm{E}-01 - 1.87446928135769\mathrm{E}+00 \, \ii $ & $ 1.52677\mathrm{E}-09 $ & $ 3.71281\mathrm{E}-04 $ \\ 
\slshape 7 & \slshape $ \mbox{P}_{ 4 } $ & $ \text{\slshape9.90706905162818}\text{\slshape{E}}-\text{\slshape01} - \text{\slshape1.93185106397859}\text{\slshape{E}}+\text{\slshape00} \, \text{\slshape{i}} $ & $ \text{\slshape2.45478}\text{\slshape{E}}-\text{\slshape09} $ & $ \text{\slshape2.35469}\text{\slshape{E}}-\text{\slshape{04}} $ \\ 
8 &  $ \mbox{P}_{ 5 } $ &  $ 9.86809861524976\mathrm{E}-01 - 2.00299291310038\mathrm{E}+00 \, \ii $ & $ 3.38590\mathrm{E}-09 $ & $ 1.68562\mathrm{E}-04 $ \\ 
9 &  $ \mbox{P}_{ 6 } $ &  $ 9.82936143590023\mathrm{E}-01 - 2.08763002827334\mathrm{E}+00 \, \ii $ & $ 4.23074\mathrm{E}-09 $ & $ 1.30181\mathrm{E}-04 $ \\ 
10 &  $ \mbox{S}_{ 1 } $ &  $ 9.79496738581810\mathrm{E}-01 - 2.18575887355621\mathrm{E}+00 \, \ii $ & $ 4.81819\mathrm{E}-09 $ & $ 1.05130\mathrm{E}-04 $ \\ 

\end{tabular*}
\caption{\label{table:spectrumAMhd}Complex phase velocity $ c $, surface energy $ E_a $, and magnetic energy $ E_b $ of the 10 least stable modes of the spectra in figure~\ref{fig:spectrumAMhd}. The energies $ E_a $ and $ E_b $ have been normalised by the total energy $ E = E_u + E_b + E_a $. If $ \Har $ is decreased to 0.1, mode~P$_4$ (in figure~\ref{fig:spectrumAMhd}, plotted using a + marker) can be continuously traced to mode~M in the $ \Prm = 10^{-5} $ part of table~\ref{table:spectrumAMhdPm}. The average steady-state velocity $ \langle U \rangle $ is, in accordance with~\eqref{eq:baseUAverage}, 0.9001, 0.9500, and 0.9800, respectively for $ \Har = 10 $, 20, and 50. Due to the alignment of the P and S branches, the classification of the sixth least stable mode for $ \Har = 10 $, the eighth least stable mode for $ \Har = 20 $, and the ninth least stable mode for $ \Har = 50 $ as members of the P family is somewhat arbitrary; these modes could equally be treated as members of the  S branch.}
\end{table}

Aside from the discrepancies associated with mode~F, however, the spectra in figure~\ref{fig:spectrumAMhd} differ from those in figure~\ref{fig:reAlphaZeroPm} in that they contain two travelling modes, labelled L$_-$ and L$_+$, which have no counterparts in the inductionless limit. As illustrated in movie~3, these modes are the outcome of a mode-conversion process, whereby the weak-field modes A$_2$ and P$_1$ separate from the A and P eigenvalue branches when $ \Har $ is increased from small values, and move towards nearly symmetric positions on the complex plane about $ \Real( c ) = 1 $. At the same time, mode~M, joins the P eigenvalue branch, and having lost the majority of its magnetic energy, behaves in the same manner as the remaining P and S modes. A principal feature of the L~modes for sufficiently strong magnetic fields, which can be observed in the $ \Har = 20 $ and $ \Har = 50 $ results in table~\ref{table:spectrumAMhd}, is a near equipartition between magnetic and kinetic energy. Because of this, we classify these modes as \emph{travelling Alfv\'en waves}. In separate numerical calculations, we have confirmed that modes of this type are also part of the spectrum when the background fluid is at rest. In channel problems with insulating walls, the pair of travelling waves becomes replaced by a single frozen-in magnetic mode \cite[][\S IX]{BetchovCriminale67}, whose phase velocity and kinetic energy are nearly equal to unity and zero, respectively. However, as with mode~M, both of the travelling and frozen-in Alfv\'en modes are removed from the spectrum when conducting boundary conditions are enforced. 

\begin{figure}
\begin{center}
\includegraphics{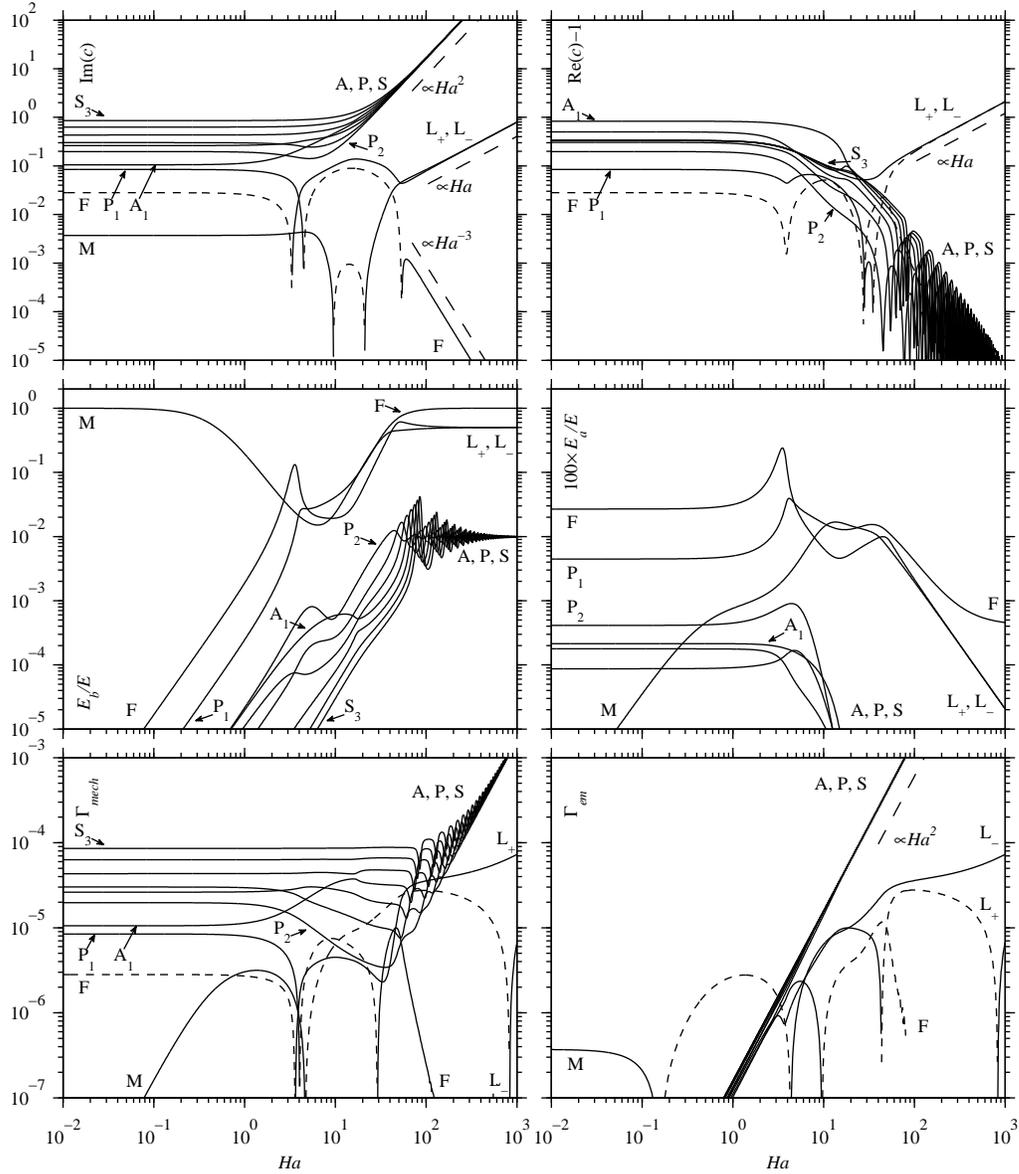}
\caption{\label{fig:harGammaAMhd}Complex phase velocity $ c $, magnetic and surface energies, $ E_b $ and $ E_a $ (normalised by the total energy $E = E_u + E_b + E_a $), and mechanical and electromagnetic energy transfer rates, $ \Gamma_{mech} $ and $ \Gamma_{em} $, for the ten least stable modes of free-surface Hartmann flow with insulating boundary conditions at $ \Rey = 6.3 \times 10^6 $, $ \alpha = 10^{-4} $, $ \Prm = 10^{-4} $, $ \Gal = 8.3 \times 10^7 $, $ \Capil = 0.7 $ and $ \Har \in [ 10^{-2}, 10^3 ] $. In the first and third-row panels, solid and dashed lines respectively correspond to negative and positive values. The weak-field ($ \Har = 0.1 $) modes~F, M, and P$_1$ respectively become converted to modes~L$_{+}$, L$_-$, and~F as $ \Har $ grows. The corresponding $ \Har $-evolution of the eigenvalues in the complex plane is shown in movie~4, available with the online version of the paper.}
\end{center}
\end{figure}

When a steady-state flow is present, the upstream-propagating wave L$_-$ may become unstable if the Alfv\'en number $ \Alf = \Rey \Prm^{1/2} / \Har $ of the flow is sufficiently large. This situation is illustrated by the eigenvalue and energy calculations in figure~\ref{fig:harGammaAMhd} and table~\ref{table:modeConversion}, evaluated at $ \Rey = 6.3 \times 10^6 $, $ \alpha = 10^{-4} $, $ \Prm = 10^{-4} $, and $ \Har \in [ 10^{-2}, 10^{3}] $, where the latter Hartmann-number interval corresponds to an Alfv\'en number decrease from $ 6.3 \times 10^6 $ ($ \Har = 10^{-2} $) to $ 63 $  ($ \Har = 10^{3} $). The resulting evolution of the eigenvalues in the complex-$ c $ plane is shown in movie~4. 

\begin{table}
\begin{scriptsize}
\begin{tabular*}{\linewidth}{@{\extracolsep{\fill}}lrrrrrr}
& \multicolumn{3}{c}{$\Har = 0.02$} & \multicolumn{3}{c}{$\Har = 1$} \\
\cline{2-4} \cline{5-7}
& F & M & P$_1$ & F & M & P$_1$ \\
$\Gamma$ &$8.449\mathrm{E}-06$ &$-8.784\mathrm{E}-06$ &$-2.535\mathrm{E}-05$ &$7.910\mathrm{E}-06$ &$-8.609\mathrm{E}-06$ &$-2.429\mathrm{E}-05$  \\ 
$C-1$ &$8.458\mathrm{E}-03$ &$-3.333\mathrm{E}-01$ &$-2.536\mathrm{E}-02$ &$8.006\mathrm{E}-03$ &$-3.226\mathrm{E}-01$ &$-2.429\mathrm{E}-02$  \\ 
\\
$|\modeu(0)/\modea|$ &$1.195\mathrm{E}-05$ &$3.334\mathrm{E}-04$ &$3.586\mathrm{E}-05$ &$1.125\mathrm{E}-05$ &$3.227\mathrm{E}-04$ &$3.435\mathrm{E}-05$  \\ 
$\angle(\modeu(0)/\modea)$ &$-2.502\mathrm{E}-01$ &$5.084\mathrm{E}-01$ &$7.500\mathrm{E}-01$ &$-2.519\mathrm{E}-01$ &$5.085\mathrm{E}-01$ &$7.500\mathrm{E}-01$  \\ 
$|\modeb(0)/\modea|$ &$2.895\mathrm{E}-07$ &$8.412\mathrm{E}-01$ &$2.886\mathrm{E}-07$ &$1.447\mathrm{E}-05$ &$1.634\mathrm{E}-02$ &$1.443\mathrm{E}-05$  \\ 
$\angle(\modeb(0)/\modea)$ &$-5.028\mathrm{E}-02$ &$1.251\mathrm{E}-01$ &$-4.934\mathrm{E}-02$ &$-4.922\mathrm{E}-02$ &$1.354\mathrm{E}-01$ &$-4.835\mathrm{E}-02$  \\ 
\\
$\mathrm{E}_u/\mathrm{E}$ &$9.987\mathrm{E}-01$ &$7.193\mathrm{E}-05$ &$9.998\mathrm{E}-01$ &$9.920\mathrm{E}-01$ &$1.524\mathrm{E}-01$ &$9.988\mathrm{E}-01$  \\ 
$\mathrm{E}_b/\mathrm{E}$ &$2.307\mathrm{E}-06$ &$6.747\mathrm{E}-01$ &$3.411\mathrm{E}-07$ &$6.130\mathrm{E}-03$ &$5.661\mathrm{E}-01$ &$9.000\mathrm{E}-04$  \\ 
$\mathrm{E}_{b+}$ &$6.572\mathrm{E}-08$ &$1.626\mathrm{E}-01$ &$1.087\mathrm{E}-08$ &$1.848\mathrm{E}-04$ &$1.408\mathrm{E}-01$ &$3.032\mathrm{E}-05$  \\ 
$\mathrm{E}_{b-}$ &$6.366\mathrm{E}-08$ &$1.626\mathrm{E}-01$ &$1.075\mathrm{E}-08$ &$1.793\mathrm{E}-04$ &$1.407\mathrm{E}-01$ &$2.999\mathrm{E}-05$  \\ 
$\mathrm{E}_a/\mathrm{E}$ &$1.328\mathrm{E}-03$ &$3.893\mathrm{E}-10$ &$2.211\mathrm{E}-04$ &$1.496\mathrm{E}-03$ &$8.936\mathrm{E}-07$ &$2.467\mathrm{E}-04$  \\ 
\\
$\Gamma_R$ &$-1.923\mathrm{E}-06$ &$-3.945\mathrm{E}-09$ &$-8.147\mathrm{E}-06$ &$-1.816\mathrm{E}-06$ &$-8.071\mathrm{E}-06$ &$-7.819\mathrm{E}-06$  \\ 
$\Gamma_M$ &$9.048\mathrm{E}-09$ &$1.012\mathrm{E}-02$ &$1.441\mathrm{E}-09$ &$2.432\mathrm{E}-05$ &$8.562\mathrm{E}-03$ &$3.842\mathrm{E}-06$  \\ 
$\Gamma_J$ &$-1.067\mathrm{E}-10$ &$2.036\mathrm{E}-09$ &$5.359\mathrm{E}-11$ &$-2.770\mathrm{E}-07$ &$4.201\mathrm{E}-06$ &$1.391\mathrm{E}-07$  \\ 
$\Gamma_\nu$ &$-2.573\mathrm{E}-05$ &$-7.107\mathrm{E}-10$ &$-1.119\mathrm{E}-05$ &$-2.493\mathrm{E}-05$ &$-1.509\mathrm{E}-06$ &$-1.070\mathrm{E}-05$  \\ 
$\Gamma_\eta$ &$-9.225\mathrm{E}-09$ &$-1.013\mathrm{E}-02$ &$-1.424\mathrm{E}-09$ &$-2.477\mathrm{E}-05$ &$-8.557\mathrm{E}-03$ &$-3.795\mathrm{E}-06$  \\ 
$\Gamma_{aU}$ &$3.610\mathrm{E}-05$ &$-8.298\mathrm{E}-12$ &$-6.012\mathrm{E}-06$ &$3.501\mathrm{E}-05$ &$-1.838\mathrm{E}-08$ &$-5.771\mathrm{E}-06$  \\ 
$\Gamma_{aJ}$ &$1.465\mathrm{E}-10$ &$-3.925\mathrm{E}-09$ &$-7.276\mathrm{E}-11$ &$3.762\mathrm{E}-07$ &$-8.070\mathrm{E}-06$ &$-1.877\mathrm{E}-07$  \\ 
$\Gamma_{mech}$ &$8.449\mathrm{E}-06$ &$-2.627\mathrm{E}-09$ &$-2.535\mathrm{E}-05$ &$7.986\mathrm{E}-06$ &$-5.397\mathrm{E}-06$ &$-2.415\mathrm{E}-05$  \\ 
$\Gamma_{em}$ &$-3.046\mathrm{E}-11$ &$-8.781\mathrm{E}-06$ &$-5.613\mathrm{E}-11$ &$-7.530\mathrm{E}-08$ &$-3.211\mathrm{E}-06$ &$-1.399\mathrm{E}-07$  \\ 

\\
& \multicolumn{3}{c}{$\Har = 15$} & \multicolumn{3}{c}{$\Har = 200$} \\
\cline{2-4} \cline{5-7}
& L$_+$ & L$_-$ & F & L$_+$ & L$_-$ & F \\
$\Gamma$ &$-3.265\mathrm{E}-05$ &$-3.101\mathrm{E}-06$ &$2.507\mathrm{E}-05$ &$-1.578\mathrm{E}-05$ &$-1.595\mathrm{E}-05$ &$-8.730\mathrm{E}-09$  \\ 
$C-1$ &$2.577\mathrm{E}-02$ &$-7.013\mathrm{E}-02$ &$-2.858\mathrm{E}-02$ &$1.163\mathrm{E}-01$ &$-1.243\mathrm{E}-01$ &$5.599\mathrm{E}-04$  \\ 
\\
$|\modeu(0)/\modea|$ &$4.160\mathrm{E}-05$ &$7.020\mathrm{E}-05$ &$3.802\mathrm{E}-05$ &$1.173\mathrm{E}-04$ &$1.254\mathrm{E}-04$ &$5.599\mathrm{E}-07$  \\ 
$\angle(\modeu(0)/\modea)$ &$-7.873\mathrm{E}-01$ &$5.141\mathrm{E}-01$ &$2.708\mathrm{E}-01$ &$-5.429\mathrm{E}-01$ &$5.406\mathrm{E}-01$ &$-5.050\mathrm{E}-01$  \\ 
$|\modeb(0)/\modea|$ &$2.168\mathrm{E}-04$ &$3.059\mathrm{E}-04$ &$2.037\mathrm{E}-04$ &$2.864\mathrm{E}-03$ &$2.865\mathrm{E}-03$ &$2.544\mathrm{E}-03$  \\ 
$\angle(\modeb(0)/\modea)$ &$-2.053\mathrm{E}-02$ &$3.179\mathrm{E}-01$ &$-3.089\mathrm{E}-02$ &$-7.608\mathrm{E}-03$ &$7.348\mathrm{E}-03$ &$-1.515\mathrm{E}-01$  \\ 
\\
$\mathrm{E}_u/\mathrm{E}$ &$9.315\mathrm{E}-01$ &$9.569\mathrm{E}-01$ &$9.159\mathrm{E}-01$ &$4.728\mathrm{E}-01$ &$5.128\mathrm{E}-01$ &$6.999\mathrm{E}-02$  \\ 
$\mathrm{E}_b/\mathrm{E}$ &$3.575\mathrm{E}-02$ &$1.527\mathrm{E}-02$ &$4.092\mathrm{E}-02$ &$1.532\mathrm{E}-03$ &$1.196\mathrm{E}-03$ &$2.653\mathrm{E}-03$  \\ 
$\mathrm{E}_{b+}$ &$1.603\mathrm{E}-02$ &$1.364\mathrm{E}-02$ &$2.122\mathrm{E}-02$ &$2.628\mathrm{E}-01$ &$2.430\mathrm{E}-01$ &$4.640\mathrm{E}-01$  \\ 
$\mathrm{E}_{b-}$ &$1.614\mathrm{E}-02$ &$1.394\mathrm{E}-02$ &$2.106\mathrm{E}-02$ &$2.628\mathrm{E}-01$ &$2.430\mathrm{E}-01$ &$4.632\mathrm{E}-01$  \\ 
$\mathrm{E}_a/\mathrm{E}$ &$5.779\mathrm{E}-04$ &$2.470\mathrm{E}-04$ &$8.664\mathrm{E}-04$ &$5.425\mathrm{E}-05$ &$5.015\mathrm{E}-05$ &$1.214\mathrm{E}-04$  \\ 
\\
$\Gamma_R$ &$2.294\mathrm{E}-06$ &$6.140\mathrm{E}-06$ &$1.046\mathrm{E}-05$ &$-8.344\mathrm{E}-08$ &$-1.100\mathrm{E}-07$ &$1.733\mathrm{E}-07$  \\ 
$\Gamma_M$ &$3.917\mathrm{E}-04$ &$2.810\mathrm{E}-04$ &$4.631\mathrm{E}-04$ &$2.859\mathrm{E}-04$ &$1.618\mathrm{E}-04$ &$4.108\mathrm{E}-04$  \\ 
$\Gamma_J$ &$-1.634\mathrm{E}-05$ &$1.662\mathrm{E}-05$ &$3.234\mathrm{E}-05$ &$1.741\mathrm{E}-04$ &$2.635\mathrm{E}-04$ &$3.999\mathrm{E}-04$  \\ 
$\Gamma_\nu$ &$-9.919\mathrm{E}-06$ &$-1.145\mathrm{E}-05$ &$-1.306\mathrm{E}-05$ &$-2.869\mathrm{E}-04$ &$-1.651\mathrm{E}-04$ &$-4.010\mathrm{E}-04$  \\ 
$\Gamma_\eta$ &$-4.178\mathrm{E}-04$ &$-2.797\mathrm{E}-04$ &$-4.406\mathrm{E}-04$ &$-2.950\mathrm{E}-04$ &$-1.709\mathrm{E}-04$ &$-4.108\mathrm{E}-04$  \\ 
$\Gamma_{aU}$ &$2.609\mathrm{E}-10$ &$-2.344\mathrm{E}-10$ &$-3.239\mathrm{E}-10$ &$5.662\mathrm{E}-89$ &$-6.087\mathrm{E}-89$ &$-3.988\mathrm{E}-90$  \\ 
$\Gamma_{aJ}$ &$1.746\mathrm{E}-05$ &$-1.577\mathrm{E}-05$ &$-2.710\mathrm{E}-05$ &$1.063\mathrm{E}-04$ &$-1.051\mathrm{E}-04$ &$9.002\mathrm{E}-07$  \\ 
$\Gamma_{mech}$ &$-2.396\mathrm{E}-05$ &$1.131\mathrm{E}-05$ &$2.975\mathrm{E}-05$ &$-1.129\mathrm{E}-04$ &$9.830\mathrm{E}-05$ &$-9.430\mathrm{E}-07$  \\ 
$\Gamma_{em}$ &$-8.692\mathrm{E}-06$ &$-1.442\mathrm{E}-05$ &$-4.675\mathrm{E}-06$ &$9.715\mathrm{E}-05$ &$-1.143\mathrm{E}-04$ &$9.327\mathrm{E}-07$  \\ 

\end{tabular*}
\end{scriptsize}
\caption{\label{table:modeConversion}Growth rate and phase velocity, velocity and magnetic field eigenfunctions, energy components, and energy transfer rates for modes F, L$_-$, L$_+$, M, and P$_1$ in figure~\ref{fig:harGammaAMhd} for representative values of the Hartmann number in the interval $ [ 0.02, 200 ] $. For the reasons stated in the caption to figure~\ref{fig:harGammaCompareF}, the relative numerical error in the energy-transfer-rate calculations for mode~F at $ \Har = 200 $ is of order 0.3.}
\end{table}

When the Hartmann number is large ($ \Har \gtrsim 200$), the travelling Alfv\'en modes are clearly distinguishable by the linear $\Har $-dependence of their decay rate $ | \Imag( c ) \alpha |$ and their phase velocity relative to the free-surface steady-state velocity, $ \Real( c ) -1 $. The nearly equal split between kinetic and magnetic energies remarked upon above is also evident, both in figure~\ref{fig:harGammaAMhd} and in the $ \Har = 200 $ results in table~\ref{table:modeConversion}. The latter also indicate that, in light of the smallness of the wavenumber employed in the calculation, the magnetic energy contained in the exterior region exceeds the internal one by more than two orders of magnitude. Interestingly, a duality between the mechanical and electromagnetic energy transfer rates appears to apply between the downstream and upstream waves, in the sense that the mechanical energy transfer rate $ \Gamma_{ mech} = -1.129 \times 10^{-4} $ for mode~L$_+$ (the downstream-propagating mode) is nearly equal to the $ \Gamma_{ em } $ term for mode~L$_-$ and likewise, the $ \Gamma_{mech} = 9.830 \times 10^{-5 } $ term for mode~L$_-$ is close to the electromagnetic energy transfer rate $ \Gamma_{em} $ for mode~F. In both cases, however, the energy transfer rate due to Maxwell stress and the current interaction, respectively $ \Gamma_M $ and $ \Gamma_J $, as well as the surface term $ \Gamma_{a J }, $ are positive, whereas the energy transfer rate due to Reynolds stress, $ \Gamma_R $, is negative. 

At smaller Hartmann numbers (equivalently, larger Alfv\'en numbers), the upstream-propagating mode tends to be advected along the direction of the basic flow, and for $ 10 \lesssim \Har \lesssim 20 $, the positive mechanical energy transfer rate, driven in part by positive Reynolds stress (see the $ \Har = 15 $ calculation in table~\ref{table:modeConversion}), exceeds the rate of energy dissipated electromagnetically, resulting in an instability. As $ \Har $ is further decreased, mode~L$_-$ is stabilised and becomes converted to mode~M, \ie its energy becomes predominantly magnetic, and its phase velocity $ C $ approaches the mean value of the steady-state flow. This mode conversion is different from the one observed in figure~\ref{fig:spectrumAMhd} and movie~3, where mode L$_-$ develops from weak-field mode A$_2$, indicating that the precursors of the Alfv\'en modes in the weak-field spectrum are not universal.

As for the downstream-propagating wave L$_+$, instead of joining the P eigenvalue branch when $ \Har $ is decreased (which would be the case for the flow parameters in figure~\ref{fig:spectrumAMhd}), becomes converted, around $ \Har = 3 $, to mode~F, \ie it has greater than unity phase velocity, and for the chosen values of the Reynolds and Galilei number, is unstable for $ \Har = 0 $. At the same time, the strong-field ($ \Har \gtrsim 80$) mode in figure~\ref{fig:harGammaAMhd} labelled F, which exhibits the asymptotic neutrality that we ascribed to mode~F in figure~\ref{fig:harGammaAZeroPm}, turns into mode P$_1$. This exchange of identity between modes~F and P$_1$ takes place at Reynolds numbers greater than the one employed in our preceding examples at $(\Rey, \alpha) = (7 \times 10^5, 2 \times 10^{-3} )$, and as can be seen in figure~\ref{fig:harGammaAMhd} and movie~4, is accompanied by a small Hartmann-number interval where both modes are stable. The latter gives rise to regions of stability in the $(\Rey, \alpha ) $ plane which would contain unstable modes in the inductionless limit (see figure~\ref{fig:reAlphaMhd}). 

The results for mode~F in figure~\ref{fig:harGammaAMhd} also show that the $ | \Gamma | \propto \Har^{-2} $ strong-field scaling observed in inductionless problems (\eg figure~\ref{fig:harGammaAZeroPm}) does not necessarily apply in nonzero-$ \Prm $ flows. In particular, while no such evidence was observed in the calculations in figure~\ref{fig:modeFMhd} for $ ( \Rey, \alpha, \Prm ) = ( 7 \times 10^5, 2 \times 10^{-3}, 10^{-5} ) $, the growth rate of mode~F for $ ( \Rey, \alpha, \Prm ) = ( 6.3 \times 10^6, 10^{-4}, 10^{-4} ) $ follows an inverse-cubic decrease. This is most likely caused by the steady-state current, since, as we have numerically confirmed, setting $ B $ to zero restores the $ | \Gamma | \propto \Har^{-2} $ scaling.   

Before closing, we note that the A, P, and S modes that do not participate in the modal interactions described above also exhibit new aspects of behaviour compared to inductionless problems. In particular, as the magnetic field strength grows their phase velocity experiences a series of oscillations about $ C = 1 $, of diminishing amplitude (\cf the inductionless calculations in figure~\ref{fig:harGammaAZeroPm}, where $ C $ monotonically approaches unity from below), and so does their magnetic energy as it settles towards a $ \Har $-independent equilibrium value. An intricate pattern of oscillation is also observed for the mechanical energy dissipation rate $ -\Gamma_{mech} $, which now exhibits an increasing trend with $ \Har \gg 1 $ instead of asymptoting towards $ \Har $-independent values. However, at least for the examined Hartmann-number interval, $ | \Gamma_{mech} | $ remains small compared to the rate of electromagnetic energy dissipation $ - \Gamma_{em} $, which, dominated by the resistive term $ \Gamma_\eta $, grows quadratically with $ \Har $. As a result, the decay rate $ - \Gamma  = -(\Gamma_{mech} + \Gamma_{em} ) $ of the A, P, and S modes exhibits to a good approximation a $ \Gamma \propto \Har^2 $ scaling for strong applied fields, as it does in the inductionless limit.       

\section{\label{sec:conclusions}Conclusions}

A numerical investigation of the stability of temporal normal modes of free-surface Hartmann flow at low magnetic Prandtl numbers ($\Prm \leq 10^{-4}$, including the inductionless limit $ \Prm \searrow 0$) has been presented. Our main objective has been to study the influence of a flow-normal magnetic field (of associated Hartmann number $ \Har \leq 1000$) on the soft and hard instability modes present in free-surface Poiseuille flow \cite[][]{DeBruin74,FloryanDavisKelly87,Yih63,Yih69}, imposing either insulating or conducting boundary conditions at the wall. 

We have confirmed that the Squire transformation for MHD \cite[][]{BetchovCriminale67} is compatible with the kinematic, stress, and magnetic field-continuity boundary conditions for free-surface problems, but found that unless the flow is driven at constant capillary and Galilei numbers, respectively parameterising the surface tension and the flow-normal gravitational force, the onset of instability as the Reynolds number $ \Rey $ grows is not necessarily governed by a two-dimensional mode.

In inductionless flows, we have observed that the critical Reynolds number $ \Reyc $ of both of the hard and soft instability modes increases monotonically with $ \Har $. In particular, except for applied fields sufficiently weak for gravity to dominate over Lorentz forces, the hard mode's critical Reynolds number, as well as its critical wavenumber $ \alphac $ and phase velocity $ C_c $, were found to be very close to the corresponding parameters of the even unstable mode in channel Hartmann flow \cite[][]{Takashima96}, reflecting the common, critical layer, nature of these two instabilities. In fact, for sufficiently large Hartmann numbers, the critical Reynolds number of both is well approximated by the linear power law $ \Reyc( \Har ) = \text{48,250}\,\Har $ computed for the critical Reynolds number of the unbounded Hartmann layer \cite[][]{LingwoodAlboussiere99}. 

As for the soft mode, we derived, using large-wavelength perturbation theory, closed-form expressions for the Reynolds number $ \Reyb $ and phase velocity $ C_b $ at the bifurcation point $ ( \Reyb, 0 ) $ in the $ ( \Rey, \alpha ) $ plane, where $ \alpha $ is the wavenumber, from which the upper and lower branches of its neutral-stability curve emanate. Even though in inductionless Hartmann flow the soft mode acquires a nonzero critical wavenumber, numerically computed values for $ \Reyc $ and $ C_c $ were found to be in close agreement with the analytical results for $ \Reyb $ and $ C_b $, respectively. From these it follows that $ \Reyc \sim ( \Gal / \Har )^{1/2} \exp( \Har ) $ increases exponentially with $ \Har \gg 1 $, where $ \Gal $ is the Galilei number, and $ C_c \sim 1 + \sech( \Har ) $ decreases from its non-MHD value of twice the free-surface steady-state velocity to unity.   

As is also the case in channel flow \cite[][]{Takashima96}, we recorded little variation in the hard mode's critical parameters between small-$ \Prm $ problems with insulating boundary conditions and the corresponding inductionless flows. On the other hand, we encountered considerable differences in the stability properties of the soft mode, manifested in the structure of eigenvalue contours in the $(\Rey, \alpha )$ plane, as well as the dependence of its critical parameters on $ \Har $ and $ \Prm $. Specifically, in problems with an insulating wall, our numerical results, supported by large-wavelength asymptotics, indicate that the $ \alphac = 0 $ axis ceases to be part of the soft mode's neutral-stability curve, and the exponential growth of the critical Reynolds number becomes, for sufficiently large $ \Har $, suppressed to a sublinearly increasing function. When conducting boundary conditions are imposed, $ \Reyc \sim ( \Gal / \Prm )^{1/2} \Har^{-1}$ becomes a decreasing function of the Hartmann number.

The observed $ \Prm $-sensitivity of the soft instability was attributed to the strong-field behaviour of the participating inductionless mode (here called mode~F), which, even though stabilised by the magnetic field, approaches neutral stability as $ \Har $ grows, and its energy content becomes almost exclusively gravitational. In particular, its decay rate and kinetic energy respectively decrease like $ \Har ^ {-2} $ and $ \Har^{-4} $, where the latter scaling is consistent with a work balance between gravitational and Lorentz forces. The resulting near equilibrium is distinct from the quadratically increasing Lorentz damping experienced by the shear modes in the A, P, and S families \cite[labelled according to the convention of][]{Mack76}. In particular, it renders mode~F susceptible to effects associated with magnetic field perturbations, which are neglected in the inductionless limit, even when the magnetic diffusivity of the working fluid is large. 

Our analysis has identified two ways that nonzero magnetic field perturbations influence the soft instability, both of which depend strongly on the wall boundary conditions. The first is through the component $ \Rey \Prm \crossp{ \basej}{ \pertb} $ of the Lorentz force associated with the steady-state current $ \basej $ and the perturbed magnetic field $ \pertb $. That force, which vanishes in the inductionless limit, results in a positive net energy transfer to mode~F, leading in turn to the observed deviation of the critical Reynolds number from its behaviour in the inductionless limit. The fact that $ \basej $ depends, through the boundary conditions, on the wall conductivity, accounts for the different $ \Reyc $ results between insulating and conducting-wall problems. In particular, when conducting boundary conditions are enforced, the magnitude of $ \basej $ increases without bound with $ \Har $, and the resulting energy transfer to mode~F causes $ \Reyc( \Har ) $ to become a decreasing function. On the other hand, in insulating-wall problems $ \basej $ becomes constant throughout the inner part of the fluid domain, which is consistent with the comparatively milder modification of the soft mode's critical parameters. In this case, however, the boundary conditions are compatible with a stable, large-wavelength mode, which, as we have confirmed by means of large-wavelength approximations, is singular in the inductionless limit $ \Prm \searrow 0 $. When $ \Prm $ is nonzero this magnetic mode couples with mode~F, causing the growth rate of the latter to become negative for sufficiently small $ \alpha $, irrespective of the value of the Reynolds number. The resulting large-wavelength instability suppression for all $ \Rey $ is the second major influence of nonzero magnetic field perturbations on the soft instability. As with the effects associated with $ \basej $, it too depends strongly on the boundary conditions. That is, the magnetic mode is absent from the spectrum when conducting boundary conditions are imposed, and in the same manner as inductionless flow, (for fixed $ \Rey > \Reyb $) the soft instability takes place for arbitrarily small wavenumbers.     
 
Besides the large-wavelength magnetic mode, the spectrum of free-surface Hartmann flow with an insulating wall was found to contain a pair of travelling Alfv\'en waves, characterised by a near-equipartition of the modal energy between the kinetic and magnetic degrees of freedom. At sufficiently high Alfv\'en numbers, the upstream-propagating wave undergoes an instability where both Reynolds and Maxwell stresses are positive. Frozen-in analogues of the travelling Alfv\'en modes  \cite[in the sense described by][]{BetchovCriminale67} were encountered in channel Hartmann flow with insulating walls, but are absent from conducting wall-problems in both free-surface and channel geometries.         

To conclude, the analysis presented in this paper highlights the important role played by the boundary conditions in free-surface Hartmann flow, and identifies potential shortcomings of the inductionless approximation. Future work will explore signatures of the differences between the linear-stability properties of inductionless and low-$ \Prm $ flows in fully nonlinear time-dependent simulations.

\section*{Acknowledgements}

We thank H.~Ji and M.~Nornberg for useful conversations. This work was supported by the Mathematical, Information, and Computational Science Division subprogram of the Office of Advanced Scientific Computing Research, and by the Office of Fusion Energy Sciences (Field Work Proposal No.~25145), Office of Science, U.S. Department of Energy, under Contract DE-AC02-0611357. D.~G.~acknowledges support from the Alexander S.~Onassis Public Benefit Foundation.

\appendix

\section{\label{app:pertTheory}Large-wavelength approximations}

\subsection{\label{app:pertZeroPm}Non-MHD and inductionless problems}

\subsubsection{\label{sec:pertTheoryFormulation}Formulation}

In non-MHD and inductionless free-surface problems, the Reynolds number $ \Reyb $ at the bifurcation point $ ( \Reyb, 0 ) $ of the soft mode's neutral-stability curve, as well as the corresponding phase velocity $ C_b $, can be evaluated in closed form using regular perturbation theory about $ \alpha = 0$. Following \cite{Yih63,Yih69} and \cite{SmithDavis82}, we start with the expansions $ \modeu(z) = u_0(z) + \alpha u_1(z) + \alpha^2 u_2(z) + \ord( \alpha^3 ) $, $ \modea = a_0 + \alpha a_1 + \alpha^2 a_2 + \ord( \alpha^3 ) $, and $ \gamma = \gamma_0 + \alpha \gamma_1 + \alpha^2 \gamma_2 + \ord( \alpha^3 ) $, which, when substituted into the modified OS equation~\eqref{eq:orrSommerfeldZeroPm}, lead to a series of ordinary differential equations of the form
\begin{equation}
\label{eq:ospertalpha}
\DD^4 u_n + \mu^2 \DD^2 u_n = s_n,\\
\end{equation}
where $ n=0,1,2,\ldots $ and $ \mu^2 := - ( \Harz^2 + \Rey \gamma_0 ) $. Here the source terms $ s_n$  vanish for $n=0$, and depend on the solutions up to order $n-1$ for $n \geq 1$. In particular, for the first two orders we have
\begin{gather}
\begin{aligned}
s_1 & := \Rey( \gamma_1 + \ii U ) \DD^2 u_0 - \ii \Rey ( \DD^2 U ) u_0 + 2 \Harx \Harz \DD u_0, \\
s_2 & := \Rey ( \gamma_1 + \ii U ) \DD^2 u_1  - \ii \Rey ( \DD^2 U ) u_1 + 4 \Harx \Harz \DD u_1 + ( \Rey \gamma_2 + 2 ) \DD^2u_0 \\
& \quad - ( \Harx^2 + \Rey \gamma_0 ) u_1.
\end{aligned}
\end{gather}
Similarly, the $ 5 =: Q $ boundary conditions \eqrefac{eq:2dNormalModesBC1} and~\eqref{eq:normalStressZeroPm} lead to
\begin{subequations}
\label{eq:pertBCZeroPm}
\begin{gather}
\tagab
u_n( -1 ) = \DD u_n( -1 ) = 0, \quad \DD^2 u_n( 0 ) = S_n^{(3)}, \\
\tagcd
\DD^3 u_n( 0 ) - \mu^2 \DD u_n( 0 ) = S_n^{(4)}, \quad u_n ( 0 ) - \gamma_0 a_n = S_n^{(5)},
\end{gather}
\end{subequations}
where $ S_0^{(3)} ,\ldots, S_0^{(5)} $ vanish, and   
\begin{equation}
\label{eq:pertSZeroPm}
\begin{gathered}
S_1^{(3)} := \ii \DD^2U( 0 ) a_0, \quad S_2^{(3)} := \ii \DD^2U( 0 ) a_1 - u_0( 0 ), \\
\begin{aligned}
S_1^{(4)} & := \Rey ( \gamma_1 + \ii U( 0 ) ) \DD u_0( 0 ) + \ii ( \Harx \Harz - \Rey \DD U( 0 ) ) u_0( 0 ), \\
S_2^{(4)} & := \Rey( \gamma_1 + \ii U( 0 ) ) \DD u_1( 0 ) + \ii ( \Harx \Harz - \Rey \DD U( 0 ) ) u_1( 0 ) \\
& \quad + ( 3 + \Rey \gamma_2 ) \DD u_0 ( 0 ) + \Gal \Rey^{-1} a_0,
\end{aligned}
\\
S_1^{(5)} := ( \gamma_1 + \ii U( 0 ) ) a_0, \quad S_2^{(5)} := ( \gamma_1 + \ii U( 0 ) ) a_1 + a_0 \gamma_2.
\end{gathered}
\end{equation}

We express the general solution to~\eqref{eq:ospertalpha} as 
\begin{equation}
\label{eq:generalpertu}
u_n = \sum_{i=0}^3 K_{ni} \homogu_i + \partintu_n,\\
\end{equation}
where $ \homogu_i $ are four linearly independent functions satisfying $ \DD^4 \homogu_i + \mu^2 \DD^2 \homogu_i = 0 $, $ K_{ni} $ are constants, and $ \partintu_n ( z ) $ are particular solutions associated with the source terms $ s_n$. The constants $ K_{ni} $, as well as the expansion coefficients $ a_n $ for the free-surface oscillation amplitude, are to be determined by systems of algebraic equations of the form 
\begin{equation}
\label{eq:pertsyst}
\mat{ A }_0 \colvec{ w }_n = \colvec{ t }_n,
\end{equation}
that follow by substituting for $ u_n $ in the boundary conditions~\eqref{eq:pertBCZeroPm} using~\eqref{eq:generalpertu}. Here
\begin{equation}
\label{eq:matA0General}
\mat{ A }_0 :=
\left(
\begin{array}{llll}
\homogu_0( -1 )  & \ldots & \homogu_3( -1 ) & 0 \\
\DD\homogu_0( -1)  & \ldots & \DD\homogu_3(- 1) & 0 \\
\DD^2\homogu_0( 0 ) & \ldots & D^2\homogu_3( 0 ) & 0 \\
( \DD^3 - \mu^2 ) \homogu_0( 0 ) &  \ldots & ( \DD^3 - \mu^2 ) \homogu_3( 0 ) & 0 \\
\homogu_0( 0 ) & \ldots & \homogu_3(0) &
\gamma_0 \\ 
\end{array}
\right)
\end{equation}
is a $5 \times 5$ matrix acting on five-element column vectors $ \colvec{ w }_n := ( K_{n0},\ldots, K_{n3} , a_n )^\mathrm{T} $. Also, the column vectors $\colvec{ t }_n$, which vanish for $n=0$, are given by
\begin{equation}
\label{eq:bn}
\colvec{ t }_n := 
\left(
\begin{array}{l}
-\partintu_n( -1 )\\
-\DD\partintu_n( -1 )\\
S^{(3)}_n-\DD^2\partintu_n( 0 )\\
S^{(4)}_n-( \DD^3 - \mu^2 \DD ) \partintu_n(0)\\
S^{(5)}_n -\partintu_n( 0 )
\end{array}
\right)
\end{equation}
for $n \geq 1$. Note that $\mat{ A }_0$ is generally a nonlinear function of $\gamma_0$, while the source vectors $\colvec{ t }_n $ are linear functions of $\gamma_n$. 

Assuming that $ \mat{ A }_0$ has a $q_0$-dimensional right nullspace, denoted by $\ker( \mat{ A }_0)$ (as discussed below, $ \gamma_0 $ will be chosen such that $ \ker( \mat{ A }_0 ) $ is non-trivial), the solution to~\eqref{eq:pertsyst} can be expressed as
\begin{equation}
\label{eq:pertgen}
\colvec{ w }_n = \mat{ R }_{A_0} \colvec{ \Pi }_n + \colvec{ w }^{(p)}_n,
\end{equation}
where $\mat{R}_{A_0}$ is a $Q \times q_0$ matrix whose columns form a basis for $\ker(\mat{A}_0)$, $\colvec{\Pi}_n$ is a $q_0$-dimensional column vector of free parameters, and $\colvec{ w }^{(p)}_n$ is a particular solution associated with the source term $ \colvec{ t }_n $. Therefore, introducing the notation $ \colvec{ v }_n := ( u_n, a_n )^\mathrm{T} $ and $ \colvec{ v }^{(p)}_{ n} := ( \partintu_n, 0 )^\mathrm{T} $, as well as the matrix
\begin{equation}
\mat{ M } := \left(
\begin{array}{llll}
\homogu_0 & \cdots & \homogu_3  & 0  \\
0 & \cdots & 0 & 1
\end{array}
\right),
\end{equation}
the solution for the velocity field and the free-surface oscillation amplitude at $n$-th order in perturbation theory becomes
\begin{equation}
\label{eq:fn}
\colvec{ v }_n = \mat{ M } \mat{ R }_{A_0} \colvec{ \Pi }_n + \mat{ M } \colvec{ w}_{n}^{(p)} + \colvec{ v }^{(p)}_{n}.
\end{equation}
In what follows, we choose the homogeneous solutions
\begin{gather}
\label{eq:homogu}
\homogu_0 := 1,\quad
\homogu_1 := z,\quad
\homogu_2 := ( 1-\cos(\mu z) ) / \mu^2 , \quad
\homogu_3 := ( \mu z - \sin(\mu z) ) / \mu^3,
\end{gather}
all of which are well-behaved in the limit $ \mu \to 0 $, giving, upon substitution into~\eqref{eq:matA0General},
\begin{equation}
\label{eq:matA0}
\mat{ A }_0 =
\left(
\begin{array}{lllll}
1 & -1 & (1-\cos\mu)/\mu^2 & (\sin\mu-\mu)/\mu^3
&  0\\
0 & 1 & -\sin\mu/\mu & (1-\cos\mu)/\mu^2 & 0 \\
0 & 0 & 1 & 0 & 0  \\
0 & \mu^2 & 0 & 1 & 0 \\
1 & 0 & 0 & 0 & - \gamma_0 \\
\end{array}
\right).
\end{equation}

At zeroth-order, the homogeneous problem $ \mat{ A}_0 \colvec{ w }_0 = 0$ has a non-trivial solution only if $ \mat{ A }_0 $ has a non-trivial nullspace, or, equivalently, 
\begin{equation}
\label{eq:zerothdisp}
\det( \mat{ A }_0 ) = \cos(\mu) \gamma_0 = 0. 
\end{equation}
The equation above has two distinct classes of roots, given by $ \gamma_0 = 0 $ and $ \gamma_0 = - ( \Harz^2 + ( 2 n + 1 )^2 \pi^2 / 4 ) $, where $ n = 0, 1, 2, \ldots $. Among these, only the zero solution can potentially be connected to a large-wavelength unstable mode, since the eigenvalues associated with the first class of roots approach zero from below as $ \alpha \searrow 0 $. Setting therefore $ \gamma_0 = 0 $ equips $ \mat{ A }_0 $ with a one-dimensional nullspace spanned by the column vector $ \colvec{ \xi } := ( 0, 0, 0, 0, 1 )^\mathrm{ T } $. Thus, the parameter vectors $ \colvec{ \Pi }_n $ become scalars, playing the role of normalisation constants and, through~\eqref{eq:fn}, we obtain $ \colvec{ v }_0 = ( u_0, a_0 ) = \Pi_0 ( 0, 1 ) $. We remark that channel problems do not admit asymptotically neutral solutions as $ \alpha \searrow 0 $; in this case $ \det( \mat{ A }_0 ) = 4 ( \sin( \mu ) - \mu \cos( \mu ) ) / \mu^4 $ tends to $ 4 / 3 $ in the limit $ \gamma_0 \ttz $. 

At higher orders in $\alpha$ one has to find solutions to inhomogeneous systems of equations of the form~\eqref{eq:pertsyst}. Here we will outline an inductive procedure which, given a zeroth-order solution, can be applied to obtain solutions at successively higher orders in $\alpha$, and can also be generalised to treat the coupled differential equations~\eqref{eq:orrSommerfeldInd} governing nonzero-$ \Prm $ problems (see appendix~\ref{app:pertMhd}).

First, assume that the eigenvalue coefficients $\gamma_0,\ldots,\gamma_{n-1}$ and the corresponding eigenvectors  $ \colvec{ v }_0,\ldots, \colvec{ v }_{n-1} $ have been evaluated to some order $n-1$, where $n \geq 1$. Moreover, assume that $ \{ \colvec{v}_i \}_{i=0}^{n-1} $ are linear and homogeneous functions of $ q_0 $ free parameters $ \Pi_{n-1,1},\ldots,\Pi_{n-1,q_0}$, \ie 
\begin{equation}
\label{eq:pertfi}
\colvec{ v }_i = \mat{ D }_{i,n-1} \colvec{ \Pi }_{n-1}
\end{equation} 
for $ 2 \times q_0 $ matrices $ \mat{ D }_{0,n-1},\ldots, \mat{ D }_{n-1,n-1}$ and a $q_0$-dimensional column vector $ \colvec{ \Pi }_{n-1} := ( \Pi_{n-1,1}, \ldots, \Pi_{n-1,q_0} )^\mathrm{ T } $. Under these conditions, the particular solution $ \partintu_{n}$, the boundary-condition source terms $S^{(1)}_{n},\ldots,S^{(5)}_{n}$ and, by construction, the elements of $\colvec{ t }_n$ at order $ n $ are also homogeneous linear functions of $ \{ \Pi_{n-1,i} \}_{i=1}^{q_0}$. That is, we can write
\begin{equation}
\label{eq:pertbn}
\colvec{ t }_{n} = \mat{ T }_{n} \colvec{ \Pi }_{n-1},
\end{equation}
where $ \mat{ T }_n $ is a $ Q \times q_0 $ matrix. 

In general, a solution $ \colvec{ w }_n$ to~\eqref{eq:pertsyst} will only exist if $ \colvec{ t }_n$ lies in the range of $ \mat{ A }_0 $, denoted by $\ran( \mat{ A }_0 ) $. According to the fundamental theorem of linear algebra \cite[e.g.][]{Strang05}, $\ran( \mat{ A }_0)$ is the orthogonal complement, in the sense of the Euclidean inner product, of the left nullspace of $ \mat{ A }_0$  (i.e.~the nullspace of $ \mat{ A }_0^\mathrm{T}$, denoted by $\ker( \mat{ A }_0^\mathrm{T})$). Thus, a solution to~\eqref{eq:pertsyst} exists if and only if $ \mat{ L }_{A_0}^\mathrm{T} \colvec{ t }_{n} = 0 $, where $ \mat{ L }_{A_0}$ is a $Q \times q_0$ matrix whose columns form a basis of $\ker( \mat{ A }_0^\mathrm{T})$.\footnote{According to the fundamental theorem, the row rank and the column rank of any matrix are equal. Thus, since $ \mat{ A }_0$ is square, the dimensions of its left and right nullspaces are equal, which implies in turn that $ \mat{ L }_{A_0}$ has size $Q \times q_0$.} It therefore follows from~\eqref{eq:pertbn} that a solvability condition for~\eqref{eq:pertsyst} is that the $q_0 \times q_0$ matrix $ \mat{ A }_n := \mat{ L}_{A_0}^\mathrm{T} \mat{ T }_n$ has a non-trivial nullspace,  \ie $\det( \mat{ A }_n)=0$. Since $ \det( \mat{ A }_n ) $ is a polynomial in $\gamma_n$ of degree no greater than $q_0$ (recall that the elements of $\colvec{ t }_n$ depend linearly on $\gamma_n$), the latter equation yields up to $ q_0 $ distinct solutions for the $n$-th order expansion coefficient $ \gamma_n $.   

Denoting the dimension of $ \ker( \mat{ A }_n ) $ corresponding to a given solution for $ \gamma_n $ by $ q_n \leq q_0 $ (the procedure can be repeated for each of the roots of $ \det( \mat{ A }_n ) = 0 $), we now express the parameter vector $ \colvec{ \Pi }_{n-1} $ as 
\begin{equation}
\label{eq:parvan}
\colvec{ \Pi }_{n-1} = \mat{ R }_{A_n} \tilde{ \colvec{ \Pi} }_n,
\end{equation}
where $ \mat{ R }_{A_n}$ is a $q_{0} \times q_n$ matrix  whose columns are basis vectors for $\ker( \mat{ A }_n)$, and $\tilde{ \colvec{ \Pi } }_n := (\tilde\Pi_{n,1}, \ldots, \tilde \Pi_{n,q_n} )^\mathrm{T} $ is an updated vector of free parameters in the solution. Upon substitution into~\eqref{eq:pertfi}, \eqref{eq:parvan} leads potentially to a decrease in the number of degrees of freedom in the eigenfunctions of order up to $n-1$, as well as in $ \colvec{ w }^{(p)}_n $ (through its dependence on $ \colvec{ t }_n $), from $q_0$ to $q_n$. Moreover, since the particular solution $\partintu_{n}$ to~\eqref{eq:ospertalpha} also depends linearly and homogeneously on $\tilde{ \colvec{ \Pi } }_n$, it is possible to recast~\eqref{eq:fn} as
\begin{equation}
\label{eq:pertgenalt}
\colvec{v}_{n} = \mat{ M } \mat{ R }_{A_0} \colvec{ \Pi }'_{n} + \tilde{ \mat{ D } }_{n} \tilde{ \colvec{ \Pi } }_{n},
\end{equation}
where $ \colvec{ \Pi }'_{n}$ is a (provisional) $q_0$-dimensional column vector of free parameters and $\tilde{ \mat{ D } }_{n}$ is a $2 \times q_n$ matrix such that $ \tilde{ \mat{  D } }_n \tilde{ \colvec{ \Pi } }_n = \mat{ M } \colvec{ w }^{(p)}_n +  \colvec{ v }^{(p)}_{n} $. Although~\eqref{eq:pertgenalt} may contain up to $q_0 + q_n$ arbitrary constants, the part of $\colvec{ \Pi }'_{n}$ that is parallel to $ \colvec{ \Pi }_{n-1}$ can be set to zero, since its only effect would be a renormalisation of the lower-order solutions. We therefore require that $ \mat{ R }_{A_n}^\mathrm{ T } \colvec{ \Pi }'_{n} $ vanishes, or, equivalently,
\begin{equation}
\label{eq:pertpiprime}
\colvec{ \Pi }'_{n} = \mat{ E }_n \hat{ \colvec{ \Pi } }_n,
\end{equation}
where $ \mat{ E }_n$ is a $q_0 \times (q_0 - q_n)$ matrix whose columns form a basis of the left nullspace of $ \mat{ R }_{A_n}$,\footnote{The column rank of the $ q_0 \times q_n $ matrix $\mat{ R }_{A_n}$ is $q_n$ by construction (its columns are linearly independent vectors). Since the row rank of a matrix is equal to its column rank, it follows that its row space (i.e.~$\ran(\mat{ R }_{A_n}^\mathrm{T})$) is $q_n$-dimensional, and its left nullspace, $ \ker( \mat{ R }_{A_n}^\mathrm{T} ), $ is $(q_0-q_n)$-dimensional.} and $\hat{\colvec{ \Pi} }_n$ is a $(q_0-q_n)$-dimensional column vector. If $q_0$ happens to equal unity, one can set $ \colvec{ \Pi }'_n $ equal to zero. Inserting~\eqref{eq:pertpiprime} into~\eqref{eq:pertgenalt}, we then obtain $ \colvec{ v }_n = \hat{ \mat{ D } }_n \hat{ \colvec{ \Pi } }_n + \tilde{ \mat{ D } }_n \tilde{ \colvec{ \Pi } }_n, $ where $\hat{ \mat{ D } }_n := \mat{ M } \mat{ R }_{A_0} \mat{ E }_n$, or 
\begin{equation}
\colvec{ v }_n = \mat{ D }_{n,n} \colvec{ \Pi }_n,
\end{equation}
where $ \mat{ D }_{n,n} := ( \hat{ \mat{  D } }_n, \tilde{ \mat{ D } }_n ) $ and $ \colvec{ \Pi }_n := ( \hat{ \colvec{ \Pi } }_n^\mathrm{ T }, \tilde{ \colvec{ \Pi} }_n^\mathrm{ T } )^\mathrm{T} $ are a $2 \times q_0$ matrix and a $q_0$-dimensional column vector, respectively. The lower-order solutions can also be written in terms of $ \colvec{ \Pi }_n$ using
\begin{equation}
\colvec{ v }_i = \mat{ D }_{i,n} \colvec{ \Pi }_n, \quad  \mat{ D }_{i,n} := (  \mat{ 0 }_{2 \times (q_0-q_n)}, \mat{ D }_{i,n-1} \mat{ R }_{A_n} ), \quad i = 0,\ldots,n-1,
\end{equation}
where $ \mat{ 0 }_{2 \times (q_0-q_n)} $ denotes the zero matrix of size $ 2 \times ( q_0 - q_n ) $.

The matrices $\mat{ D }_{0,n},\ldots, \mat{ D }_{n,n}$ fully specify the perturbative expansion to $n$-th order, which, as assumed above, is a linear homogeneous function of $ q_0 $ parameters. The procedure can be repeated for successively higher orders in $\alpha$, but, since the solutions for the expansion coefficients rapidly become complicated beyond first order, it is convenient to implement it in a symbolic-mathematics programming language so as to minimise effort and error probability.

\subsubsection{\label{app:pertResultsZeroPm}Results at first and second order}

We now apply the procedure described in \S\ref{sec:pertTheoryFormulation} to evaluate the first and second-order corrections to the zeroth-order solution, $ ( \gamma_0, u_0, a_0 ) = ( 0, 0, \Pi_0 )$. Setting $ U( z )$ equal to the Hartmann velocity profile (equation~\eqrefa{eq:generalUB} with $ C_0 = C_1 = 0 $), and the streamwise component of the applied magnetic field equal to zero (\ie $ \Harx = 0 $ and $ \Harz = \Har $) the first-order expansion terms for the eigenvalue and the velocity eigenfunction are found to be   
\begin{equation}
\label{eq:gamma1ZeroPmHartmann}
i \gamma_1 = 1 + \sech( \Har ), \quad u_1( z ) = -\ii \Pi_0 \sech( \Har ) \sinh^2( \Har (1 + z) / 2 ) / \sinh^2(\Har/2),
\end{equation}
while the coefficient for the free-surface amplitude, $ a_1 $, vanishes. We remark that since $ \gamma_1 $ is purely imaginary, the $ \alpha = 0 $ axis is part of the neutral-stability curve $ 0 = \Imag( c ) = \Real( \gamma_1 ) + \alpha \Real( \gamma_2 ) + \ord( \alpha^2 ) $ in the $ ( \Rey, \alpha ) $ plane. Moreover, the leading-order phase velocity  $ C_0 := i \gamma_1 $ is a monotonically decreasing function of the Hartmann number, with $ C_0 \searrow 1 $ as $ \Har \tti $. As for the eigenfunction $ u_1( z ) $, it varies exponentially with the flow-normal coordinate, and like the Hartmann velocity profile, it possesses a boundary layer of thickness $ \ord( 1 / \Har ) $ near the no-slip wall. In the vanishing magnetic field limit ($ \Har \searrow 0 $), $ C_0 $ is twice the steady-state velocity at the free surface, and $ u_1 $ reduces to the quadratic function $ u_1( z ) = - i \Pi_0 ( 1 + z )^2 $, as computed by \cite{Yih63,Yih69} for free-surface Poiseuille flow.

Proceeding now with the second-order approximation, 
\begin{multline}
\label{eq:gamma2ZeroPmHartmann}
\gamma_2 = \frac{ \Rey \coth(\Har/2) \sech^3( \Har ) (2 \Har (2 + \cosh(2 \Har)) - 3 \sinh(2 \Har)) }{\Har^2  (\cosh( \Har ) -1 )}\\
- \frac{8 \Gal \sinh^2( \Har/ 2) ( \Har - \tanh( \Har ) )}{ \Har^3 \Rey (\cosh( \Har ) -1 )}
\end{multline}
is the leading-order coefficient for $ \gamma $ with nonzero real part, and governs the modal stability in the limit $ \alpha \searrow 0 $. In particular, setting $ \gamma_2 $ equal to zero and solving for $ \Rey $ leads to~\eqrefa{eq:criticalSoft}, quoted in the main text for the Reynolds number $ \Reyb $ at the bifurcation point $ ( \Reyb, 0 ) $. That is, $ \gamma_2 $ is negative for $ 0 < \Rey < \Reyb $ but becomes positive for all $ \Rey > \Reyb $.  For weak magnetic fields $ \Reyb / \Gal^{1/2} = ( 5 / 8 )^{1/2} + 191 / ( 168 \times 10^{1/2 } ) \Har^2 + \ord( \Har^4 ) $ grows quadratically with $ \Har $, but when the Hartmann number is large, $ \Reyb / \Gal^{1/2}  \sim  \exp( \Har ) / \Har^{1/2} $ increases exponentially. Moreover, since $ \gamma_1 $ is independent of $ \Rey $, the phase velocity $ C_b $ at the bifurcation point is equal to the zeroth-order phase velocity $ C_0 $, in accordance with~\eqrefb{eq:criticalSoft} in the main text. We remark that the result $ \Reyb = ( 5 \Gal / 8 )^{1/2} $ for zero magnetic field strength is consistent with equation~(38) in the paper by \cite{Yih63}, under the proviso that $ \Rey $ is replaced by $ \Rey' := 2 \Rey / 3 $ (Yih chooses the mean steady-state velocity as the characteristic velocity for reduction to non-dimensional form), and the inclination angle $ \theta $ is substituted for $ \Gal $ using~\eqref{eq:reTheta}. 

In order to assess the relative importance of the formation of the Hartmann velocity profile versus the Lorentz force in the behaviour of $ \Reyb $ and $ C_b $, we have carried out similar large-wavelength calculations for (physically unrealistic) problems with (i) the Hartmann velocity profile, but no Lorentz force (\ie $ \Har $ set to zero in \eqref{eq:ospertalpha}--\eqref{eq:pertSZeroPm}), and (ii) the Lorentz force included, but the velocity profile set to the $U(z)=1-z^2$ Poiseuille solution. The results for the perturbation-expansion coefficients for the complex growth rate $ \gamma $ are
\begin{subequations}
\label{eq:gammaHydroHartmann}
\begin{align}
i \gamma_1 & = 1 + (\Har / 2)^2 / \sinh^2( \Har / 2 ), \\
\gamma_2 & = - \frac{ \Gal }{ 3 \Rey } + \frac{ \Rey ( 432 - 24 \Har^2 - 22 \Har^4 + 5 \Har^6 - 432 \cosh( \Har ) + 240 \Har \sinh( \Har ) )}{ 192 \Har^2 \sinh^4( \Har / 2 ) },
\end{align}
\end{subequations}
and
\begin{subequations}
\label{eq:gammaZeroPmPoiseuille}
\begin{align}
i \gamma_1 & = 1 + 2 ( 1 - \sech( \Har ) ) / \Har^2, \\
\nonumber
\gamma_2 & = - \Gal( \Har - \tanh( \Har ) ) / ( \Rey \Har^3 )  + \Rey ( 72 + 12 \sech^2( \Har ) ( 4 + \Har \tanh( \Har ) ) \\
\label{eq:gamma2zeroPmPoiseuille}
& \quad - \sech( \Har ) ( 3 ( 40 + 9 \Har^2 ) + \Har( 2 \Har^2 - 3 ) \tanh( \Har ) ) ) / (  6 \Har^6 ),
\end{align}
\end{subequations}
respectively for cases~(i) and~(ii). As above, the phase velocity $ C_b = C_0 = - i \gamma_1 $ at the bifurcation point follows directly from the first-order coefficients, while setting~\eqrefb{eq:gammaHydroHartmann} and~\eqrefb{eq:gammaZeroPmPoiseuille} to zero and solving for $ \Rey $ leads to the expressions
\begin{align}
\label{eq:reycHydroHartmann}
\Reyb & = \frac{ 4 \Har \Gal^{1/2} ( \cosh( \Har ) -1 ) }{ ( ( 18 + 5  \Har ^2 ) ( 24 - 8 \Har^ 2 + \Har^ 4 ) - 432 \cosh( \Har ) + 240 \Har \sinh( \Har ) )^{1/2} }
\end{align}
and
\begin{align}
\nonumber
\Reyb & = 2 \Har ( 3 \Gal \cosh( \Har ) ( \Har \cosh( \Har ) - \sinh( \Har ) ) )^{1/2} \\
\nonumber
& \quad \times ( -3 ( 40 + 9 \Har^2 + ( 40 + 9 \Har^ 2) \cosh( 2 \Har ) - 12 \cosh( 3 \Har ) - 8 \Har \sinh( \Har ) ) \\
\label{eq:reycZeroPmPoiseuille}
& \quad + \cosh( \Har ) ( 204 + 2 \Har ( 3 - 2 \Har^2 ) \sinh( \Har ) ) )^{-1/2}, 
\end{align}
for the position of the bifurcation point on the $ \alpha = 0 $ axis. These two types of test problems have different strong-field behaviour, with 
\begin{subequations}
\label{eq:strongFieldHydroHartmann}
\begin{equation}
\tagac
\Reyb \sim \Gal^{1/2}( \Har / 15 )^{1/2} \exp( \Har / 2 ), \quad C_b - 1 \sim ( \Har / 2 )^2 \exp( - \Har ),  \quad \gamma_2 \sim - \Gal / 3 \Rey
\end{equation}
\end{subequations} 
and
\begin{subequations}
\label{eq:strongFieldZeroPmPoiseuille}
\begin{equation}
\tagac
\Reyb  \sim \Gal^{1/2} \Har^2 / 3, \quad C_b - 1 \sim 2 / \Har^2, \quad \gamma_2 \sim - \Gal / ( \Rey \Har^2 ),
\end{equation}  
\end{subequations}
respectively, for $ \Har \gg 1 $.

\subsection{\label{app:pertMhd}Nonzero-$ \Prm $ problems}

The method described in \S\ref{sec:pertTheoryFormulation} can be used to study the large-wavelength limit of the coupled OS and induction equations~\eqref{eq:orrSommerfeldInd}, with the addition that apart from the expansions for $ \gamma $, $ \modeu $, and $ \modea $, we write $ \modeb(z) = b_0(z) + \alpha b_1(z) + \alpha^2 b_2(z) + \ord( \alpha^3 ) $ for the magnetic field eigenfunction. Moreover, \eqref{eq:ospertalpha} are replaced by coupled differential equations, which, in the special case with flow-normal external magnetic field, have the form
\begin{equation}
\label{eq:pertMhd}
\begin{aligned}
\DD^4 u_n - \Rey \gamma_0 \DD^2 u_n + \Har \Prm^{-1/2} \DD^3 b_n &= s^{(u)}_n, \\
\quad \DD^2 b_n - \Reym \gamma_0 b_n + \Har \Prm^{1/2} \DD u_n &= s^{(b)}_n.
\end{aligned}
\end{equation}
As in \S\ref{app:pertZeroPm}, we are interested in perturbation order $ n \leq 2 $, where $ s_0^{(u)} = s_0^{(b)} = 0 $ and
\begin{equation}
\begin{aligned}
s_1^{(u)} / \Rey & := ( \gamma_1 + \ii U) \DD^2 u_0 - \ii ( \DD^2 U ) u_0 + \ii ( \DD^2 \basebx) b_0 - \ii \basebx \DD^2 b_0, \\
s_1^{(b)} / \Reym & := ( \gamma_1 + \ii U ) b_0 - \ii \basebx u_0, \\
s_2^{(u)} / \Rey & := ( \gamma_1 + \ii U ) \DD^2 u_1 - \ii ( \DD^2 U ) u_1 + \ii ( \DD^2 \basebx ) b_1 - \ii \basebx \DD^2 b_1 \\
& \quad ( \gamma_2 + 2 \Rey^{-1} ) \DD^2 u_0 - \gamma_0 u_0, \\ 
s_2^{(b)} / \Reym & := ( \gamma_1 + \ii U ) b_0 - \ii \basebx u_1 + ( \gamma_2 + \Reym^{-1} ) b_0.
\end{aligned}
\end{equation} 
Among the boundary conditions, which are now $ 7 =: Q $, the no-slip, shear-stress, and kinematic conditions, respectively~\eqrefac{eq:2dNormalModesBC1} have the same expansions as~(\ref{eq:pertBCZeroPm}\,{\it a,\,b,\,d}), but the boundary condition for the normal stress~\eqrefd{eq:2dNormalModesBC1} now yields
\begin{equation}
\label{eq:pertBCMhdNormalStress}
\DD^3 u_n( 0 ) - \Rey \gamma_0 \DD u_n( 0 ) + \Har \Prm^{-1/2} \DD b_n( 0 ) = S_n^{(4)},
\end{equation}    
where $ S_0^{(4)} = 0 $ and
\begin{equation}
\begin{aligned}
S_1^{(4)} / \Rey & := ( \gamma_1 - \ii U( 0 ) ) \DD u_0( 0 ) - \ii (\DD U ( 0 ) )u_0( 0 ) + \ii ( \DD \basebx ( 0 ) ) b_0( 0 ), \\
S_2^{(4)} / \Rey & := ( \gamma_1 + \ii U( 0 ) ) \DD u_1( 0 ) - \ii ( \DD U( 0 ) ) u_1( 0 ) + \ii ( \DD \basebx( 0 ) )( 0 ) b_1( 0 ) \\
& \quad + ( 3 \Rey^{-1} + \gamma_2 ) \DD u_0( 0 ) + \Gal \Rey^{-2} a_0 + \Har \Prm \Rey^{-1} b_0( 0 ). 
\end{aligned}
\end{equation}  
In problems with an insulating wall, the magnetic field boundary conditions~\eqref{eq:2dNormalModesInsulatingFreeSurf} lead in addition to
\begin{subequations}
\label{eq:pertBCMhdInsulating}
\begin{equation}
\tagab
\DD b_n( -1 ) = S_n^{(6)} \quad \mbox{and} \quad \DD b_n( 0 ) = S_n^{(7)},
\end{equation}
\end{subequations}
respectively, where $ S_0^{(6)} = S_0^{(7)} = 0 $ and, for $ n \in \{1, 2\} $, $ S_n^{ (6) } := b_{n-1}(-1) $ and $ S_n^{(7)} := \ii \DD \basebx( 0 ) a_{n-1} - b_{n-1}(0) $. When the wall is conducting, the boundary condition at $ z= -1 $ is replaced with $ b_n( -1 ) = 0 $, in accordance with~\eqref{eq:conductingWall2DNormalModes}. 

Following an analogous approach as in~\eqref{eq:generalpertu}, we express the general solution to~\eqref{eq:pertMhd} at order $ n $ as
\begin{equation}
\label{eq:generalpertub}
( u_n, b_n ) = \sum_{i=0}^5 K_{ni} ( \homogu_i, \homogb_i ) + ( \partintu_n, \partintb_n ),\\
\end{equation}
where $ \{ ( \homogu_i, \homogb_i ) \}_{i=0}^5 $ are six linearly independent solutions to the homogeneous parts of~\eqref{eq:pertMhd}, and $ ( \partintu_n , \partintb_n ) $ are particular solutions dependent on the source terms $ ( s^{(u)}_n, s^{(b)}_n ) $.  Due to the high order of the differential equations involved, instead of seeking expressions for $ (\homogu_i,\homogb_i) $ for arbitrary $ \gamma_0 $, we shall set $ \gamma_0 = 0 $ from the outset, which is the solution of interest for large-wavelength instabilities. We will verify subsequently that the resulting matrix $ \mat{ A }_0 $ (the $ \Prm > 0 $ analogue of~\eqref{eq:matA0}) possesses a non-trivial nullspace. Choosing then
\begin{equation}
\begin{gathered}
( \homogu_0( z ), \homogb_0( z ) ) := ( 1, 0 ), \quad
( \homogu_1( z ), \homogb_1( z ) ) := ( z, - \Har \Prm^{1/2} z^2/ 2), \\
( \homogu_2( z ), \homogb_2( z ) ) :=
\left( ( \cosh( \Har \, z ) - 1 ) / \Har^2, - \Prm^{1/2} ( \sinh( \Har \, z ) - \Har \, z )/ \Har^2   \right), \\ 
( \homogu_3( z ), \homogb_3( z ) ) :=
\left( \frac{ \sinh( \Har \, z ) - \Har \, z }{ \Har^3 }, - \Prm^{1/2} \frac{ \cosh( \Har \, z ) - 1 - ( \Har \, z )^2 / 2 }{ \Har^3 } \right), \\
( \homogu_4( z ), \homogb_4( z ) ) := ( 0, 1 ), \quad 
( \homogu_5( z ), \homogb_5( z ) ) := ( 0, z )
\end{gathered}
\end{equation}
as a set of linearly independent solutions to the homogeneous part of~\eqref{eq:pertMhd}, valid for $ \gamma_0 = 0 $, and substituting $ ( u_0, b_0 ) = \sum_{i=0}^5 K_{0i} ( \homogu_i, \homogb_i ) $ into the zeroth-order boundary conditions, leads to the homogeneous algebraic equations $ \mat{ A }_0 \colvec{ w }_0 = 0 $, where $ w_0 := ( K_{00}, \ldots, K_{05}, a_0 )^\mathrm{T} $ is a seven-element column vector, and $ \mat{ A }_0 $ is a $ 7 \times 7 $ matrix. In problems with an insulating wall, $ \mat{ A }_0 $ is given by
\begin{multline}
\label{eq:matA0insulating}
\mat{ A }_0 = \\
\left(
\begin{array}{lllllll}
1 & -1 & (\cosh( \Har ) - 1 ) / \Har^2 & ( \Har - \sinh( \Har ) ) / \Har^3 & 0 & 0 & 0 \\
0 & 1 & - \sinh( \Har ) / \Har & ( \cosh( \Har ) - 1 ) / \Har^2 & 0 & 0 & 0 \\
0 & 0 & 1 & 0 & 0 & 0 & 0 \\
0 & - \Har^2 & 0 & 1 & 0 & 0 & 0 \\
1 & 0 & 0 & 0 & 0 & 0 & 0 \\
0 & \Har \Prm^{1/2} & - \Prm^{1/2} ( \cosh( \Har ) - 1 ) / \Har & ( \sinh( \Har ) - \Har ) / \Har ^ 2 & 0 & 1 & 0 \\
0 & 0 & 0 & 0 & 0 & 1 & 0 
\end{array}
\right), 
\end{multline} 
where rows 1--7 respectively result from~\eqrefab{eq:pertBCZeroPm}, \eqref{eq:pertBCMhdNormalStress}, \eqrefab{eq:pertBCMhdInsulating}, and \eqrefd{eq:pertBCZeroPm}. In conducting-wall problems, $ \mat{ A }_0 $ has the same form as above but with the sixth row, originating from the wall boundary condition for $ \modeb $, replaced by
\begin{multline}
\label{eq:matA0conducting}
( \mat{ A } _{ 6, j} ) = \\ \left(  0, - \frac{ \Har \Prm^{1/2}}{2}, \Prm^{1/2} \frac{ \sinh( \Har ) - \Har } { \Har^2 }, \Prm^{1/2} \frac{ 2 + \Har^2 - 2 \cosh( \Har ) }{ 2 \Har^3 }, 1, -1, 0 \right).
\end{multline}
The presence of all-zero columns in~\eqref{eq:matA0insulating}, as well as in the matrix resulting from~\eqref{eq:matA0insulating} with the substitution~\eqref{eq:matA0conducting}, confirms our earlier assertion that setting $ \gamma_0 $ equal to zero endows $ \mat{ A }_0 $ with a non-trivial nullspace. In the insulating-wall case that nullspace has dimension $ q_0 := 2 $, and is spanned by the column vectors $ \colvec{ \xi }_1 := ( 0, 0, 0, 0, 0, 0, 1 )^\mathrm{ T } $ and $ \colvec{ \xi }_2 := ( 0, 0, 0, 0, 1, 0, 0 )^\mathrm{T} $, respectively corresponding to a free-surface displacement with zero velocity and magnetic field perturbations (as in inductionless problems), and a uniform flow-normal magnetic field perturbation with no change in either of $ \modeu $ or $ \modea $. In problems with a conducting wall the latter option is not available since the magnetic field must vanish at the wall; here $ \ker( \mat{ A }_0 ) $ is one-dimensional, and spanned by $ \colvec{ \xi }_1 $.

According to the discussion in \S\ref{sec:pertTheoryFormulation}, the two-dimensionality of $ \ker( \mat{ A }_0 ) $ implies that in insulating-wall problems there exist up to two solutions for the first-order expansion coefficient $ \gamma_1 $, and carrying out the procedure outlined in that section establishes that this is indeed the case. However, unlike inductionless problems, the resulting expressions for $ \gamma_1 $, which we denote by $ \gamma_1^{(F)} $ and $ \gamma_1^{(M)} $, both possess negative real parts for all $ \Har > 0 $. Therefore, when $ \Prm $ is nonzero all unstable inductionless modes acquire negative growth rate for sufficiently small $ \alpha > 0 $, and the $ \alpha = 0 $ axis is no longer part of the neutral-stability curve $ \Imag( c ) = 0 $. 

Explicit expressions for $ \gamma_1^{(F)} $ and $ \gamma_1^{(M)} $ in terms of $ \{ \Rey, \Har, \Prm \} $ are complicated and not particularly illuminating. However, their salient properties are revealed by means of the series approximations
\begin{subequations}
\label{eq:gamma1Har}
\begin{align}
\taga 
\gamma_1^{(F)} &= - \frac{ 32 \Har^2 \Reym }{ 15 ( 9 + 4 \Reym^2 ) } - \ii\left( 2 - \frac{13 \Har^2}{90} - \frac{ 16 \Har^2 }{ 5 ( 9 + 4 \Reym^2 ) } \right) + \ord( \Har^4 ), \\
\tagb
\gamma_1^{(M)} &= - \frac{ 2( 3 + \Har^2 ) }{ 3 \Reym } + \frac{ 32 \Har^2 \Reym }{15(9 + 4 \Reym^2 )} - \ii \left( \frac{ 2 }{ 3 } + \Har^2 \left( \frac{ 1 }{90} + \frac{ 16}{5(9 + 4 \Reym^2)} \right) \right) + \ord( \Har^4 )
\end{align}
\end{subequations}
and
\begin{subequations}
\label{eq:gamma1Pm}
\begin{align}
\nonumber
\gamma_1^{(F)} &= - \frac{ \Reym( ( \cosh( 4 \Har ) - 1 - 16 \Har^2 - 8 \Har^2 \cosh( 2 \Har ) ) \tanh( \Har ) + 16 \Har^3 ) }{ 64 \Har^3 \sinh^4( \Har / 2 ) \cosh^2( \Har ) } \\
\taga
& \quad -\ii ( 1 + \sech( \Har ) ) + \ord( \Prm^2 ), \\
\tagb
\gamma_1^{(M)} &= -\frac{ 2 \Har \coth( \Har ) }{ \Reym } - \ii \frac{  \cosh( \Har ) -2 \Har \csch( \Har ) + \sech( \Har ) }{ \cosh( \Har ) -1 } + \ord( \Prm ),
\end{align}
\end{subequations}    
respectively valid for small $ \Har $ and $ \Prm $. Inspecting~\eqref{eq:gamma1Har} and~\eqref{eq:gamma1Pm}, we deduce that among the two solutions $ \gamma_1^{(M)} $ is singular in the inductionless limit $ \Prm \searrow 0 $, whereas $ \gamma_1^{(F)} $ tends to the result~\eqref{eq:gamma1ZeroPmHartmann} for the first-order expansion coefficient for mode~F obtained in appendix~\ref{app:pertResultsZeroPm}. For small Hartmann numbers the mode associated with $ \gamma_1^{(M)} $, referred to in \S\ref{sec:mhdTopEnd} as the magnetic mode, has negative growth rate and  propagates with phase velocity close to the $ \langle U \rangle = 2/3 $ mean value of the Poiseuille profile. On the other hand, mode~F becomes neutral as $ \Har \searrow 0 $ (that is, $ \Real( \gamma_1^{(F)} ) \nearrow 0$), and propagates at twice the steady-state velocity at the free surface irrespective of the value of the magnetic Prandtl number. A similar procedure applied for problems with $ U = B = 0 $, but $ \Har \geq 0 $, yields
\begin{equation}
\label{eq:gamma1FS}
\gamma_1^{(F)} = 0, \quad \gamma_1^{(M)} = - 2 \Har \coth( \Har ) / \Reym,
\end{equation}
indicating that the magnetic mode is stable even in the absence of a steady-state flow, whereas mode~F becomes neutral to first order in $ \alpha $.

Because in insulating-wall problems $ \Real( \gamma_1^{(F)} ) $ is negative for all $ \Prm > 0 $ and $ \Har > 0 $, the second-order approximation in $ \alpha $ is not relevant for the stability of mode~F in the limit $ \alpha \searrow 0 $, and for this reason we will not pursue it here. On the other hand, the analysis for problems with a conducting wall, where, as discussed above, $ \ker( \mat{ A}_0 ) $ is one-dimensional, leads to the result that, as in inductionless problems, the leading-order coefficient in the expansion for $ \gamma $ with nonzero real part is $ \gamma_2 $, which, in addition to $ \Rey $, $ \Gal $, and $ \Har $, now depends on $ \Prm $. Solving for $ \Rey $ in the equation $ \gamma_2( \Rey, \Gal, \Har, \Prm ) = 0 $, we obtain 
\begin{multline}
\label{eq:criticalReMhdConducting}
\Reyb = 8 \cosh( \Har ) \sinh(\Har/2)^2 \Gal^{1/2} (\Har \cosh(\Har) - \sinh(\Har))^{1/2} \\
/ (\Har (2 (3 + (7 + 17 \Har^2) \Prm) \cosh(\Har) + ((13 \Har^2-12) \Prm -6) \cosh( 3 \Har) \\
+ (\Har^2 -1) \Prm \cosh(5 \Har) + \Har (2 (6 + 7 \Prm) \sinh(\Har) + (4 + 5 \Prm) \sinh(3 \Har) \\
- \Prm \sinh(5 \Har))))^{1/2}
\end{multline}
for the Reynolds number $ \Reyb $ at the bifurcation point of the neutral-stability curve, which in this case includes the $ \alpha = 0 $ axis. Moreover, the first-order coefficient $ \gamma_1 $ and, in turn, the phase velocity $ C_b $ at the bifurcation point, are found to be given by the same expressions as in inductionless problems, namely~\eqref{eq:gamma1ZeroPmHartmann} and~\eqrefb{eq:criticalSoft}, respectively. For small Hartmann numbers, and provided that $ \Prm $ is also small,  $ \Reyb / \Gal^{1/2} = (5 / 8 )^{1/2} + (191-25 \Prm ) \Har^2 / (168 \times 10^{1/2}) + \ord( \Har^4) $ increases quadratically with $ \Har $, but for strong magnetic fields $ \Reyb \sim ( \Gal / \Prm )^{1/2} \Har^{-1} $ becomes inversely proportional to the Hartmann number (\cf the exponentially increasing $ \Reyb $ in inductionless problems).

\bibliographystyle{jfm}
\bibliography{bibliography}

\end{document}